\documentclass[%
 superscriptaddress,
 amsmath,amssymb,
 aps,
 prb,
 twocolumn,
 longbibliography %, linenumbers
]{revtex4-2}

\usepackage{graphicx}% Include figure files
\usepackage{dcolumn}% Align table columns on decimal point
\usepackage{bm}% bold math
\usepackage{xcolor}
\usepackage{tikz}
\usepackage{xfrac}
\usepackage{blindtext}
\usepackage{times}
\usepackage{multirow}
\usepackage{scalerel}
\usepackage{siunitx}[=2021-04-09]

\usepackage{color}
\usepackage[colorlinks=true, urlcolor=blue, linkcolor=blue, citecolor=blue, pdftex]{hyperref}

\usepackage[normalem]{ulem}

\definecolor{darkblue}{HTML}{004D6B}
\definecolor{darkred}{HTML}{8c1515}
\definecolor{darkgreen}{HTML}{006400}
\usepackage{hyperref}
\hypersetup{
	colorlinks=true,
	urlcolor=darkred,
	citecolor=darkblue,
	linkcolor=darkred,
	breaklinks
}

\usepackage[caption=false]{subfig}

\graphicspath{{figures/}}

\def\bowtie{\scalerel*{\includegraphics{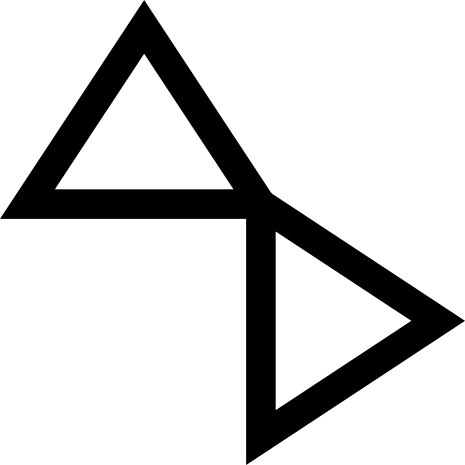}}{\sum}}
\newcommand{\avg}[1]{\left< #1 \right>} % for average
\renewcommand{\vec}[1]{\ensuremath{\mathbf{#1}}}

\begin{document}

%%%%%%%%%%%%%%%%%%%%%%
\title{Non-Coplanar Magnetic Orders in Classical Square-Kagome Antiferromagnets}
%\title{Non-Coplanar Magnetic Orders on the Shuriken Lattice}

%%%%%%%%%%%%%%%%%%%%%%
\author{Martin Gemb\'e}
\affiliation{Institute for Theoretical Physics, University of Cologne, 50937 Cologne, Germany}
\author{Heinz-J\"{u}rgen Schmidt} 
\affiliation{Fachbereich Physik, Universit\"{a}t Osnabr\"{u}ck, 49069 Osnabr\"{u}ck, Germany}
\author{Ciar\'an Hickey}
\affiliation{Institute for Theoretical Physics, University of Cologne, 50937 Cologne, Germany}
\affiliation{School of Physics, University College Dublin, Belfield, Dublin 4, Ireland}
\affiliation{Centre for Quantum Engineering, Science, and Technology, University College Dublin, Dublin 4, Ireland}
\author{Johannes Richter}
\affiliation{Institut f\"{u}r Physik, Otto-von-Guericke-Universit\"{a}t Magdeburg, 39016 Magdeburg, Germany}
\affiliation{Max-Planck-Institut f\"{u}r Physik Komplexer Systeme, N\"{o}thnitzer Stra{\ss}e 38, 01187 Dresden, Germany}
\author{Yasir Iqbal}
\affiliation{Department of Physics and Quantum Centre of Excellence for Diamond and Emergent Materials (QuCenDiEM), Indian Institute of Technology Madras, Chennai 600036, India}
\author{Simon Trebst}
\affiliation{Institute for Theoretical Physics, University of Cologne, 50937 Cologne, Germany}

\date{\today}

%%%%%%%%%%%%%%%%%%%%%%
\begin{abstract}
Motivated by the recent synthesis of a number of Mott insulating square-kagome materials, we explore the rich phenomenology 
of frustrated magnetism induced by this lattice geometry, also referred to as the squagome or shuriken lattice. 
On the classical level, square-kagome antiferromagnets are found to exhibit extensive degeneracies, order-by-disorder, 
and non-coplanar ordering tendencies, which we discuss for an elementary, classical Heisenberg model with nearest-neighbor 
and cross-plaquette interactions.
Having in mind that upon introducing quantum fluctuations non-coplanar order can melt into chiral quantum spin liquids, 
we provide detailed information on the multitude of non-coplanar orders, including some which break rotational symmetry 
(possibly leading to nematic quantum orders), as well as a number of (incommensurate) spin spiral phases. 
Using extensive numerical simulations, we also discuss the thermodynamic signatures of these phases, which often show 
multi-step thermal ordering. Our comprehensive discussion of the classical square-kagome Heisenberg model, 
often drawing comparisons to the conventional kagome antiferromagnet, sets the stage for future explorations of
quantum analogs of the various phases, either conceptually such as in quantum spin-1/2 generalizations of our model or experimentally such as in the Cu-based candidate materials.
\end{abstract}

\maketitle

%%%%%%%%%%%%%%%%%%%%%%
\section{Introduction}

Classical Heisenberg spin models with frustrated interactions are known to host a rich variety of magnetic orders such as collinear, coplanar, or helimagnetic states~\cite{Messio-2011,Ghosh-2019,Balla-2020}. 
Of particular interest are non-Bravais lattices (with more than one atom per unit cell) 
which offer the possibility of stabilizing {\em non-coplanar} magnetic ordering~\cite{Messio-2011,Iqbal-2019}. 
Such non-coplanar ground states distinguish themselves from other magnetic orders by exhibiting a scalar spin chirality.
Notably, the spontaneous breaking of such a $\mathbb{Z}_2$ (chiral) symmetry manifests itself in a finite-temperature phase transition -- even in two spatial dimensions~\cite{Domenge-2005}, while other types of magnetic order are, for the Heisenberg models of interest here, subject to the Mermin-Wagner theorem \cite{Mermin-1966}. The latter implies that, for the thermodynamic limit of infinite system size, any fluctuation-driven phase transition in two spatial dimensions, which only breaks the continuous spin symmetry, occurs at zero temperature. Finite systems (often explored in numerical simulations) will, however, exhibit a thermal crossover to magnetic order at some finite temperature, with the accompanying entropy release leading to a peak in the specific heat.
This should be distinguished from cooperative paramagnetic phases~\cite{Villain-1979}, which defy magnetic ordering tendencies due to the existence of substantial residual entropies, even at temperatures orders of magnitude below the coupling scales and for infinitely large systems. In classical Heisenberg models, the thermodynamics of such cooperative paramagnetic phases, also referred to as classical spin liquids, is typically signified by a plateau in the specific heat~\cite{Zhitomirsky-2008}. 

 The two themes, the formation of non-coplanar magnetic order and cooperative paramagnetic phases, are conceptually tied when looking at their quantum mechanical counterparts. By melting non-coplanar magnetic order via quantum fluctuations, 
 e.g. by going to small spins such as $S=1/2$, one could possibly restore spin rotational symmetry, i.e., realize a non-magnetic quantum ground state. If the chiral symmetry breaking present in the parent classical magnetic order would persist (at some finite temperature scale), 
 one would realize a much sought after chiral quantum spin liquid phase~\cite{Bieri-2016,Hickey2016,Hickey2017}. 
 Similarly, the inclusion of quantum fluctuations on cooperative paramagnetic ground states provides another promising route towards realizing unconventional quantum phases such as quantum spin liquids, valence bond crystals~\cite{Schmoll-2022,Astrakhantsev2021,Richter-2023_mag,Ralko-2015,Rousochatzakis-2013}, or spin and lattice nematics~\cite{Lugan-2019}.

\begin{figure}[t]
	\centering
	\includegraphics[width=1.0\columnwidth]{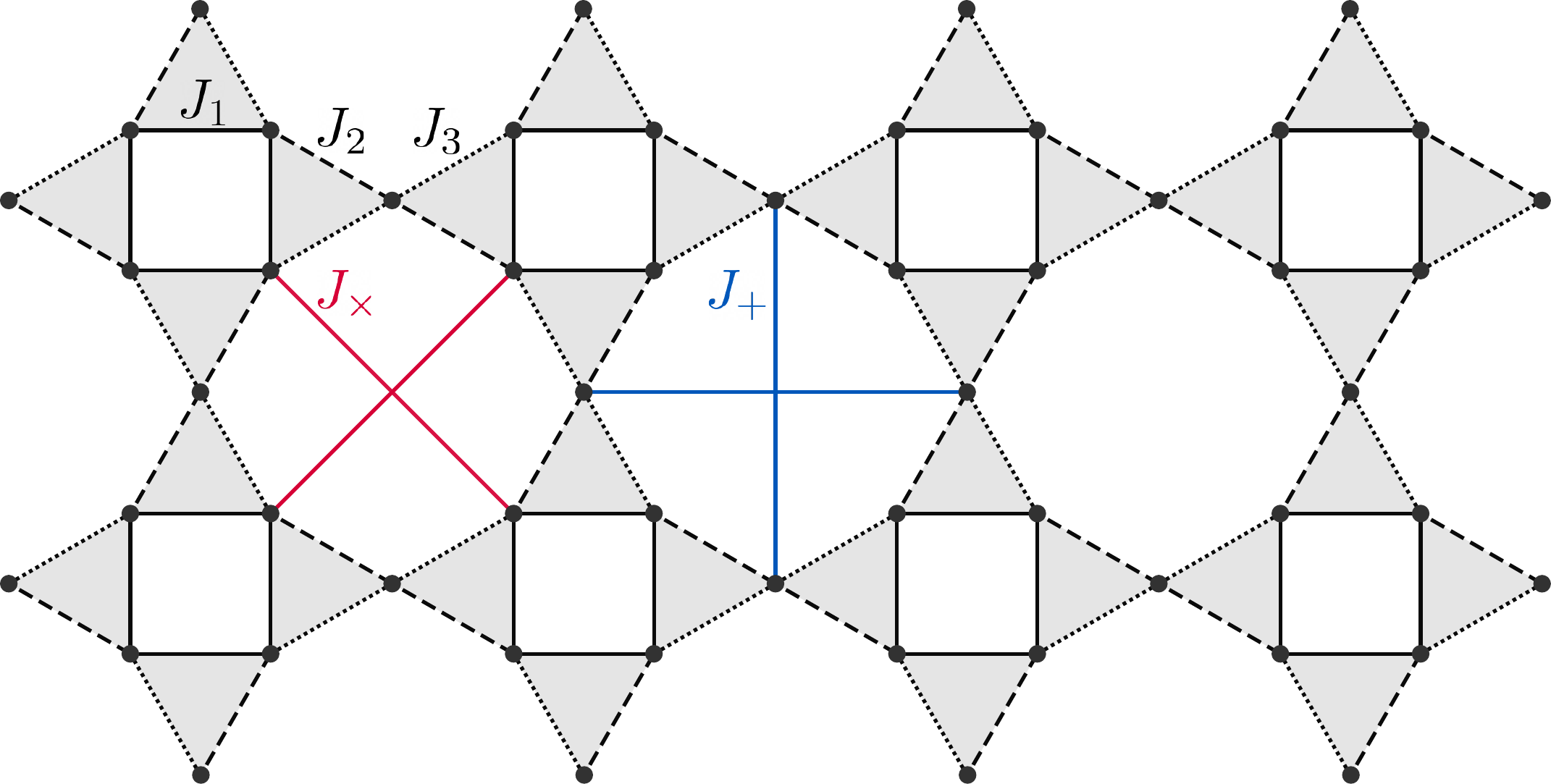}
	\caption{\textbf{Square-kagome lattice and interactions.} 
	The square-kagome lattice, also referred to as the squagome or shuriken lattice in the literature, 
	consists of two sets of topologically distinct sites -- square sites and bow-tie sites. 
	Nearest-neighbor interactions are referred to as $J_1$ (square) and $J_2$, $J_3$ (bow-tie), respectively. 
	Additionally, we introduce further neighbor cross-plaquette $J_+$-bonds ($J_{\times}$-bonds) as indicated in blue (red). 
	The unit cell contains six spins, namely the four square sites and two bow-tie sites of a single \textit{shuriken} star.}
	\label{Fig:lattice}
\end{figure}

An ideal playground to explore this physics in experiment has come in 
the arrival of materials based on the novel square-kagome lattice geometry~\cite{Siddharthan2001}, whose potential to host intricately textured magnetic ground states or quantum spin liquid phases 
is currently under much investigation~\cite{Yakubovich-2021}. 
Indeed, no sign of long-range magnetic order down to 50 mK has been observed in the spin $S=1/2$ Cu$^{2+}$ based materials KCu$_6$AlBiO$_4$(SO$_4$)$_5$Cl~\cite{Fujihala2020} and Na$_6$Cu$_7$BiO$_4$(PO$_4$)$_4$[Cl,(OH)]$_3$~\cite{Liu-2022} despite having large negative Curie-Weiss temperatures of $\SI{-237}{\kelvin}$ and $\SI{-212}{\kelvin}$, respectively. On the other hand, their 
sister compounds KCu$_7$(TeO$_4$)(SO$_4$)$_5$Cl and NaCu$_7$(TeO$_4$)(SO$_4$)$_5$Cl develop antiferromagnetic order~\cite{Markina-2022,Murtazoev-2023}, while related compounds Rb$_7$(TeO$_4$)(SO$_4$)$_5$Cl and Cs$_7$(TeO$_4$)(SO$_4$)$_5$Cl do not show sign of magnetic order down to 2K~\cite{Murtazoev-2023}. 
In general, the model Hamiltonians for these materials can host up to three symmetry inequivalent couplings on the three sides of the elementary triangles as well as potential longer-range Heisenberg couplings across the octagonal plaquettes, see Fig.~\ref{Fig:lattice}, whose presence could be resolved by ab-initio density functional theory calculations.
The latter are, in contrast to diagonal square couplings, a key ingredient towards stabilizing non-coplanar magnetic orders on the classical level and, potentially, chiral spin liquids in the quantum realm. This is in a spirit similar to the diagonal couplings across hexagons on the kagome lattice which are known to yield non-coplanar spin structures dubbed cuboc orders~\cite{Domenge-2005,Janson-2008,Janson-2009,Messio-2011}. The details of these non-coplanar states are, by their very nature, rather sensitive to the underlying lattice geometry with unique features expected for the square-kagome lattice geometry at hand. 

The purpose of this manuscript is to set the staging ground for future explorations of square-kagome antiferromagnets by providing a comprehensive discussion of their physics in the classical realm and identifying its unique features.
To this end, we investigate the ground state and thermodynamics of the classical Heisenberg model on the square-kagome lattice in the presence of nearest-neighbor $(J_{1}, J_{2}, J_{3})$ couplings as well as cross-plaquette interactions inside the octagons, $J_\times$ and $J_+$, as indicated in Fig.~\ref{Fig:lattice}. Our analysis is based on extensive classical Monte Carlo simulations and an analytical construction of ground states (beyond the Luttinger-Tisza approach), which is shown to be rendered exact for some orders. As summarized in the phase diagram of Fig.~\ref{Fig:phase_diagram_afm} below, we find a rich variety of non-coplanar magnetic orders with cuboctohedral symmetry, including types of cuboc order not found on the kagome lattice (or any other known lattice geometry). In addition, we report a multitude of non-coplanar incommensurate spirals, in addition to commensurate coplanar orders. 
Exploring the quantum analogs of these phases in the future, either conceptually such as in quantum spin-1/2 generalizations of our model or experimentally such as in the Cu-based candidate materials, might prove fruitful in identifying chiral quantum spin liquids. \\

The remainder of this manuscript is structured as follows. To begin with, we discuss the nearest neighbor Heisenberg model on the square-kagome lattice in Sec.~\ref{sec:nn_model}, where we analyze its ground states, finite-temperature physics, and spin-spin correlations. Afterwards, in Sec.~\ref{sec:full_model}, we introduce additional further-neighbor cross-plaquette interactions and study the resulting rich phase diagram for fixed, antiferromagnetic nearest-neighbor interactions, looking at each phase separately in great detail. Finally, we briefly discuss results for both mixed and pure ferromagnetic nearest-neighbor interactions.

%%%%%%%%%%%%%%%%%%%%%%%%%%%%%%%%%%%%%%%%%%%%%%%%%%%%%%%%%%%%%%%%%%%%%%%
\section{Nearest-Neighbor Model}
\label{sec:nn_model}

With just nearest-neighbor Heisenberg interactions the square-kagome model shares much of the same physics as the nearest-neighbor kagome Heisenberg model. Below, we briefly summarize some of the known results for the ground states of the square-kagome model that can be inferred from the conventional kagome antiferromagnet. 
We then move on to discuss its finite temperature physics and explore critical fluctuations, going beyond what has been studied for the conventional kagome scenario.

\begin{figure*}[t]
	\centering
	\includegraphics[width=1.0\linewidth]{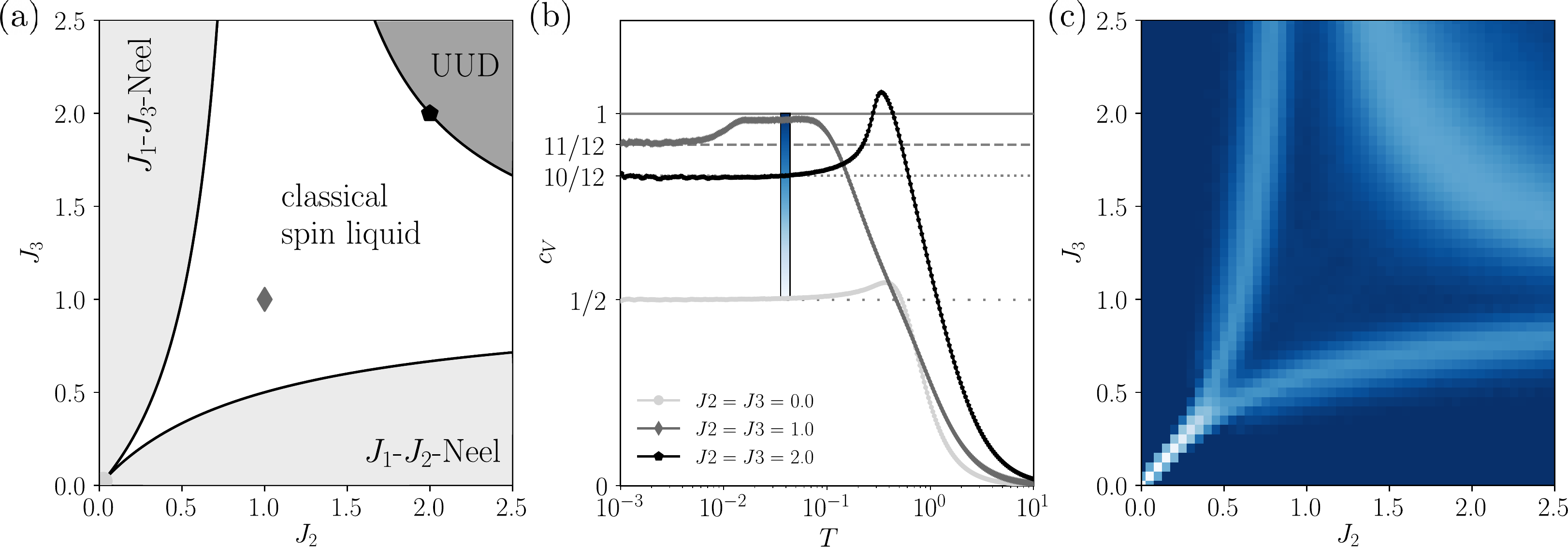}
	\caption{\textbf{The nearest-neighbor model.}
	(a) Zero-temperature phase diagram as a function of nearest-neighbor couplings $J_2$ and $J_3$, 
		reproduced from Ref.~\cite{Morita-2018}. 
	(b) Specific heat traces for three representative points along the $J_2 = J_3$ diagonal in the phase diagram of panel (a).
	     The one for the isotropic point, $J_1=J_2=J_3=1$, closely resembles the well-studied specific heat trace 
	     of the kagome AFM \cite{Chalker-1992,Zhitomirsky-2008}.
	(c) Specific heat scan across the phase diagram of panel (a) at fixed, low temperature of $T=0.04$ (in units of $J_1=1$).
	}
	\label{Fig:classical_phase_diagram}
\end{figure*}

%%%%%%%%%%%%%%%%%%%%%%%%%%%%%%%%%%%%%%%%%%%%%%
\subsection{Ground states}

The Hamiltonian with nearest-neighbor couplings can be written as
\begin{align}
  \mathcal{H} = \sum_{\avg{i,j}\in a} J_a\,  \vec{S}_i \cdot  \vec{S}_j \,,
\end{align}
where $a=\{1,2,3\}$ runs over the three different types of bonds, as in Fig.~\ref{Fig:lattice}. However, it can be more easily understood by rewriting it as
\begin{align}
    \notag \mathcal{H} = \sum_{i,j,k\in\triangle} & \left[  \frac{1}{2} \left( \sqrt{\frac{J_1 J_3}{J_2}} \vec{S}_i + \sqrt{\frac{J_1 J_2}{J_3}} \vec{S}_j + \sqrt{\frac{J_2 J_3}{J_1}} \vec{S}_k \right)^2 \right.   \\
& \left. \,\,\,\, - \frac{1}{2} \left( \frac{J_1 J_3}{J_2} + \frac{J_1 J_2}{J_3} + \frac{J_2 J_3}{J_1} \right)  \right] \,,
\end{align}
where the sum is now over all four types of elementary triangles of the lattice and we have assumed all couplings to be antiferromagnetic, $J_a > 0$ (such that all of the arguments of the square roots are positive). The Hamiltonian of the kagome Heisenberg model can be written in the same form, but with a key distinction being the number and nature of elementary triangles that are summed over. The new form of the Hamiltonian allows us to easily construct a special class of classical ground states. These are the states that satisfy the constraint
\begin{equation}
	\left(  \sqrt{\frac{J_1 J_3}{J_2}} \vec{S}_i + \sqrt{\frac{J_1 J_2}{J_3}} \vec{S}_j + \sqrt{\frac{J_2 J_3}{J_1}} \vec{S}_k \right) = 0\,. 
	\label{eqn:NNconstraint}
\end{equation}
where $\vec{S}_i, \vec{S}_j$ are the spins on the squares and $\vec{S}_k$ is the spin on the bow-ties. To see that such states are indeed ground states note that the Hamiltonian, for a given set of parameters, is the sum of a squared term, which is always positive, and a constant term. The minimal possible energy is thus obtained when the squared term is precisely zero, i.e., the constraint above. However, there is the additional constraint that the spins at each site are all properly normalized, $\vert\vec{S}_i\vert=1 \,\,\forall \,\,i$. Satisfying both of these constraints is only possible within a restricted region of parameter space, namely 
\begin{equation}
	-2 \leq  \frac{J_2 J_3}{J_1^2} - \left( \frac{J_2 }{J_3} + \frac{J_3 }{J_2} \right) \leq +2 \,.
\end{equation}
Thus, within this parameter region, all spin configurations satisfying the constraint (\ref{eqn:NNconstraint}) on each and every triangle are guaranteed to be bona fide ground state spin configurations. This includes both globally coplanar and non-coplanar spin configurations as, though the three spins in each triangle are constrained to lie within the same plane, it is not necessary that the planes for different triangles are the same. The resulting unusual and highly degenerate phase is the classical spin liquid indicated in the phase diagram of Fig.~\ref{Fig:classical_phase_diagram}(a) and previously reported in Ref.~\cite{Morita-2018}, which shares the same qualitative physics as the classical spin liquid found in the distorted kagome version of the model~\cite{PhysRevB.76.094421,Yavorskii-2007,Schnyder-2008,Hering-2022}.    

At the isotropic point, $J_1=J_2=J_3$, the spins in each triangle lie within the same plane at an angle of exactly $120^\circ$ away from one another. On the other hand, away from the isotropic point, e.g.~along the diagonal line $J_2=J_3$, the coplanar spin configuration for a single triangle obeying the constraint can be written as
\begin{align}
\notag \vec{S}_i &= (-J_2/(2J_1),\,-\sqrt{4-(J_2/J_1)^2}/2,\,0), \\
\notag \vec{S}_j &= (-J_2/(2J_1),\,+\sqrt{4-(J_2/J_1)^2}/2,\,0) ,\\
\vec{S}_k &= (1,\,0,\,0) \,, \label{eqn:sampleconfig}
\end{align}
where we have fixed the spins to lie in the $xy$-plane for simplicity. As one increases $J_2$ from $J_2=0$ to $J_2=2J_1$, the angle $\theta$ between $\vec{S}_k$ (the spin on the bow-tie) and the other two spins (on the squares) increases from $\pi/2$ to $\pi$, passing through $2\pi/3$ exactly at the isotropic point $J_2=J_1$~\cite{Rousochatzakis-2013}. 

% figure moved for placement in text
\begin{figure*}[t]
	\centering
	\includegraphics[width=1.0\linewidth]{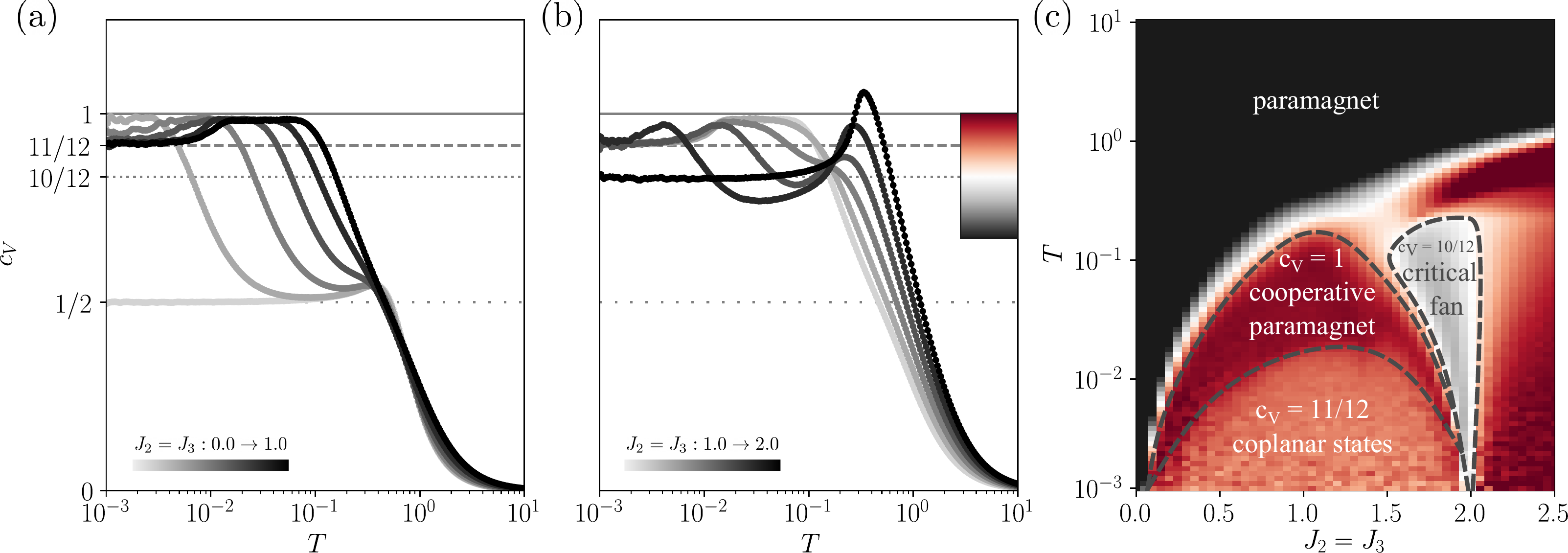}
	\caption{\textbf{The nearest-neighbor model II.}
    (a) Specific heat traces for $0.0 \leq J_2 = J_3 \leq 1.0$ (in steps of 0.2).
	(b) Specific heat traces for $1.0 \leq J_2 = J_3 \leq 2.0$. Of special interest is the curve for $J_2 = J_3 = 2.0$ which is the only one which converges to a value of $c_V = 10/12$.
	(c) Finite temperature specific heat along the diagonal $J_2 = J_3$ illustrating the $c_V = 1$ plateau of the cooperative paramagnet (dark red area), the coplanar phase with $c_V = 11/12$ (light red area), the critical fan with $c_V \leq 10/12$ (grey area), and the paramagnetic phase (black region).
	}
	\label{Fig:cv_finite_T}
\end{figure*}
For globally coplanar spin configurations, which are the relevant configurations at the lowest temperatures as we will see in the next section, as well as the constraints already mentioned, there is one additional form of constraint \cite{PhysRevB.76.094421}. It is related to how the spins in the triangles around the square and octagonal plaquettes of the lattice are arranged. For globally coplanar ground state spin configurations, we can define on each triangle a chirality variable, $\eta_a=\pm1$, which encodes whether the spins rotate clockwise or anti-clockwise as one goes from say $i$ to $j$ to $k$ \cite{Chandra1993} (keep in mind that the angles between the spins in each triangle are fixed, e.g.~all three angles are fixed to $2\pi/3$ at the isotropic point). Now, starting from an initial spin which points in some specific direction, if one travels in a closed loop on the lattice then one must return back to that same initial spin pointing in that same specific direction. However, as one travels along each bond the chirality variables dictate in which direction the spins rotate, and so in order to get back to the same initial spin there is a constraint on the sum of the chirality variables along the closed loop. At the isotropic point, we require $\sum_{a} \eta_a = 0$ for the four triangles surrounding a square, and $\sum_{a} \eta_a = 0,\pm 6$ for the eight triangles surrounding an octagon. Away from the isotropic point there will in general be more stringent constraints as the angle between spins is no longer $2\pi/3$, but instead some angle incommensurate with respect to $2\pi$. This results, for a given lattice size, in a smaller number of allowed coplanar ground state spin configurations, just as in the analogous kagome case \cite{PhysRevB.76.094421}.    

Outside of the classical spin liquid region there are two distinct N\'eel states in the phase diagram of Fig.~\ref{Fig:classical_phase_diagram}(a), depending on whether $J_2$ or $J_3$ dominates. In each phase, spins are arranged antiferromagnetically along the bonds with the dominant coupling, and ferromagnetically along the weaker bonds. On the other hand, for dominant $J_2=J_3$, there is an up-up-down (UUD) state,
sometimes also referred to as a Lieb ferrimagnet \cite{LiebMattis1962}, in which the spins on the squares point along one direction while the spins on the bow-ties point in the opposite direction 
(this can be seen from Eq.~\eqref{eqn:sampleconfig} with $J_2=2J_1$, the critical point between the classical spin liquid and UUD state). 
 
%%%%%%%%%%%%%%%%%%%%%%%%%%%%%%%%%%%%%%%
\subsection{Finite-temperature physics}
\label{sec:II_finite_T}

The unusual nature of the classical spin liquid phase is naturally revealed by examining how the specific heat behaves as a function of temperature. For two-dimensional Heisenberg models with finite-range exchange interactions, true long-range magnetic order cannot set in at any non-zero temperature, as laid out in the Mermin-Wagner theorem~\cite{Mermin-1966}. However, it is possible for quasi-long-range order to develop, signaled by a peak in the specific heat. Our discussion, in the following, of such quasi-long-range ``orders'' is based on an analysis of the symmetry of spin-spin correlations at distances shorter than the correlation radius. In the case of non-coplanar, i.e., chiral orders, however, a true thermal phase transition associated with the breaking of discrete symmetries occurs. 

At the lowest temperatures, one generically expects that, in the thermodynamic limit, the specific heat per site $c_V \rightarrow 1$ as $T\rightarrow 0$. This is because each spin is free to fluctuate about its ordered ground state in two orthogonal directions on the unit sphere. These two quadratic modes each contribute $(1/2)\cdot T$, as dictated by classical equipartition, to the energy and thus $1/2$ to the specific heat (setting $k_B=1$). However, as first discussed in the context of the kagome antiferromagnet \cite{Chalker-1992}, this simple counting breaks down within the classical spin liquid region, as well as in an extended finite temperature fan about the critical lines in the phase diagram. The breakdown is due to the entropic selection of a subset of ground state spin configurations, those which carry the largest entropy and thus the lowest free energy at finite temperature. Classical fluctuations about this favored subset include one or more zero modes at the harmonic level, which contribute $1/4$ (i.e.~they are quartic modes), rather than $1/2$, to the specific heat. The deviation of the low-temperature specific heat from $c_V \rightarrow 1$ thus serves as a signature of this phenomenon of thermal order-by-disorder. 

In Fig.~\ref{Fig:classical_phase_diagram}(b), we show the specific heat as a function of temperature for three special parameter points along the diagonal line $J_2=J_3$, with a number of curves in between shown in Fig.~\ref{Fig:cv_finite_T}(a) and (b). (i) First, starting with  $J_2=J_3=0$, we have the trivial limit of fully disconnected squares. Alternatively, one can think of this limit as consisting of decoupled four-site periodic Heisenberg chains. The spins order in a simple antiferromagnetic arrangement within each square. There are 8 quadratic modes per square, minus two due to the global rotational symmetry of the antiferromagnetic moment. This leaves us with 6 independent quadratic modes, and a contribution to the specific heat per site as $c_V \rightarrow [6\cdot (1/2)]/6 = 1/2$ as $T\rightarrow 0$ \cite{Schmidt2022b}. (ii) At the isotropic point, there are three distinct regimes, which share the same physics as the isotropic kagome model at finite temperatures~\cite{Zhitomirsky-2008} (the similarity even extends to the spin-$1/2$ quantum case \cite{Richter-2022}). There is the usual high-temperature paramagnetic region, followed by a cooperative paramagnetic regime and finally a coplanar state at the lowest temperatures. These three regimes can be observed throughout the classical spin liquid phase. The cooperative paramagnet is clearly distinguished by a plateau in the specific heat with $c_V \approx 1$. Within this temperature window, the system fluctuates between the full (extensive) number of states within the ground state manifold that satisfy the constraint in Eq.~\eqref{eqn:NNconstraint}. At lower temperatures, within the coplanar states, fluctuations select the subset of globally coplanar states within the ground state manifold via the entropic-driven order-by-disorder mechanism. This is accompanied by $c_V \rightarrow  [10\cdot (1/2) + 2\cdot (1/4)]/6 = 11/12$ as $T\rightarrow 0$ due to the presence of one zero mode per triangle (thus two zero modes per unit cell) within the spectrum of classical harmonic fluctuations (very similar to the conventional kagome case \cite{Chalker-1992}). (iii) At $J_2=2J_1$ we have the transition between the classical spin liquid and UUD state. Precisely at this critical point an additional zero mode per triangle leads to $c_V \rightarrow [8\cdot (1/2) + 4\cdot (1/4)]/6 = 10/12$ as $T\rightarrow 0$.     

\begin{figure*}[t]
	\centering
	\includegraphics[width=1.0\linewidth]{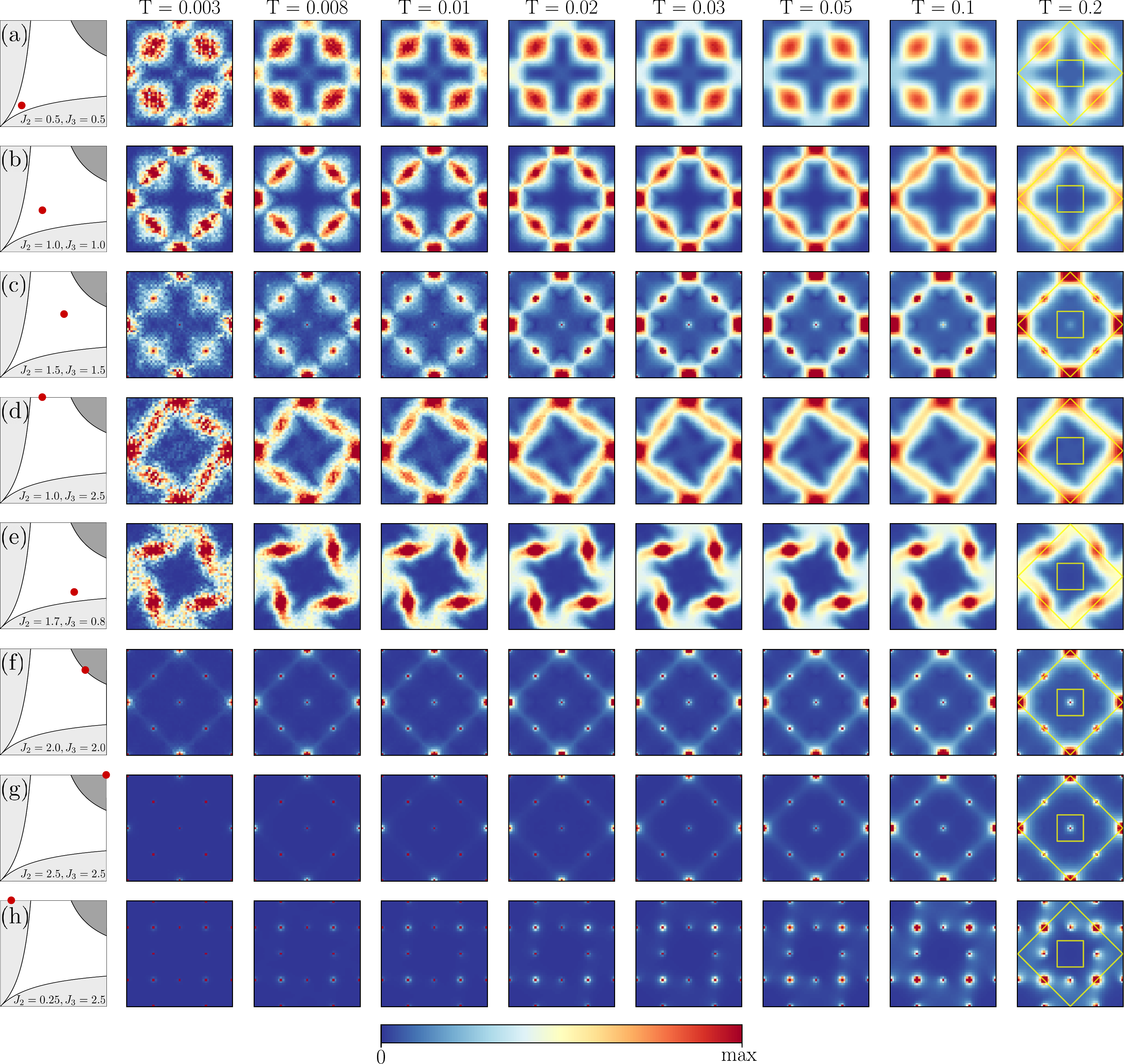}
	\caption{\textbf{Structure factors of the $J_1$-$J_2$-$J_3$ model.} Structure factors for a system with 864 Spins ($L = 12$) for different temperatures between T = 0.003 and T = 0.2 (left to right) and for different values of $J_2$ and $J_3$ as indicated in the phase diagram in the left column (top to bottom). The shown squares extent in reciprocal space from $-4\pi$ to $4\pi$ in both dimensions. Rows (a) to (c) show structure factors on the diagonal $J_2 = J_3$ within the classical spin liquid regime, especially for the isotropic point $J_1 = J_2 = J_3 = 1.0$ in row (b), which agrees well with the result obtained via large-$N$ analysis in \cite{Astrakhantsev2021}. Rows (d) and (e) as well show structure factors in the classical spin liquid regime, but off the diagonal with different values for $J_2$ and $J_3$. All structure factors within the classical spin liquid regime show sharp pinch point-like features upon entering the cooperative paramagnetic regime at around $T = 0.01$, which broaden with increasing temperature. The statistical noise at lower temperatures is due to freezing in Monte Carlo sampling. The lower three rows (f) to (h) display structure factors outside the classical spin liquid regime, namely directly on the transition line to the UUD phase, $J_2 = J_3 = 2.0$, in (f), within the UUD phase, $J_2 = J_3 = 2.5$, in (g), and in the $J_1$-$J_3$-N\'{e}el ordered phase, $J_2 = 0.25$ and $J_3= 2.5$, in (h).} 
	\label{Fig:structure_factors}
\end{figure*}

The evolution of the specific heat between these three special points, as one increases $J_2=J_3$, is shown in Fig.~\ref{Fig:cv_finite_T}. Starting from the trivial limit, $J_2=J_3=0$, the width of the plateau that develops at $c_V \approx 1/2$ shrinks as the interaction scale that couples squares, $J_2=J_3$, grows. At low temperatures, the characteristic two temperature regimes emerge, with the cooperative paramagnet and coplanar regimes 
giving rise to plateaus at $c_V=1$ and $11/12$, respectively. The $1/2$-feature disappears completely at the isotropic point. 
As one moves past the isotropic point, the width of the cooperative paramagnet plateau starts to decrease, until it eventually disappears completely at the critical point $J_2=2J_1$. 
At the same time, in the temperature regime above the cooperative paramagnet, there is a substantial drop in the specific heat as a result of increasing fluctuations on approaching the critical point (which damps the entropy loss in this temperature regime and thus the specific heat). 
This can be more clearly seen in Fig.~\ref{Fig:cv_finite_T}(c), where the drop manifests itself as a {\em finite temperature fan}, emanating from the zero-temperature critical point -- reminiscent of the fan-like structure at quantum critical points \cite{Vojta-2003}. 

%%%%%%%%%%%%%%%%%%%%%%%%%%%%%%
\subsection{Spin-spin correlations}

To explore the formation of quasi-long-range order and cooperative paramagnetic phases in the absence of any true thermal phase transitions, we turn to the static spin structure factor, i.e.\ the Fourier transform of the equal-time real-space spin-spin correlations. Sharp peak-like features indicate the formation of quasi-long-range order, while the tell-tale signatures of cooperative paramagnets are pinch points in momentum space, which map to algebraically decaying correlations in real space  \cite{Henley-2010}. 
In Fig.~\ref{Fig:structure_factors}, we provide a comprehensive overview of the static structure factor over a wide range of temperatures and parameter points. In some cases, such as rows (g) and (h), the ground state is quasi-long-range-ordered and sharp peaks are clearly visible at the corresponding ordering wavevectors. In other cases, i.e.~within the classical spin liquid parameter region, we observe a complex redistribution of weight as the system passes through the three distinct finite temperature regimes. 

For the isotropic point ($J_{2}=J_{3} = 1.0$), the $S(\mathbf{q})$ at the highest temperature shown in Fig.~\ref{Fig:structure_factors}(b) only displays a broad diffuse profile corresponding to a weakly correlated thermal paramagnet. Inside the cooperative paramagnetic regime, at $T=0.1$ and $0.05$, pinch points (with a finite-width set by the inverse correlation length) appear between the square and lobed shaped regions of  stronger relative intensity. Their presence signifies the approximate fulfilment of the $\mathbf{S}_{\triangle}=0$ constraint of Eq.~\ref{eqn:NNconstraint} and thus the onset of strong correlations between spins within triangular plaquettes. Upon cooling further, at $T=0.03$ and $0.02$, the intensity is redistributed to the centers of squares and lobes located at $\mathbf{q}=(4\pi,0)$ and $\mathbf{q}=(2\pi,2\pi)$ (and symmetry related points), respectively. Since one is still in the cooperative paramagnetic regime characterized by a dipolar $\sim1/r^2$ decay of spin-spin correlations (at distances smaller compared to the correlation length), these high intensity features do not correspond to Bragg peaks. Finally, once order-by-disorder kicks in at $T\lesssim 0.01$ selecting coplanar states, we notice the disappearance of spectral weight at the location of the pinch point as well as the absence of narrow necks connecting the squares with the lobes. The presence of well-defined maxima at the aforementioned points indicates enhanced correlations of the $120^{\circ}$ $\mathbf{q}=\mathbf{0}$ type order~\cite{richter2009heisenberg,Astrakhantsev2021}, in contrast to the conventional kagome antiferromagnet which favors $\sqrt{3}\times\sqrt{3}$ correlations \cite{Huse-1992,Harris-1992,Chern-2013}. While the kagome antiferromagnet develops long-range dipolar magnetic order of the $\sqrt{3}\times\sqrt{3}$ type in the limit $T\to0$~\cite{Chern-2013},  it remains to be established on the square-kagome lattice whether true long-range $\mathbf{q}=\mathbf{0}$ dipolar ordering of $120^{\circ}$ type asymptotically develops as $T\to 0$.

Away from the isotropic point, but still on the line $J_{2}=J_{3}$, we first note that the cooperative paramagnetic and coplanar temperature regimes are pushed down to smaller $T$ and shrink in extent [see Fig.~\ref{Fig:cv_finite_T}]. At $J_{2}=J_{3}=0.5$, within the cooperative paramagnetic regime ($0.008\lesssim T\lesssim 0.05$), the pinch points are seen to be present, and with the principal spectral weight at the center of the lobes $\mathbf{q}=(2\pi,2\pi)$ (and symmetry related points) which progressively increases, together with a relatively weaker signal at the centers of the squares $\mathbf{q}=(4\pi,0)$ (and symmetry related points) [see Fig.~\ref{Fig:structure_factors}(a)]. However, upon entering the coplanar regime, at $T=0.003$, an equally strong maximum develops at the $\mathbf{q}=(4\pi,0)$ type points, but in contrast to the isotropic point, these are not indicative of enhanced $\mathbf{q}=\mathbf{0}$ correlations of the $120^{\circ}$ type. Similar observations hold true at $J_{2}=J_{3}=1.5$ [see Fig.~\ref{Fig:structure_factors}(c)] with the noticeable difference being the presence of a finite spectral weight at the Brillouin zone centre, being more pronounced in the intermediate temperature, i.e., cooperative paramagnetic regime. Moving away from the symmetric line, i.e., $J_{2}\neq J_{3}$ but still inside the degenerate manifold region, one observes that the $S(\mathbf{q})$ are only rotationally invariant possibly reflective of the underlying symmetries of the incipient magnetic order in the limit $T\to 0$. Upon cooling, a progressive redistribution of spectral weight occurs leading to the appearance of new soft maxima in the coplanar regime at $(J_{2},J_{3})=(1.0,2.5)$ [see Fig.~\ref{Fig:structure_factors}(d)] while at $(J_{2},J_{3})=(1.7,0.8)$ [see Fig.~\ref{Fig:structure_factors}(e)] interestingly a similar intensity distribution prevails across all temperatures. Finally, inside the magnetically ordered regions, the $S(\mathbf{q})$ become more sharply peaked, as expected, and interestingly the $S(\mathbf{q})$ at $(J_{2},J_{3})=(2.0,2.0)$ [see Fig.~\ref{Fig:structure_factors}(f)] and $(J_{2},J_{3})=(2.5,2.5)$ [see Fig.~\ref{Fig:structure_factors}(g)] resemble those inside the disordered regime at $(J_{2},J_{3})=(1.5,1.5)$ seen in Fig.~\ref{Fig:structure_factors}(c) along the $J_{2}=J_{3}$ axis as if pre-empting the UUD order which onsets for $J_{2}=J_{3}\geq2.0$. Inside the N\'eel phase at $(J_{2},J_{3})=(0.25,2.5)$, in Fig.~\ref{Fig:structure_factors}(h) there is no spectral weight at the centre of the Brillouin zone as expected, and instead we find dominant peaks at the $(2\pi,2\pi)$ (and symmetry related) points, and subdominant peaks at $(2\pi,0)$ (and symmetry related) points.

%%%%%%%%%%%%%%%%%%%%%%%%%%%%%%%%%%%%%%%%%%%%%%%%%%%%%%%%%%%%%%%
% grand phase diagram figure
\begin{figure*}[th!]
	\centering
	\includegraphics[width=1.0\linewidth]{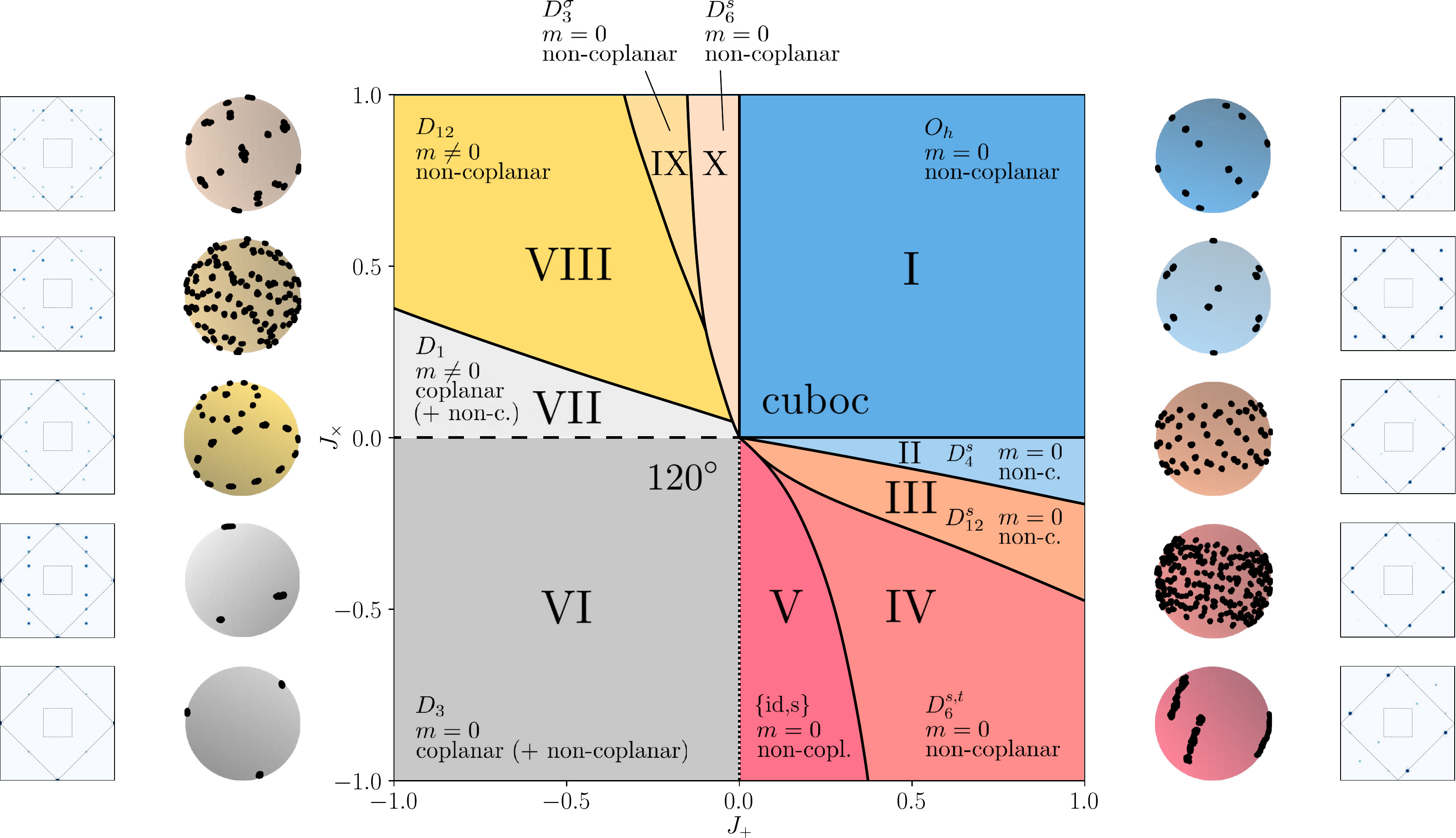}
	\caption{\textbf{Phase diagram in the presence of cross-octagonal couplings.} 
	At the center we show the phase diagram with ten different phases (labeled I-X) as a function of $J_+$ and $J_{\times}$
	for a nearest neighbor model with isotropic, antiferromagnetic interactions, i.e.\ $J_1 = J_2 = J_3 = 1$. (A companion 
	phase diagram for ferromagnetic nearest neighbor couplings is shown in Fig.~\ref{Fig:phase_diagram_fm} below.)
	Besides indicating the phase boundaries (solid, dashed and dotted lines), the phases are described by 
	symmetry (with $D_n$ referring to the dihedral group of order $n$ and $O_h$ to full octahedral symmetry), coplanarity, 
	and magnetization of their ground states.
	As a quick visualization of their distinct nature, we show common origin plots and spin structure factors for each phase
	left and right around the phase diagram, obtained from Monte Carlo simulations for systems with linear system size $L=12 \ (N = 864)$. 
	The only coplanar orders are found in form of $120^\circ$ order (VI) in the lower left quadrant, 
	i.e. ferromagnetic $J_+$ and $J_{\times}$ (darker gray phase), 
	as well as  a distorted version of this (VII) for sligthly positive $J_{\times}$ (lighter gray phase).
	All other phases are non-coplanar, which besides a rigid phase with cuboctahedral order (I) in the upper right quadrant 
	and a distorted version of this (II) includes a variety of different spirals (III-V and VIII-X). 
	Phase boundaries of special interest are found upon exiting the coplanar $120^\circ$ order, 
	indicated by the dashed and dotted lines, along which  the low temperature specific heat $c_V(T\rightarrow 0)$ 
	takes values of $11/12$ and $23/24$, respectively (cf. Fig.~\ref{Fig:cv_on_axes}).}
	\label{Fig:phase_diagram_afm}
\end{figure*}
%%%%%%%%%%%%%%%%%%%%%%%%%%%%%%%%%%%%%%%%%%%%%%%%%%%%%%%%%%%%%%%

%%%%%%%%%%%%%%%%%%%%%%%%%%%%%%%%%%%%%%%%%%%%%%%%%%%%%%%%%%%%%%%
\section{Octagon-plaquette Interactions}
\label{sec:full_model}

We now augment the nearest-neighbor model, discussed in the previous Section, with the cross octagon-plaquette interactions $J_+$ and $J_{\times}$ (cf. Fig.~\ref{Fig:lattice}), which microscopically arise upon the inclusion of longer-range Heisenberg couplings.
Conceptually, they are interesting as they are expected to stabilize non-coplanar magnetic orders, akin to the cross-hexagonal interactions in the conventional kagome case, and in distinction to the square-diagonal couplings. 

Indeed, we find that the cross octagon-plaquette interactions induce a plethora of non-coplanar orders as summarized
in the global phase diagram of Fig.~\ref{Fig:phase_diagram_afm}. 
One can, in fact, distinguish {\em ten} different phases, indicated by the different colors in the phase diagram, 
as one varies the relative coupling strength of the two couplings,  $J_+$ and $J_{\times}$, starting from an  isotropic nearest-neighbor model (i.e. for fixed $J_1=J_2=J_3=1$).
The distinct nature of these phases can be easily visualized by representative common origin plots for each phase, i.e.\ the collapse of an extended ground-state real-space configuration of spins to a single unit sphere by placing all spins at a joint origin. These, obtained by Monte Carlo simulations for systems with a linear system size of $L = 12 \ (N = 864)$, are shown around the phase diagram. In addition, we show their respective spin structure factors. 
Let us briefly go through these phases here, before providing a much more detailed description in the remainder of this Section.
The only phases with coplanar order come in the form of a $120^\circ$ ordered phase (VI) in the lower left quadrant, i.e. for ferromagnetic $J_+$ and $J_{\times}$ (indicated by dark gray in Fig.~\ref{Fig:phase_diagram_afm}), 
as well as a distorted version of this $120^\circ$ order (VII) for slightly positive $J_{\times}$ (indicated by light gray in the phase diagram).	
In the plain-vanilla $120^\circ$ ordered phase spins on elementary triangles form mutual angles of $2\pi/3$,
while this angle is increased beyond $2\pi/3$ in the distorted phase as discussed in Sec.~\ref{sec:120_deg_order}.
All other phases of the phase diagram exhibit non-coplanar order, which come in commensurate and incommensurate forms.
The simpler, commensurate variant of such non-coplanar order is the extended phase with cuboctahedral order (I) in the upper right quadrant and a distorted version of this (II), which we discuss in depth in Sec.~\ref{sec:cuboc_order}. 
Somewhat more complex non-coplanar orders come in incommensurate spin spiral order, which we find for the remaining six
phases (III-V and VIII-X). Remarkably, however, these can still be described by a semi-analytical ground-state construction~\cite{Sklan-2013,Grison-2020,Rastelli-1979,Lyons-1962,Freiser-1961,Lyons-1960,Kaplan-2007}, 
which we describe in Sec.~\ref{sec:spiral_orders}.

%%%%%%%%%%%%%%%%%%%%%%%%%%%%%%%%%%%%%%%%%%%%%%%%%%%
\subsection{Physics on the axes}
\label{sec:physics_axes}

But before we dive into the various magnetic orders that can be stabilized by the combined effects of cross octagon-plaquette interactions, we first consider their exclusive effect, i.e.\ we consider the horizontal and vertical axes in the middle of the 
phase diagram of Fig.~\ref{Fig:phase_diagram_afm}.
Adding {\em only one} of the two cross octagon-plaquette interactions ($J_\times$ or $J_+$ only), it turns out that it is still possible to locally satisfy the constraint of Eq.~\eqref{eqn:NNconstraint} provided that the added interaction is ferromagnetic in nature. 

\begin{figure}[t]
	\centering
	\includegraphics[width=1.0\columnwidth]{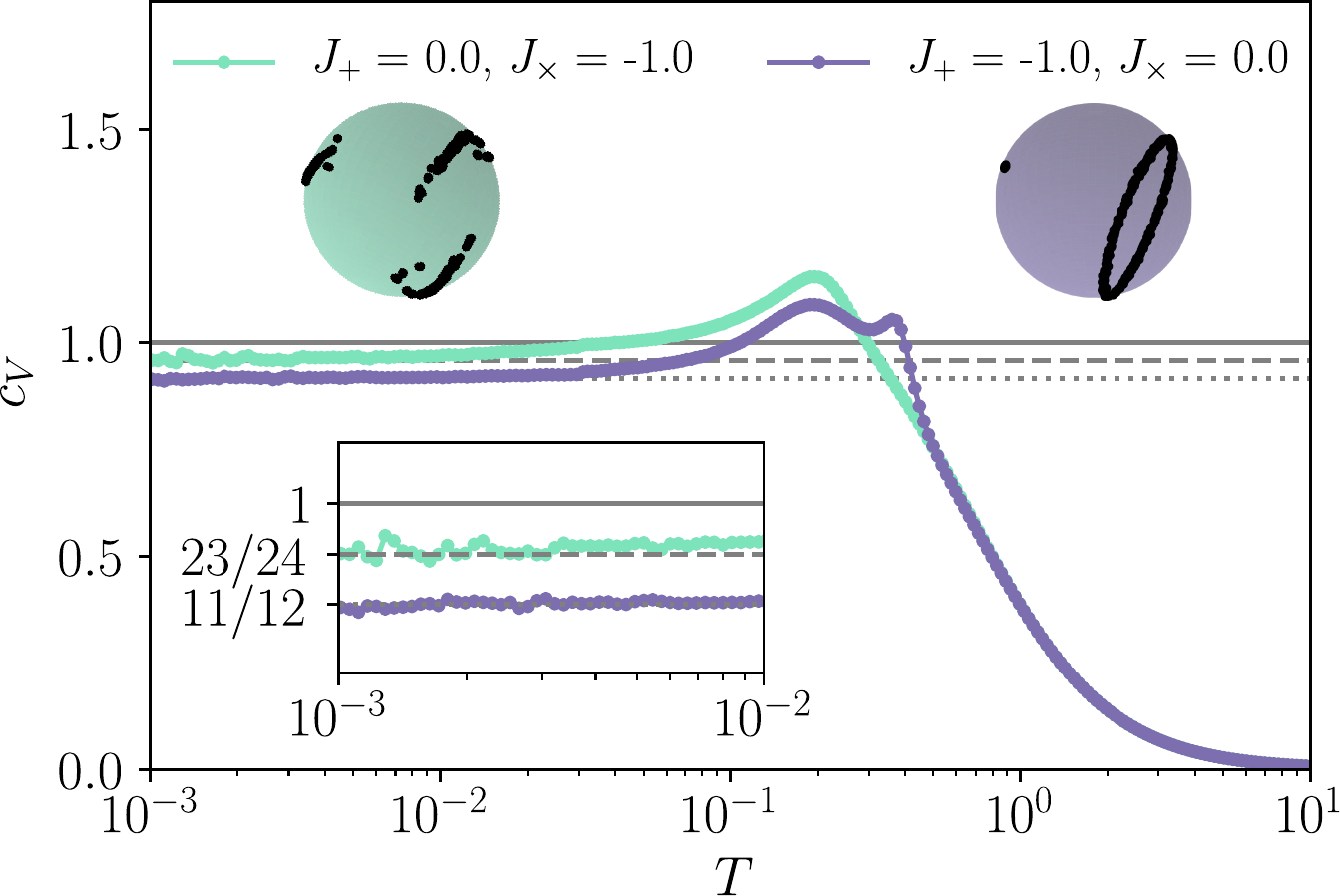}
	\caption{\textbf{Specific heat and ground states of the extended model on the axes.} For either only $J_+$ or $J_{\times}$ turned on and being ferromagnetic, one again finds special values of $c_V(T\rightarrow 0)$ that differ from 1. For For $J_{\times} = 0.0, J_{+} < 0.0$, the value is 11/12 (cf. purple curve with $J_{+} = -1.0$), and for $J_+ = 0.0, J_{\times} < 0.0$, we find a value of 23/24 (cf. mint curve with $J_{\times} = -1.0$). The upper insets show common origin plots of the corresponding ground states.}
	\label{Fig:cv_on_axes}
\end{figure}

For FM $J_+$, the spins located on the bow-ties become locked together and are all ferromagnetically aligned, meaning that $\vec{S}_k$ in the constraint (\ref{eqn:NNconstraint}) is fixed in each and every triangle to be the same. Denoting this fixed direction by $\vec{M}$, the remaining two spins (located on the squares) are thus subject to the local constraint on each triangle 
\begin{equation}
\vec{S}_i + \vec{S}_j = -\vec{M}. 
\end{equation}
This results in a global spin configuration in which $1/3$ of the spins point along $\vec{M}$ while the other $2/3$ point along a ring at an angle of $2\pi/3$ away from $\vec{M}$. This removes the possibility of having an extensive number of both globally coplanar and globally non-coplanar ground states, and thus there are only two distinct finite temperature regimes. In other words, entropic-driven selection of globally coplanar configurations would result in a regular $120^\circ$ ordered state. At high temperatures there is the usual paramagnetic regime, and at low temperatures a crossover into the ground state manifold with an accompanying specific heat $c_V \rightarrow  11/12$ as $T\rightarrow 0$, due to the continued existence of one zero mode per triangle.  This is precisely what we find in finite-temperature Monte Carlo simulations as shown in Fig.~\ref{Fig:cv_on_axes}. 

\begin{figure}[t]
	\centering
	\includegraphics[width=1.0\columnwidth]{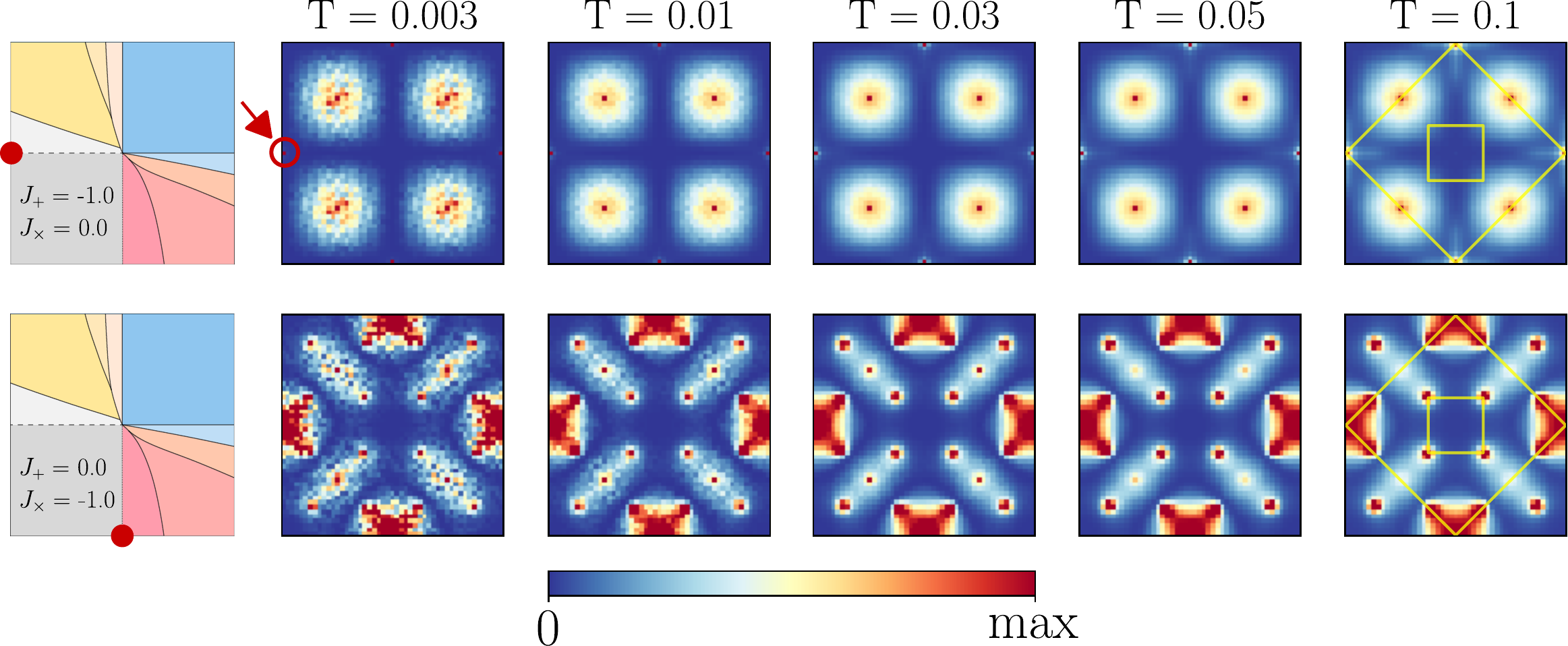}
	\caption{\textbf{Structure factors of the extended model on the axes.} Structure factors for different temperatures between T = 0.003 and T = 0.1 (left to right) on the axes ($J_+ = -1.0, J_{\times} = 0.0$ on the top and $J_+ = 0.0, J_{\times} = -1.0$ on the bottom). Note the sharp maxima at $(-4\pi, 0)$ and symmetry related momenta in the top row (indicated by red circle). The shown squares extent in reciprocal space from $-4\pi$ to $4\pi$ in both dimensions. For $T = 0.1$, the first Brillouin zone and the extended Brillouin zone are indicated.}
	\label{Fig:structure_factors_axes}
\end{figure}

For FM $J_\times$, the spins within each octagon coupled by $J_\times$ become locally ferromagnetically aligned, but the spins from one octagon to the next are not. This locks neighboring triangles together, resulting in one zero mode per unit cell, as opposed to one zero mode per triangle, and thus a low-temperature specific heat $c_V \rightarrow  [11\cdot(1/2) + 1\cdot (1/4)]/6 = 23/24$ as $T\rightarrow 0$.  This, again, is in perfect agreement with finite-temperature Monte Carlo simulations as shown in Fig.~\ref{Fig:cv_on_axes}. 

In comparison to the spin structure factors, summarized for the nearest-neighbor model in Fig.~\ref{Fig:structure_factors}, 
the spin-spin correlations discussed above lead to a deformation of $S({\bf q})$ as shown in Fig.~\ref{Fig:structure_factors_axes}. The $S(\mathbf{q})$ for ($J_+ = -1.0, J_{\times} = 0.0$) show maxima at positions where the Bragg peaks of the incipient $\mathbf{q}=\mathbf{0}$ order of the $120^{\circ}$ type would show up as $T\to0$~\cite{Astrakhantsev2021}. In contrast, for ($J_+ = 0.0, J_{\times} = -1.0$), the $S(\mathbf{q})$ display maxima at the expected locations for $\sqrt{3}\times\sqrt{3}$ order~\cite{Astrakhantsev2021}.

%%%%%%%%%%%%%%%%%%%%%%%%%%%%%%%%%
% 120 degree order
%%%%%%%%%%%%%%%%%%%%%%%%%%%%%%%%%
\subsection{$\bf 120^\circ$ order}
\label{sec:120_deg_order}

Turning to the magnetically ordered states of our phase diagram in Fig.~\ref{Fig:phase_diagram_afm}, 
we start with the $120^\circ$ order found in the lower left quadrant where both cross octagon-plaquette couplings
are ferromagnetic. While at zero temperature the ground state is degenerate and coplanar states as well as non-coplanar states with the same energy exist, this degeneracy is lifted at small, finite temperatures and coplanar states are selected over non-coplanar states by a thermal order by disorder mechanism. Therefore, the following discussion concentrates on these coplanar states, whereas the non-coplanar states are discussed briefly in Appendix~\ref{app:120_noncoplanar}.

\begin{figure}[t]
	\centering
	\includegraphics[width=1.0\columnwidth]{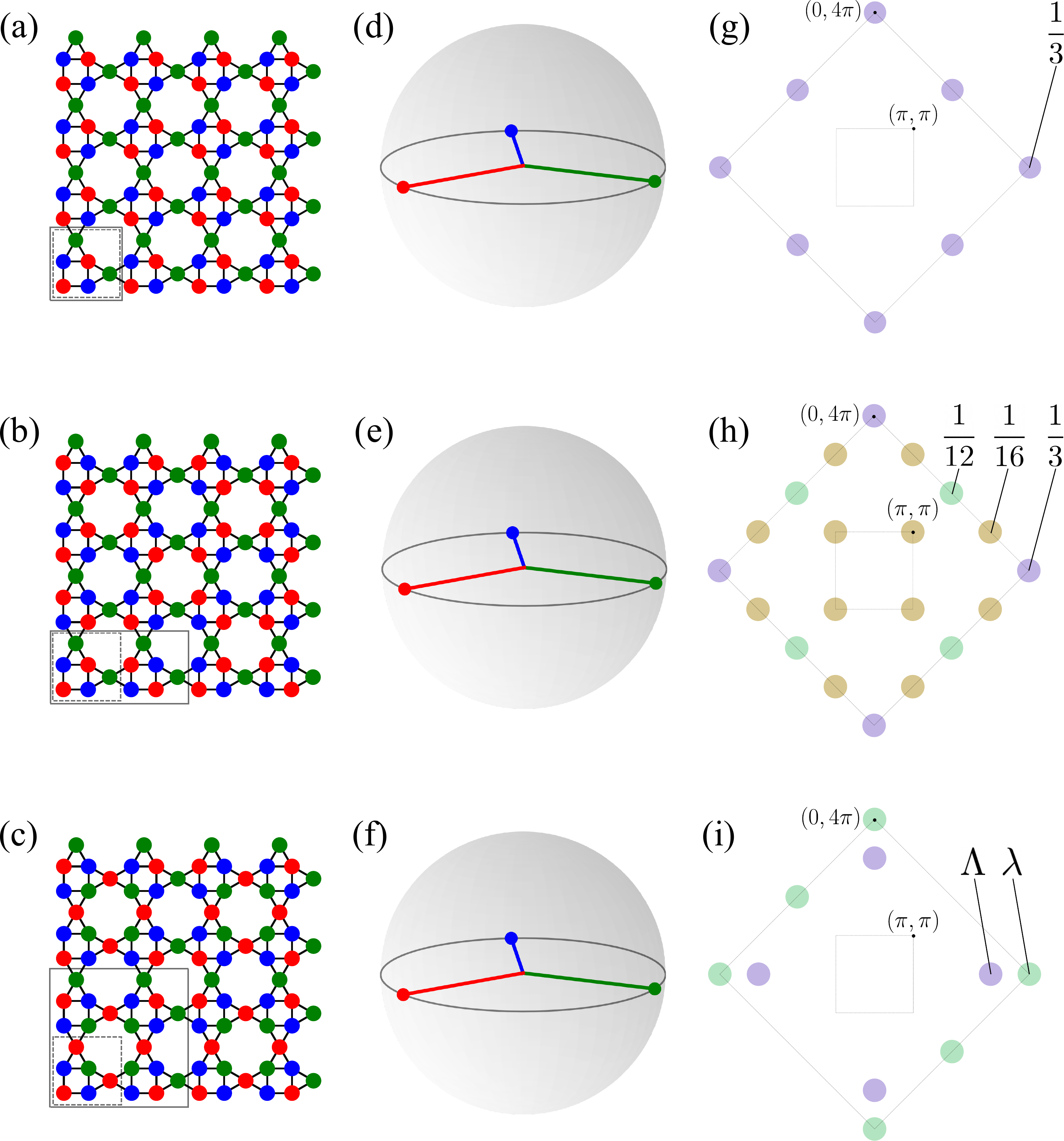}
	\caption{\textbf{Coplanar $\bf 120^\circ$ orders.} (a), (b), and (c) show real space arrangements of spins in the three different coplanar $120^\circ$ ordered states, with (a) corresponding to regular $\mathbf{q} =(0,0)$ order, (b) corresponding to $\mathbf{q} =(0,0)$ and $\mathbf{q} =(\pi,\pi)$ and (c) having no q-vectors at all. The states have either a 6-site magnetic unit cell that coincides with the geometric unit cell (a), a 12-site magnetic unit cell (large gray rectangle), which is two times larger than the geometric unit cell (small square) (b), or a 24-site magnetic unit cell (c), respectively. The magnetization of the square diagonal spins, $m_{\text{diag}}$, is $1.0$ for state (a), $0.5$ for state (b), and $0.25$ for state (c). (d)+(e)+(f) There are each three sublattices (each one corresponding to one color). Each sublattice of spins points towards a different corner of an equilateral triangle such that each neighboring pair of spins forms an angle of $2\pi/3$. (g)+(h)+(i) First and extended Brillouin zones of the square-kagome lattice showing the positions of the corresponding dominant and subdominant Bragg peaks. In (i), the ratio of the weight of the subdominant peaks $\lambda$ to the weight of the dominant peaks $\Lambda$ is $\lambda/\Lambda \approx 85 \%$. Arising from these three coplanar orders, there are three one-parameter families of non-coplanar orders, whose details are shown in Fig.~\ref{Fig:120_noncoplanar}.}
	\label{Fig:120_degree_order}
\end{figure}

\begin{figure}[b]
	\centering
	\includegraphics[width=1.0\columnwidth]{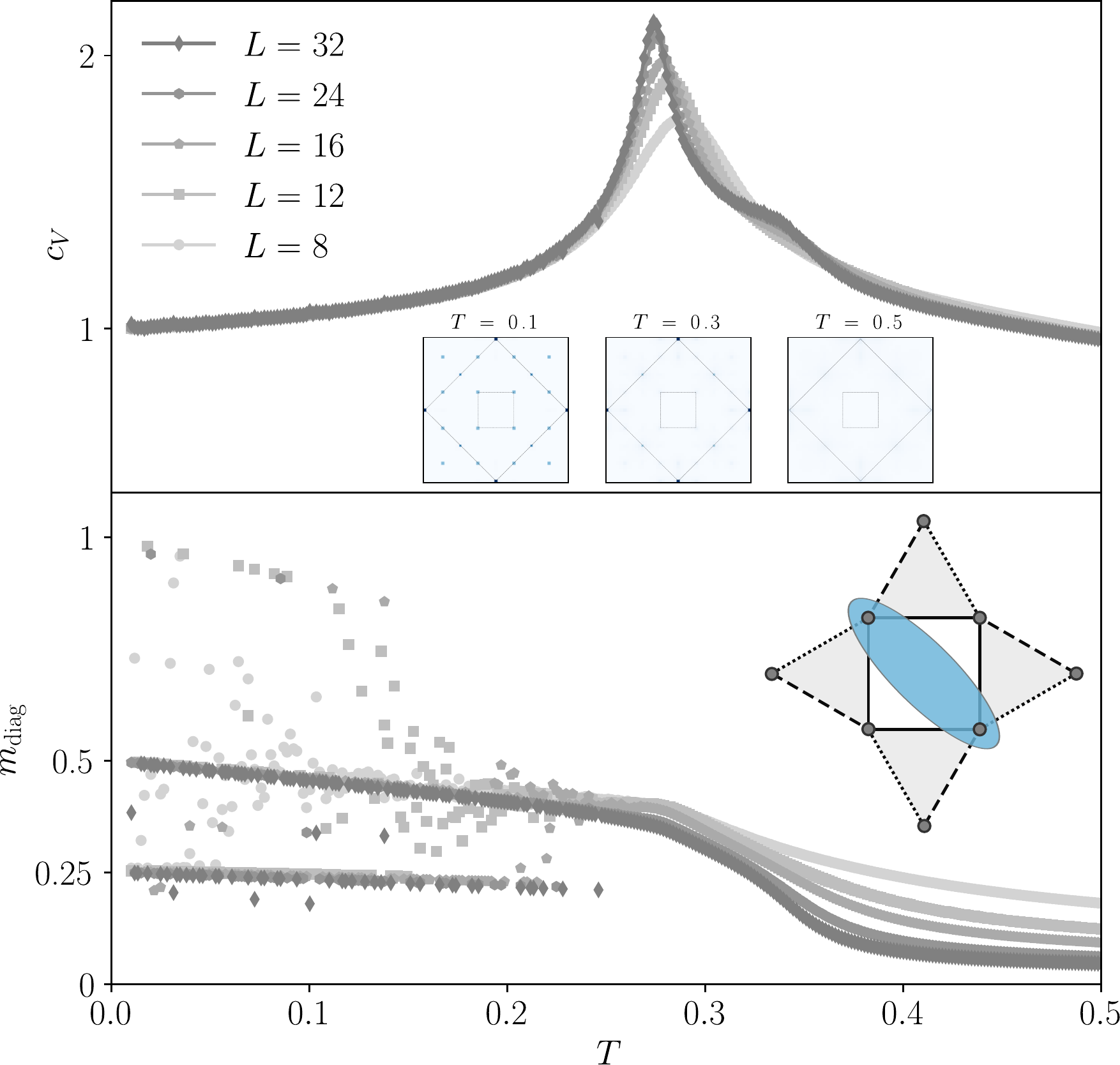}
	\caption{\textbf{Thermodynamics of $\bf 120^\circ$ order.} (top) The specific heat of the $120^\circ$ phase shows a sharp feature at $T = 0.27(3)$ and a subtle bump slightly at $T = 0.33(2)$. The subtle bump can be associated with the build-up of $120^\circ$ order, whereas at the sharp feature a specific $120^\circ$ order is selected -- in the example shown the structure factor at $T = 0.1$ (inset) coincides with the analytical structure factor in Fig.~\ref{Fig:120_degree_order}(h), whereas the intermediate structure factor can be obtained by averaging the real space correlations of all possible $120^\circ$ orders, i.e. those shown in Fig.~\ref{Fig:120_degree_order}(g)+(h)+(i) plus all possible rearrangements with the same order. The bottom panel shows the magnetization of the square-diagonal spins $m_\text{diag}$, i.e. the two spins on the main diagonal of the square of each unit cell (inset). The displayed square-diagonal magnetization starts to build up right at the bump-like feature slightly above $T = 0.3$. Below $T = 0.3$, at the sharp peak in the specific heat, the square-diagonal magnetization splits up into different branches that converge to the values of $m_{\text{diag}} = 1.0$, $m_{\text{diag}} = 0.5$, and $m_{\text{diag}} = 0.25$, corresponding to the $120^\circ$ orders shown in Fig.~\ref{Fig:120_degree_order}(a), (b), and (c), respectively.
}
 \label{Fig:fss_120}
\end{figure}

For the square-kagome lattice geometry at hand, one can, in principle, distinguish three different types of $120^\circ$ coplanar orders, each with three sublattices where the spins on each sublattice point to a different corner of an equilateral triangle such that each neighboring pair of spins forms an angle of $2\pi/3$. 
As illustrated in Fig.~\ref{Fig:120_degree_order}, the magnetic unit cell coincides either with the geometric 6-site unit cell as in the regular $\mathbf{q}=0$ order shown in Fig.~\ref{Fig:120_degree_order}(a), or contains either 12 sites and is two times larger than the geometric unit cell for the type shown in Fig.~\ref{Fig:120_degree_order}(b), or 24 sites for the type in Fig.~\ref{Fig:120_degree_order}(c), with the difference between the two latter types coming down to the bow-tie spins. While these bow-tie spins are all pointing in the same direction in the order of Fig.~\ref{Fig:120_degree_order}(b) forming a bow-tie ferromagnet, the bow-tie spins in the second type of $120^\circ$ order form stripy antiferromagnets with alternating rows/columns of up and down-pointing spins
as shown in Fig.~\ref{Fig:120_degree_order}(c). The magnetization of the spins on the main diagonal of the squares, $m_{\text{diag}}$, is different for all three types of order with $m_{\text{diag}} = 1.0$ for order Fig.~\ref{Fig:120_degree_order}(a), $m_{\text{diag}} = 0.5$ for order Fig.~\ref{Fig:120_degree_order}(b), and $m_{\text{diag}} = 0.25$ for order Fig.~\ref{Fig:120_degree_order}(c) (see also Fig.~\ref{Fig:fss_120}). 
A schematic of the static spin structure factors corresponding to these types of real-space $120^\circ$ order is shown
for an extended Brillouin zone of the square-kagome lattice in Fig.~\ref{Fig:120_degree_order}(g), (h), and (i), respectively. 
All three spin configurations exhibit the symmetry of the dihedral group $D_3$. Their net magnetization vanishes, $m = 0$, and the energy per site can be calculated as
\begin{equation}
	E_{\text{120}^\circ} = -1 + \frac{1}{3} (J_+ + J _{\times}) \,,
	\label{eqn:E_120}
\end{equation}
for varying strengths of the two couplings.

To elucidate the thermodynamics associated with these $120^\circ$ orders we show, in Fig.~\ref{Fig:fss_120}, specific heat traces
for the point in the lower left corner, $J_+, J_{\times} = -1$ (i.e.\ deep in the phase), for different system sizes between $L = 8$ and $L = 32$. Next to a sharp peak at $T = 0.27(3)$, there is a second, more subtle feature slightly at $T = 0.33(2)$. This smaller bump can, in fact, be associated with the build-up of quasi-long-range $120^\circ$ order -- but without selecting one of the types of $120^\circ$ order. Accordingly, the structure factor for the regime between the two features in the specific heat $c_V$, e.g. for $T = 0.3$, can be obtained by averaging the real space correlations of the  possible $120^\circ$ order arrangements. Also the net magnetization of the square-diagonal spins, 
shown in the lower panel of Fig.~\ref{Fig:fss_120}, builds up at this feature at higher temperatures. 
The lower temperature peak at  $T=0.27(3)$ then corresponds to the spontaneous selection of one of the types $120^\circ$ orders, resulting in a sharp feature in the specific heat. Notably, all three types of orders appear with different probabilities, as indicated by histograms of $m_\text{diag}$ (see Appendix~\ref{app:120_histograms}).

A deformed version of the $120^\circ$ order, henceforth termed $120^\circ$-d, exists in a small region touching the conventional $120^\circ$ order phase in the upper left quadrant of the phase diagram, indicated as phase VII in Fig.~\ref{Fig:phase_diagram_afm}. As in the non-deformed case, there exist non-coplanar states (discussed in Appendix~\ref{app:120_noncoplanar}) and coplanar states with the same energy, with the latter being selected by thermal order by disorder at small, but finite temperature. Therefore, the following discussion, again, concentrates on the coplanar deformed $120^\circ$ order. Its deformation --  two of the three mutual angles between neighboring spins take a value of $\alpha > 2\pi/3$, while the third one becomes smaller than $2\pi/3$ -- can be seen in the common origin plot of Fig.~\ref{Fig:phase_diagram_afm}. It can be derived by elementary geometric considerations: For instance, let the angle between the blue and the red sublattices and between the blue and the green sublattices shown in Fig.~\ref{Fig:120_degree_order} increase to $\alpha > 2\pi/3$, while the third angle, between the red and the green sublattices, becomes $2\pi - 2\alpha < 2\pi/3$. The angle $\alpha$ is found to vary as  $\cos \alpha = -(2+J_{\times})^{-1}$ and the ground-state energy per site takes the value
\begin{equation}
	E_{\text{120}^\circ\text{-d}} = -\frac{6+J_{\times}^2-2J_+-J_{\times}(J_+-4)}{3(J_{\times}+2)}  \,.
	\label{eqn:E_120d}
\end{equation}
The ground state symmetry is reduced to the symmetry of the dihedral symmetry group $D_1$ and has a non-zero, yet small magnetization $m_{120^\circ\text{-d}} = \tfrac{J_{\times}}{3(2+J_{\times})}$, which only depends on the values of $J_{\times}$ (see also Fig.~\ref{Fig:mag_cuts_afm} in Appendix~\ref{sec:energy_cuts}).

%%%%%%%%%%%%%%%%%%%%%%%%%%%%%%%%%
% cuboc order
%%%%%%%%%%%%%%%%%%%%%%%%%%%%%%%%%
\subsection{Cuboc order}
\label{sec:cuboc_order}

\begin{figure}[t]
	\centering
	\includegraphics[width=1.0\columnwidth]{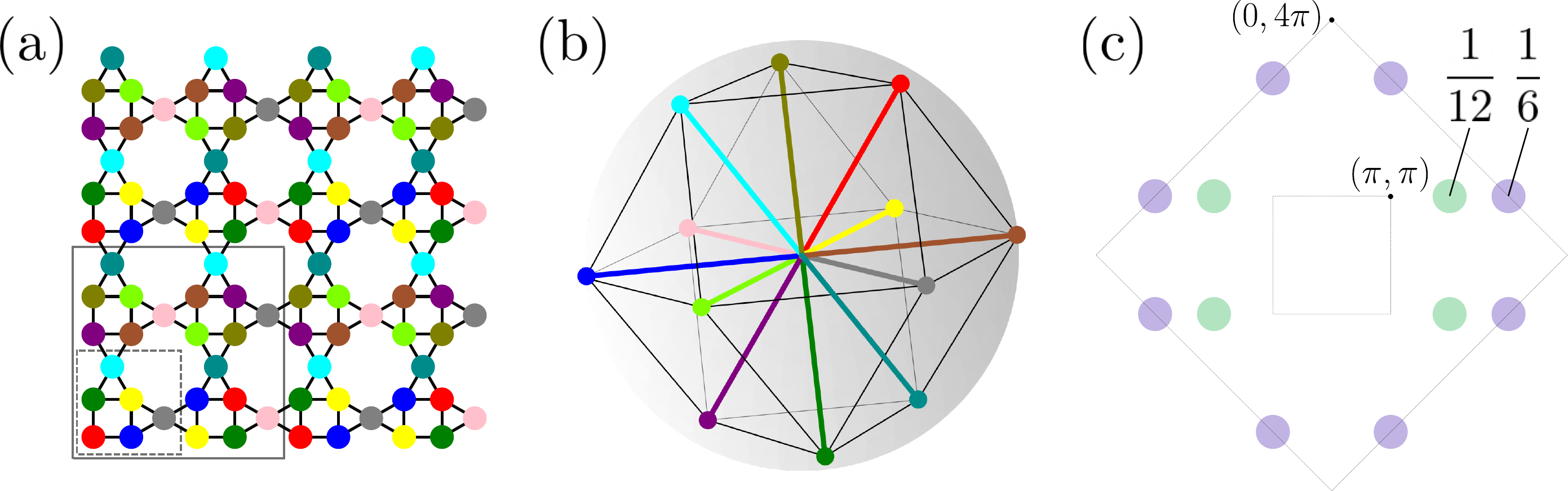}
	\caption{\textbf{Cuboctahedral order \textit{cuboc1}.} (a) Real space arrangement of spins in a cuboctahedral (cuboc) ordered state. There are 12 sublattices (each one corresponding to one color) with a 24-site magnetic unit cell (large gray square), which is four times larger than the geometric unit cell (small square). (b) Each sublattice of spins points towards a different corner of a cuboctahedron such that each neighboring pair of spins forms an angle of $2\pi/3$. This order corresponds to the \textit{cuboc1} state in \cite{Messio-2011}. (c) First and extended Brillouin zones of the square-kagome lattice showing the positions and the fractions of total spectral weight of the corresponding Bragg peaks. The order breaks $C_4$ symmetry.}
	\label{Fig:cuboc1_order}
\end{figure}

Let us now turn to the first instance of exclusively non-coplanar order 
and consider the cuboctahedral (cuboc) order 
found in the upper right quadrant of the phase diagram,  where both cross-octagon plaquette couplings are antiferromagnetic.
In real-space this order is described by 12 sublattices where the spins on each sublattice point towards a different corner of a cuboctahedron, as illustrated in Fig.~\ref{Fig:cuboc1_order}(b). The underlying magnetic unit cell contains 24 sites and is thus four times larger than the geometric unit cell, see Fig.~\ref{Fig:cuboc1_order}(a). 
All neighboring spins form an angle of $2\pi/3$, which corresponds to the \textit{cuboc1} state, discussed for the kagome lattice with cross-hexagonal couplings~\cite{Janson-2008,Messio-2011}. 
Note that there are eight possible ways to arrange cuboc1 order on the square-kagome lattice as illustrated in Fig.~\ref{Fig:cuboc1_orders} of Appendix \ref{sec:cuboc1_orders}, each of which breaks $C_4$ rotation symmetry.
The remaining symmetry of this ordered state is the full octahedral symmetry group $O_h$. 
It has zero magnetization, $m = 0$, and its energy per site is given by
\begin{equation}
	E_{\text{cuboc}} = -1 - \frac{1}{3} (J_+ + J _{\times}) \,. 
	\label{eqn:E_cuboc1}
\end{equation}

The real-space correlations of the cuboc1 order give rise to a spin structure factor as schematically visualized for one of the eight possible real-space configurations in Fig.~\ref{Fig:cuboc1_order} (c), which shows the positions of the corresponding Bragg peaks within the extended Brillouin zone of the square-kagome lattice together with the associated fraction of total spectral weight.
Averaging over the real-space correlations of all eight possible cuboc1 arrangements, restores $C_4$ symmetry in the structure factor (cf. Appendix~\ref{sec:cuboc1_orders}). 
 
\begin{figure}[t]
	\centering
	\includegraphics[width=1.0\columnwidth]{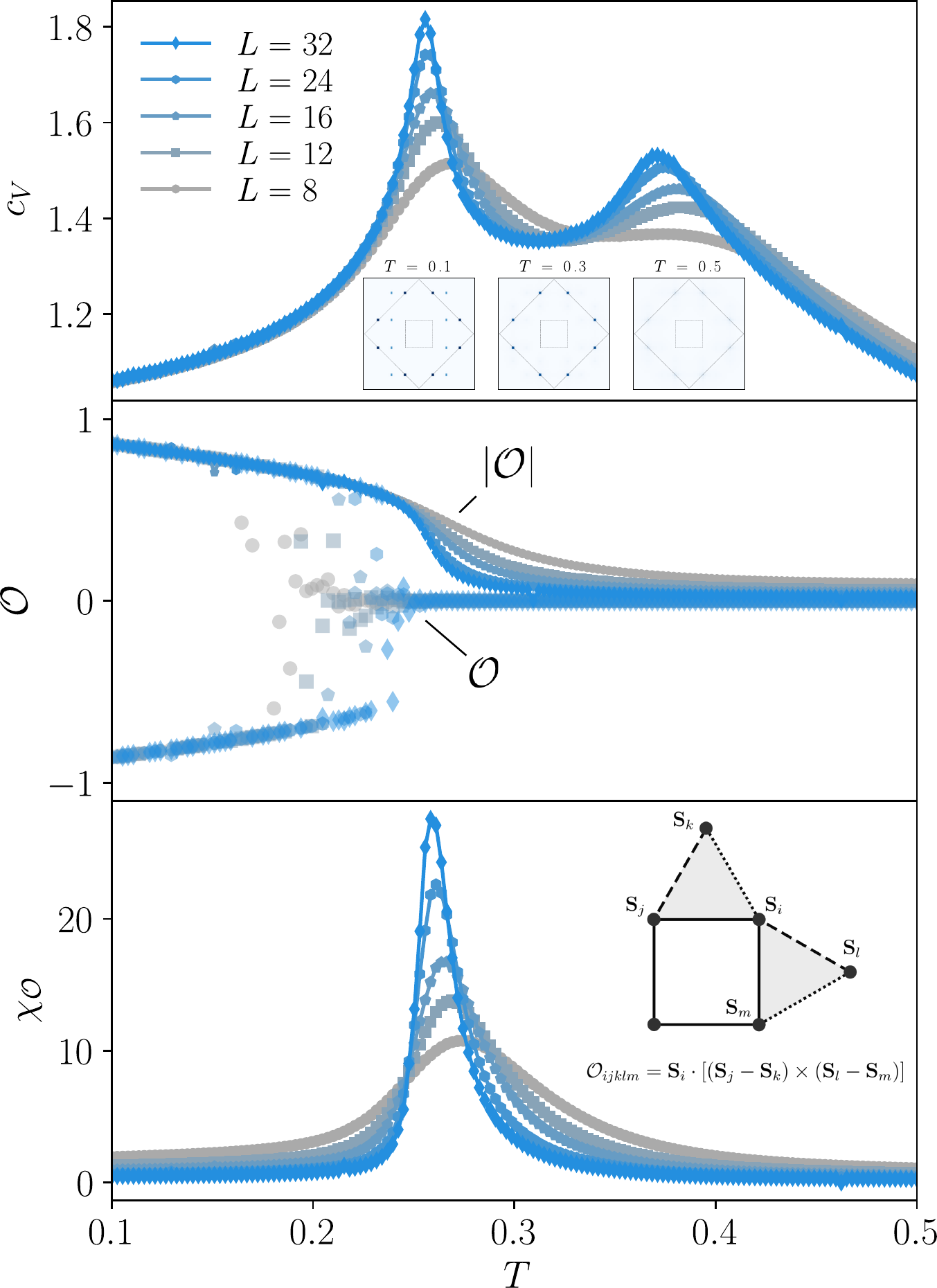}
	\caption{\textbf{Thermodynamics of cuboc1 phase.} 
	The specific heat (top panel) of the cuboc1 phase displays a double-peak structure.
	 The high-temperature peak at $T = 0.371$ can be associated with the initial build-up of 
	 coexisting cuboc1 orders. 
	 At the low-temperature peak (at $T = 0.256$), one specific realization of cuboc1 order is selected
	  (cf. Appendix~\ref{sec:cuboc1_orders}). This can be seen from the structure factors (inset). While the structure factor at $T = 0.1$ corresponds to one specific realization of cuboc1 order, the intermediate structure factor at $T = 0.3$ conincides with the analytical structure factor of the superposition of all possible cuboc1 order realizations (see also Fig.~\ref{Fig:cuboc1_orders}). Middle and bottom panels show the cuboc1 order parameter $\mathcal{O}$ (Eq. \ref{eq:cuboc1_op}), and the corresponding susceptibility $\chi_{\mathcal{O}}$ (Eq. \ref{eq:cuboc1_op_susc}). The lower inset shows how $\mathcal{O}_{ijklm}$ is being calculated on a single skew bow-tie. This quantity is then averaged over all skew bow-ties in order to calculate $\mathcal{O}$.}
	\label{Fig:fss_cuboc}
\end{figure}
 
Turning to the thermodynamics of the cuboc phase,  we show, in Fig.~\ref{Fig:fss_cuboc}, the specific heat, the cuboc1 order parameter $\mathcal{O}$, and its associated susceptibility $\chi_{\mathcal{O}}$ (both introduced in Appendix~\ref{sec:cuboc1_orders}) for different system sizes between $L = 8$ and $L = 32$. The specific heat displays a clearly visible double-peak structure which, similar to our discussion of the coplanar $120^\circ$ order, can be rationalized by the coexistence of multiple possible cuboc1 arrangements on the square-kagome lattice: At the high-temperature peak in $c_V$ (at $T=0.371$), the system builds up cuboc1 order, but does not select a specific arrangement out of the eight  possible realization, as can be seen from the $C_4$ symmetric structure factor in the inset. 
At the low-temperature peak (at $T = 0.256$) a specific cuboc1 order is then spontaneously selected. At this temperature, the order parameter $\mathcal{O}$ (Eq. \ref{eq:cuboc1_op}), which takes values of $\pm 1$ for different specific cuboc1 arrangements, builds up and the corresponding susceptibility $\chi_{\mathcal{O}}$ (Eq. \ref{eq:cuboc1_op_susc}) diverges as shown in the lowest panel of Fig.~\ref{Fig:fss_cuboc}.

A deformed version of cuboc1 order (denoted as phase II in our phase diagram), termed cuboc-d, extends to a part of the lower right quadrant in the phase diagram Fig.~\ref{Fig:phase_diagram_afm} with  $J_{\times} < 0$ and $J_+ \geq 0$, which can be derived by applying the generalized Luttinger-Tisza method of Ref.~\cite{Schmidt2022}, see also the Mathematica files in the Supplement \cite{Supplement}.
In this cuboc-d phase, the antipodal squares of the cuboctahedron are deformed into 
antipodal squares with modified $z$-values while the square in the equatorial plane remains undeformed.
Its symmetry is thereby reduced to the symmetry of the dihedral symmetry group $D_4^s$ where the superscript $s$ denotes additional mirror symmetry on the $xy$-plane, i.e. $s = \text{diag}(1,1,-1)$.
 The ground state energy per site can be calculated to be 
\begin{equation}
	E_{\text{cuboc-d}} = \frac{3+J_{\times}^2+J_+-J_{\times}(J_++3)}{3(J_{\times}-1)} .
	\label{eqn:E_cubocd}
\end{equation}
The cuboc-d ground state still has zero magnetization, $m = 0$.  

%%%%%%%%%%%%%%%%%%%%%%%%%%%%%
% Spin spiral orders
%%%%%%%%%%%%%%%%%%%%%%%%%%%%%
\subsection{Spiral orders}
\label{sec:spiral_orders}

The upper left and lower right quadrant of our phase diagram are occupied by spin spiral phases, coming in the form of six different variants (labeled III-V and VIII-X, respectively).
The complexity of these incommensurate, non-coplanar orders becomes immediately clear when looking at their common origin plots, whose intricate patterns point to magnetic unit cells of hundreds of spins. This renders any direct analytical description of these phases rather elusive, but it turns out that one can, in fact, deduce a {\em semi-analytical description}~\cite{Sklan-2013,Grison-2020,Rastelli-1979,Lyons-1962,Freiser-1961,Lyons-1960,Kaplan-2007} of these phases from low-temperature numerical simulation data. As we will discuss below, this approach provides us with a symmetry-optimized description of these spin spirals including explicit expressions of their ground-state energy as function of the coupling parameters. The latter then allows us to establish sharp phase boundaries between these complex spin spiral phases as depicted in the phase diagram of Fig.~\ref{Fig:phase_diagram_afm}.

%%%%%%%%%%%%%%%%%%%%%%%%%%%%%
\subsubsection*{Semi-analytical approach}
\label{sec:semi-analytical}

\begin{figure}[b]
	\centering
	\includegraphics[width=1.0\columnwidth]{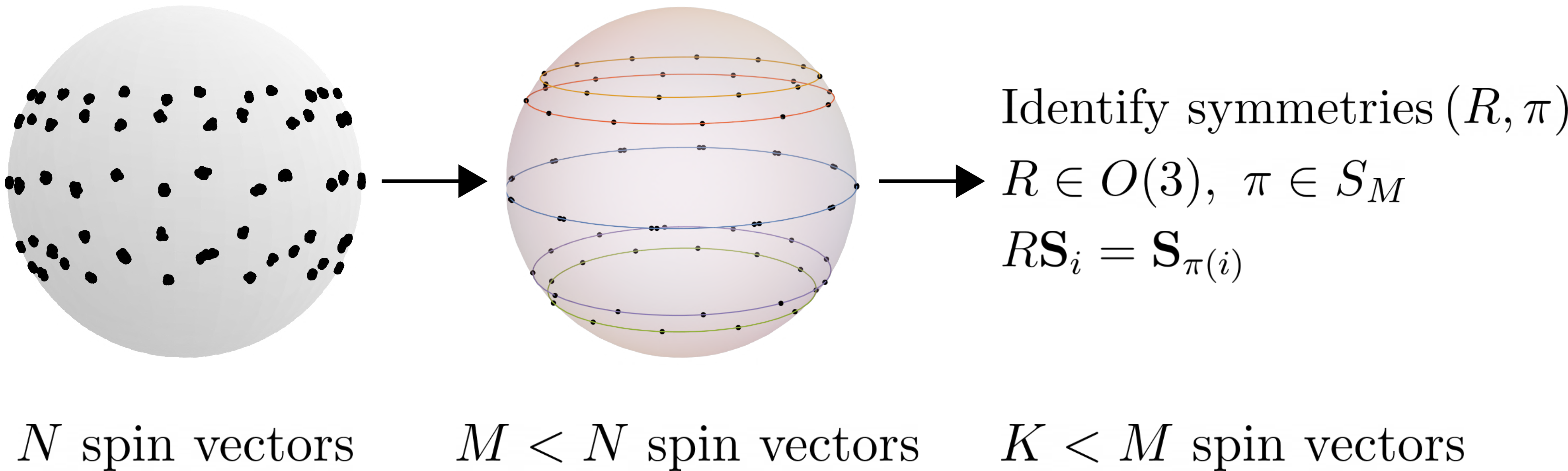}
	\caption{\textbf{Semi-analytical scheme for spin spiral phases.} 
	Starting point is a common origin plot of the $N$ spin vectors of a ground-state spin configuration sampled 
	in Monte Carlo simulations at ultra-low temperatures $T = 10^{-4}$, as shown on the left.
	In a second step, we identify $M <N$ unique spin directions by grouping spins that point approximately in the same direction 
	(middle). 
	From these unique spin vectors, we then identify symmetries that further reduce the number of unique spin vectors to $K < M$ 
	and allow us to describe the spiral phase analytically (right). 
	The data shown is for spin spiral phase III of the lower right quadrant with couplings $J_+ = +1$ and $J_{\times} = -0.2$.
	The initial common origin plot has $N = 864$ spins corresponding to a system size of 
	$L = 12$. The initial reduction leads to $M = 72$ points on five circles (as indicated in the middle panel). 
	The symmetry group $G$ of the example is generated by rotations of $\pi/6$ about the vertical symmetry axis 
	as well as by a reflection at the equatorial plane, i.e.\ $G = D_{12}^s$, and therefore $K = 4$.}
	\label{Fig:semi_analytical_method}
\end{figure}

The starting point of our semi-analytical approach is numerical data in the form of a common origin plot of the $N$ spin vectors of a ground-state spin configuration sampled in Monte Carlo simulations at ultra-low temperatures $T = 10^{-4}$, typically explored in conjunction with a parallel tempering scheme. Example input for the spin spiral phase III in the lower right quadrant is shown on the left in the schematic illustration of Fig.~\ref{Fig:semi_analytical_method}.

In a second step, we then identify a smaller number of $M < N$ unique spin vectors by grouping spins in the initial common origin plot that point approximately in the same direction, see the middle panel of Fig.~\ref{Fig:semi_analytical_method}. In practice, we say that two spins $\vec{S}_i$ and $\vec{S}_j$ point approximately in the same direction if $\vec{S}_i \cdot \vec{S}_j \geq \gamma$, where the exact value of $\gamma$ slightly varies from case to case, but typically $\gamma \approx 0.995$. 

For these $M$ spin directions we then identify all possible symmetries, which allows us to further reduce the number of unique spin vectors to $K < M$. A symmetry in the aforementioned sense is a tuple $(R,\pi)$ with $R \in O(3)$ and $\pi \in S_M$ where $S_M$ is the permutation group of $M$ elements, such that $R\vec{S}_{i} = \vec{S}_{\pi(i)}$ for all spin vectors $\vec{S}_{i}$. From the remaining $K$ spin vectors, all spin vectors can be generated by applying these symmetries. 

In total, this approach allows us to describe the ground state by at most $2K-1$ parameters -- maximally two parameters per spin minus one parameter due to a global rotation around the symmetry axis, but less if some polar or azimuthal angles of the ground state assume fixed values. This compact representation is summarized in Table \ref{Tab:Summary} for all six spin spiral phases of our phase diagram.
Having such an analytical representation at hand, we can then explicitly calculate various observables such as the magnetization or ground-state energy for arbitrary couplings $J_+$ and $J_{\times}$ for all phases, which in turn allows us to determine the phase boundaries shown in the phase diagram Fig.~\ref{Fig:phase_diagram_afm}. Several cross-checks can be used to validate this approach, including a comparison of the analytical ground-state energy and the Monte Carlo result as well as the determination of the phase boundaries, which we compare to scans of derivatives of the Monte Carlo energy, as shown in Appendix~\ref{sec:energy_cuts}. In general, we find excellent agreement.

\begin{table}[h]
\begin{tabular}{c|c|c|ccc|c}
	\multirow{2}{*}{phase} & \multirow{2}{*}{symmetry} & \multirow{2}{*}{$\mathbf{q}$ vectors} & \multicolumn{4}{c}{semi-analytical} \\
	  &   &   & $N$ & $M$ & $K$ & \# parms\\
	\hline \hline
	I 	& $O_h$	& $(0,\pi), (\pi, \pi)$ & --& --& --&--\\
 	II 	& $D_4^s$	& $(0,0), (\pi, \pi)$ & --& --& --&--\\
  \hline
  	III 	& $D_{12}^s$ & $(-5\pi/6,-5\pi/6), (\pi, \pi)$ & 864 & 72 & 4 & 3\\
   	IV 	& $D_{6}^{s,t}$ & -- & 864 & 228 & 13 & 19\\
    	V 	&	$\{ {\rm id, s} \}$ & -- & 864 & 48 & 24 & 47\\
    \hline
    VI 		& $D_3$	& $(0,0),(\pi, \pi)$ & --& --& --&--\\
    VII 	& $D_1$	& $(0,0),(0,\pi)$ &-- &-- & --&--\\
    \hline
    VIII 	& $D_{12}$ & $(\pi/6,\pi)$ & 864 & 37 & 3 & 3\\
    IX 		& $D_{3}^\sigma$ & -- & 864 & 108 & 18 & 26\\
    X 		& $D_{6}^s$ & $(\pi/3,\pi), (\pi, \pi)$ & 864 & 32 & 4 & 3\\
\end{tabular}
\caption{{\bf Symmetry characterization of the ten ground-state phases} of the phase diagram Fig.~\ref{Fig:phase_diagram_afm}.
		%according to the used method (a: analytical, sa: semi-analytical), 
		Given are the ground state symmetry (second column), and the $\mathbf{q}$ vectors of each phase (if any, third column).
		 For the six semi-analytically described phases (III-V, VIII-X), the compression of the parametrization of the spin spirals 
		 via clustering and symmetrization (Fig.~\ref{Fig:semi_analytical_method}) is given in the four columns on the right.
		 Technically, this semi-analytical description is obtained by starting with a common origin plot with $N=864$ spins sampled 
		 at $T=10^{-4}$ for a linear system size $L=12$, and is then given in terms of $M$, $K$, and the number of needed 				parameters (last column) to describe the phase.
}
\label{Tab:Summary}
\end{table}
\subsubsection*{Example: Spiral phase III with $D_{12}^s$ symmetry}
Let us illustrate this semi-analytical approach and its validity for an explicit example, picking the spiral phase III in the lower right quadrant of the phase diagram Fig.~\ref{Fig:phase_diagram_afm} for $J_+ = +1$ and $J_{\times} = -0.2$. Our schematic illustration of the semi-classical approach in Fig.~\ref{Fig:semi_analytical_method} also uses this example. 
The common origin plot on the left consists $N = 864$ spin vectors of the numerical ground state (at $T = 10^{-4}$) of a system of linear length $L=12$. By grouping spins that point to the same direction, using the criterion $\vec{S}_i \cdot \vec{S}_j \geq 0.999$, we find that there are only $M = 72$ unique spin directions which are shown in the middle panel. Performing a symmetry analysis, one finds that the symmetry group $G$ of this state is generated by rotations of $\pi/6$ about the vertical symmetry axis as well as by a reflection at the equatorial plane, i.e.\ $G = D_{12}^s$. This leaves us with just $K=4$ representative spins that describe the entire spin spiral configuration -- a significant reduction compared to the $N=864$ spins in the original real-space configuration. 
The $K=4$ representative spins can be written as functions of three parameters $\alpha, z_1, z_2$ in the following way
\begin{align}
\notag {\mathbf S}_1=&\left(0,\sqrt{1-z_1^2} ,z_1 \right)\;, \\
\notag {\mathbf S}_2=&\left(\frac{\sqrt{1-z_2^2}}{\sqrt{2}},\frac{\sqrt{1-z_2^2}}{\sqrt{2}} ,z_2 \right)\;, \\
 {\mathbf S}_{3,4}=&\left(\cos{\left(\frac{\pi}{4}\pm \alpha\right)}, \sin{\left(\frac{\pi}{4}\pm \alpha\right)},0 \right) 
 \label{eqn:rep_phase_III}
 \;.
\end{align}

With this compact representation at hand, the energy per site can now be explicitly calculated as
\begin{align}
	\notag E_{\text{III}} = &\frac{1}{12} \bigg[J _{\times} \cos 2\alpha + \sqrt{3} J _{\times} \sin 2\alpha \\
    & \notag - 2\Big(\sqrt{3} J_+-(\sqrt{3} - 2) J_+ z_1^2 + 4z_1z_2 \\
    & \notag + (\sqrt{6}-\sqrt{2})\sqrt{(1-z_1^2)(1-z_2^2)} + J _{\times}(2z_2^2 -1) \\
    & \notag + 2 (\sqrt{3-3z_2^2}-\sqrt{2-2z_1^2}) \sin \alpha \\
    & + 2 (\sqrt{2-2z_1^2}+\sqrt{1-z_2^2}) \cos \alpha \Big)\bigg] \,.
	\label{eqn:E_spiral}
\end{align}
Numerical minimization of this energy with $J_+ = +1$ and $J_{\times} = -0.2$ then leads to $E_{\text{III},\text{semi-analytical}} = -1.34515$ which is in excellent agreement with and (as expected) slightly below the Monte Carlo result $E_{\text{III},\text{MC}} = -1.345$, which is shifted upwards by finite-temperature fluctuations commensurate with a temperature of $T=10^{-4}$. 

We present explicit results of this semi-classical approach for all other spin spiral phases in Appendix~\ref{sec:spiral}. It turns out that the six semi-analytically described phases can be divided into three regular ones that possess $\mathbf{q}$ vectors and three irregular ones, possessing no $\mathbf{q}$ vectors. A compact summary that characterizes all ten phases of the phase diagram Fig.~\ref{Fig:phase_diagram_afm} by the used method, the ground state symmetry, and corresponding $\mathbf{q}$ vectors, if any, is given in Table~\ref{Tab:Summary}.

\begin{figure}[t]
	\centering
	\includegraphics[width=1.0\columnwidth]{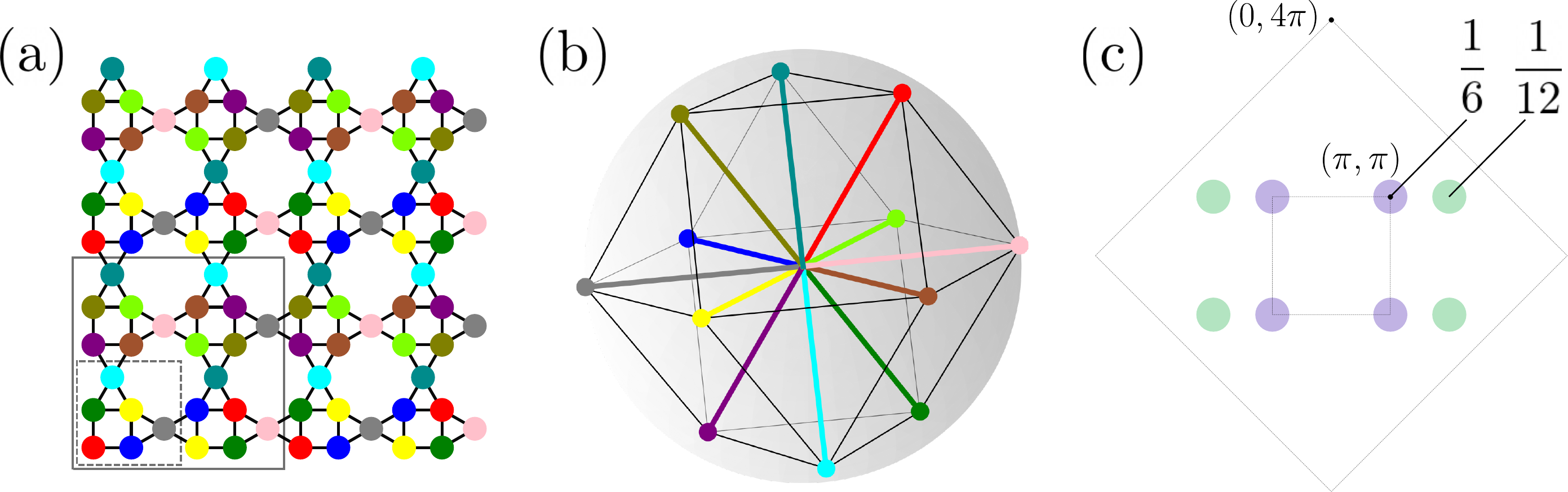}
	\caption{\textbf{Cuboctahedral order \textit{cuboc3}.}
	 This variant of a non-coplanar cuboctahedral order is found when flipping the bow-tie interactions $J_2$ and $J_3$ of the nearest 
	 neighbor model to ferromagnetic (while keeping the square interactions $J_1$ antiferromagnetic). 
	 It is obtained from the original cuboc order (Fig.~\ref{Fig:cuboc1_order}) via a local spin transformation (see main text).
	 (a) Real space arrangement of spins in the cuboc3 ordered state. 
	 There are 12 sublattices (each one corresponding to one color) with a 24-site magnetic unit cell (large gray square), 
	 four times larger than the geometric unit cell (small square). 
	 (b) Each sublattice of spins points towards a different corner of a cuboctahedron 
	 such that each neighboring pair of spins forms an angle of $2\pi/3$ on the squares and an angle of $\pi/3$ on the triangles,
	 making it different from the  \textit{cuboc1} and \textit{cuboc2} orders discussed in the literature \cite{Messio-2011}.
	 (c) First and extended Brillouin zones of the square-kagome lattice showing the positions of the corresponding Bragg peaks. 
	 The order breaks $C_4$ symmetry.}
	\label{Fig:cuboc3_order}
\end{figure}

\begin{figure}[b]
	\centering
	\includegraphics[width=1.0\columnwidth]{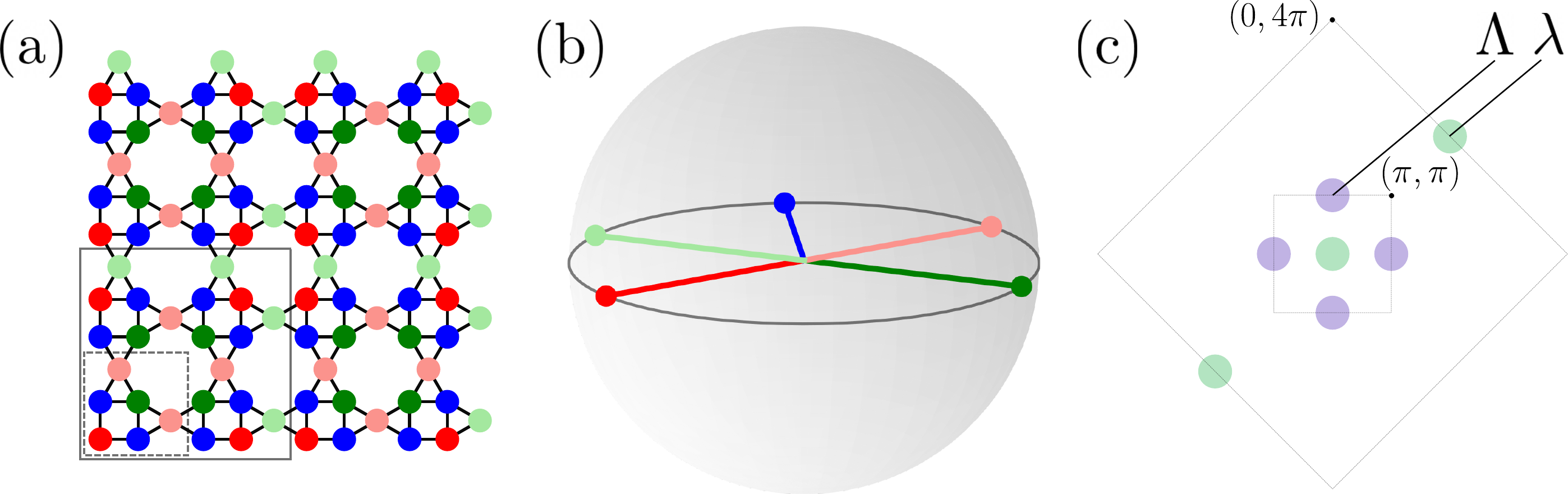}
	\caption{\textbf{Pentagonal order.}  
	Coplanar pentagonal order arises when flipping the bow-tie interactions $J_2$ and $J_3$ of the nearest neighbor model to ferromagnetic 
	(while keeping the square interactions $J_1$ antiferromagnetic). 
	It is obtained from the original $120^\circ$ order (Fig.~\ref{Fig:120_degree_order}(c)) via a local spin transformation (see main text).
	 (a) Real space arrangement of spins in the coplanar pentagonal state. 
	 There are five sublattices (each one corresponding to one color) with a 24-site magnetic unit cell (large gray square), 
	 four times larger than the geometric unit cell (small square). 
	 (b) The spins on the five sublattices point to five out of the six corners of a hexagon. 
	 (c) First and extended Brillouin zones of the square-kagome lattice showing the positions of the corresponding dominant and subdominant Bragg peaks. The ratio of the weight of the subdominant peaks $\lambda$ to the weight of the dominant peaks 
	 $\Lambda$ is $\lambda/\Lambda \approx 85 \%$.}
	\label{Fig:penta_order}
\end{figure}

\subsection{Cuboc3 and  pentagonal order for mixed interactions}
\label{sec:mixed_interactions}
Finally, we note that one could also consider variations of the model at hand where one changes the sign of the interactions in the original nearest-neighbor $(J_1, J_2, J_3)$ Heisenberg model (see Fig.~\ref{Fig:lattice} for a reminder of the coupling geometries). Flipping the sign of the bow-tie interactions to ferromagnetic couplings, i.e. $J_2 = J_3 = -1$,
while keeping the square interactions antiferromagnetic, i.e. $J_1 = 1$, 
yields exactly the same phase diagram as shown in Fig.~\ref{Fig:phase_diagram_afm}, up to local spin transformations.
Specifically, since the spins on the bow-ties are coupled via $J_2$ and $J_3$ to their nearest neighbors and via $J_+$ to other bow-tie spins, the total energy remains unchanged if all bow-tie spins are inverted simultaneously while the sign of the triangular couplings is changed, $J_2 \rightarrow -J_2$ and $J_3 \rightarrow -J_3$. 

Performing these local spin transformations on the orders of our original phase diagram yields qualitatively new types of orders:
The original cuboctahedral order (Fig.~\ref{Fig:cuboc1_order}) becomes a new cuboctahedral order \textit{cuboc3} (Fig.~\ref{Fig:cuboc3_order}), where each neighboring pair of spins forms an angle of $2\pi/3$ on the squares and an angle of $\pi/3$ on the triangles, which is different to \textit{cuboc1} order (where all neighboring pairs of spins form an angle of $2\pi/3$) and to \textit{cuboc2} order  (where all neighboring pairs of spins form an angle of $\pi/3$) \cite{Messio-2011}. The $120^\circ$ orders (Fig.~\ref{Fig:120_degree_order}) are transformed into new coplanar orders in an analogous way, e.g.~into a coplanar {\em pentagonal} order (Fig.~\ref{Fig:penta_order}), that results from flipping the bow-tie spins in the $120^\circ$ order Fig.~\ref{Fig:120_degree_order}(c). Note that, just as in the original $120^\circ$ case, at zero temperature there are both coplanar and non-coplanar states degenerate in energy but the coplanar states are selected at finite temperature via thermal order-by-disorder (see Appendix \ref{app:120_noncoplanar} for further details).

%%%%%%%%%%%%%%%%%%%%%%%%%%%%%%%
\subsection{Octagonal and conical orders for FM interactions}
\label{sec:fm_interactions}

\begin{figure}[b]
	\centering
	\includegraphics[width=1.0\linewidth]{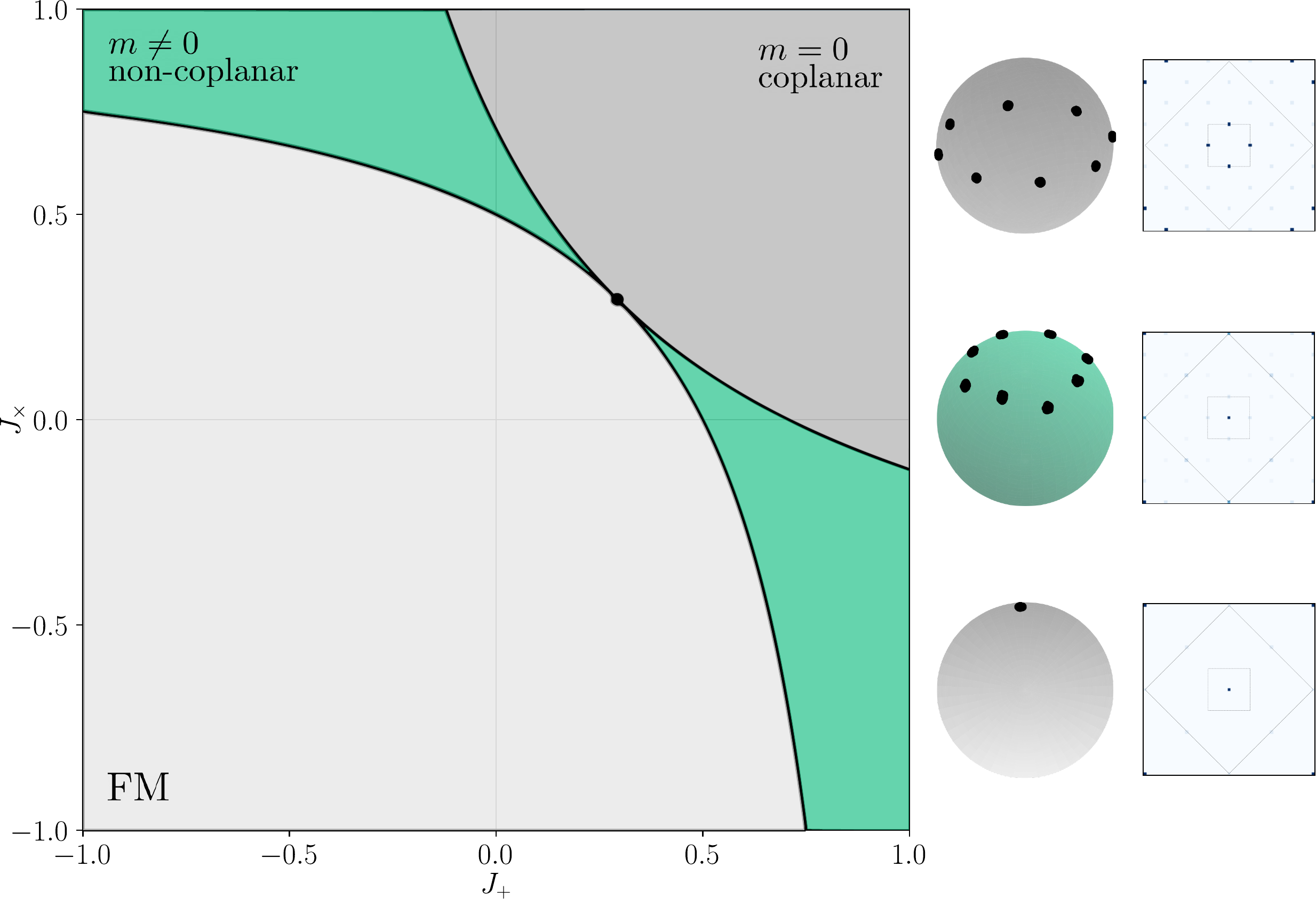}
	\caption{\textbf{FM phase diagram.} 
	For fixed $J_1 = J_2 = J_3 = -1$, the model shows a large ferromagnetic phase (light gray), a coplanar phase with octagonal order (dark gray) and non-coplanar umbrella-like double cone states (turquoise) that are an interpolation between octagonal and ferromagnetic states. Ferromagnetic and coplanar phases touch at the single point $J_+ = J_{\times} = 1 - \tfrac{1}{\sqrt 2}$. On the right hand side are corresponding common origin plots and structure factors from Monte Carlo simulations.} 
	\label{Fig:phase_diagram_fm}
\end{figure}

\begin{figure}[t]
	\centering
	\includegraphics[width=1.0\columnwidth]{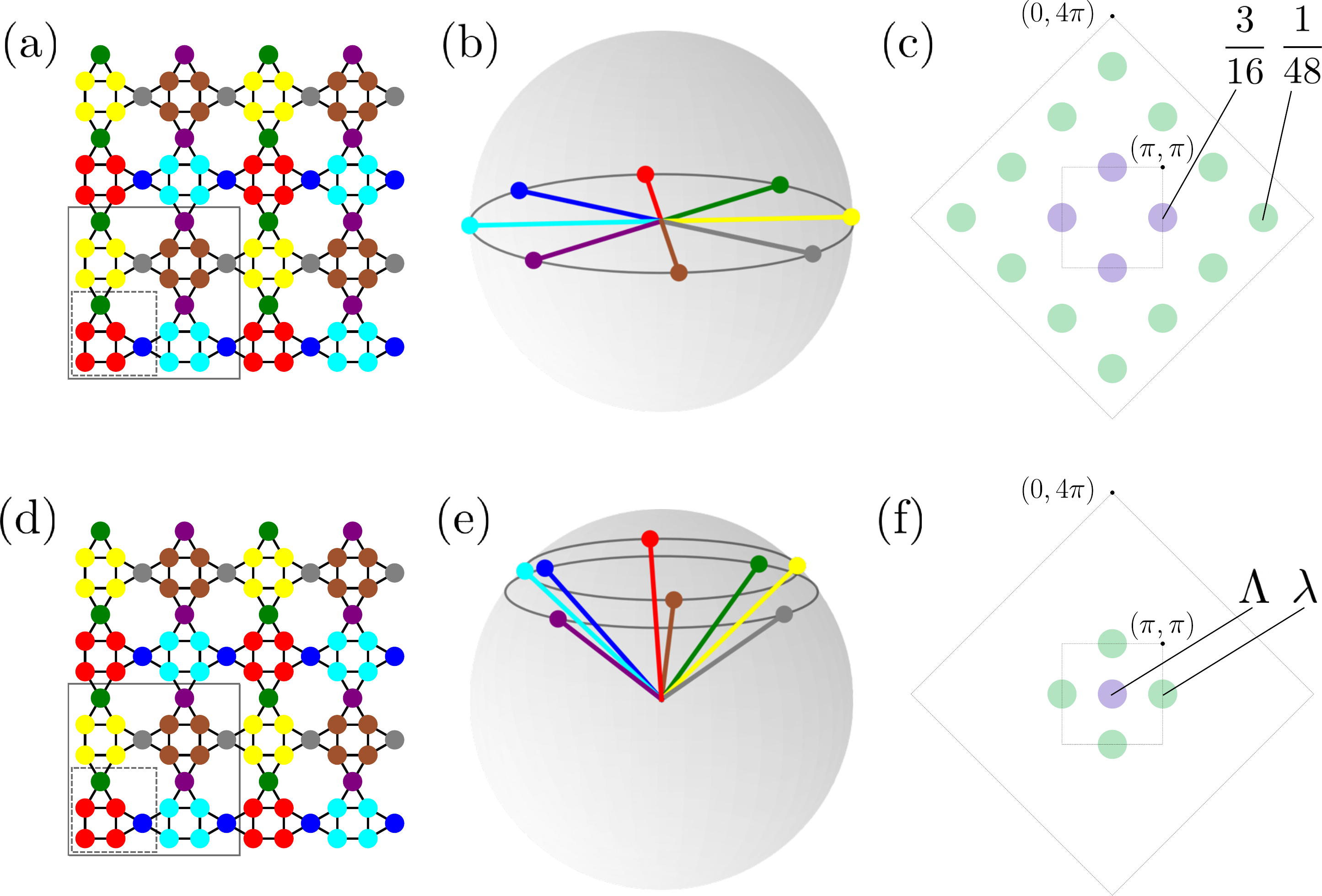}
	\caption{\textbf{Octagonal and double-conical orders.} 
	(a) Real space arrangement of spins in the coplanar octagonal ordered state 
	stabilized in the purely ferromagnetic nearest-neighbor model. 
	There are eight sublattices with a 24-site magnetic unit cell (large gray square), four times larger than the geometric unit cell (small square). 
	(b) Each sublattice of spins points towards a different corner of an octagon such that each pair of neighboring spins forms an angle of $\pi/4$. 
	(c) First and extended Brillouin zones showing the positions and the fractions of total spectral weight of the corresponding Bragg peaks. 
	(d) Real space arrangement of spins in the non-coplanar double cone state. 
	The eight sublattices coincide with those of the octagonal state shown in (b). 
	(e) The directions in which the spins on the sublattices point form two cones. 
	The first cone is formed by the four sublattices of spins located on the squares with mutual angles of $\pi/2$,
	whereas the second cone is formed by the four sublattices of bow-tie spins, again with mutual angles of $\pi/2$ but rotated $\pi/4$ with respect to the first cone. 
	(f) First and extended Brillouin zones showing the positions of the corresponding Bragg peaks. The ratio of the weight of the subdominant peaks $\lambda$ to the weight of the dominant peaks $\Lambda$ depends on the two polar angles of the cones and therefore on $J_+$ and $J_{\times}$.}
	\label{Fig:fm_states}
\end{figure}

Flipping the sign of all three couplings to ferromagnetic in the underlying nearest neighbor Heisenberg model, i.e.\ $J_1 = J_2 = J_3 = -1$,
the phase diagram for varying cross octagon-plaquette interactions changes its topology substantially. 
As depicted in Fig.~\ref{Fig:phase_diagram_fm} there are only three distinct phases. Trivially, there is a large ferromagnetic phase when $J_+, J_{\times} \leq 0$, which also extents to the other three quadrants in the phase diagram. 
Its ground state energy per site is given by
\begin{equation}
	E_{\text{FM}} = -2 + \frac{J_{\times}+J_+}{3} \,. 
	\label{eqn:E_fm}
\end{equation}
It is bound by the hyperbola defined by $(J_{\times } -1)(J_+ -1)= \tfrac{1}{2}$. 
Bound by a second hyperbola, given by $(J_{\times } + \tfrac{1}{\sqrt{2}})(J_+ + \tfrac{1}{\sqrt{2}})= 1$, there is a coplanar ordered phase with eight sublattices where the spins on each sublattice point to the corners of an octagon. The magnetic unit cell contains 24 sites and is four times larger than the geometric unit cell (cf. Fig.~\ref{Fig:fm_states} (a) - (c)). Its ground state has zero magnetization, $m = 0$, and its energy per site is given by
\begin{equation}
	E_{\text{octagonal}} = -\frac{1}{3} (2+2\sqrt{2}+J_++J _{\times}) \,. 
	\label{eqn:E_octa}
\end{equation}  
Between these two phases, which touch each other only at the point $J_+ = J_{\times} = 1 - \tfrac{1}{\sqrt 2}$, there exist non-coplanar umbrella-like states that smoothly interpolate between octagonal and FM order (cf. Fig.~\ref{Fig:fm_states}(d)-(f)). The magnetic unit cell of this phase coincides with the one of the octagonal phase, but the spins are no longer coplanar, but rather the directions in which the spins on the sublattices point form two cones. The first cone is formed by the four sublattices of spins located on the squares with mutual angles of $\pi/2$, whereas the second cone is formed by the four sublattices of bow-tie spins, again with mutual angles of $\pi/2$ but rotated $\pi/4$ with respect to the first cone. The two polar angles that describe these two cones depend on $J_+$ and $J_{\times}$, the ground state energy per site for those states can be calculated analytically and is given by
\begin{align}
	\notag E_{\text{conical}} = &-\frac{1}{6J_+J _{\times}} \bigg( J_+ + J _{\times} + 4 J_+J _{\times} \\
	& \pm (J_+ - J _{\times})\sqrt{1-12J_+J _{\times}+4J_+^2J _{\times}^2} \bigg) \,,
	\label{eqn:E_umbrella}
\end{align} 
where the plus sign applies to the left domain with $J_+ < 1 - \tfrac{1}{\sqrt 2}$, and the minus sign to the right domain with $J_+ > 1 - \tfrac{1}{\sqrt 2}$. While the ferromagnetic phase has uniform magnetization and the octagonal phase has zero magnetization, the magnetization in the conical phase depends on $J_+$ and $J_{\times}$. Analytically, one finds
\begin{align}
    \notag m_{\text{conical}} &= [(2 (J_{\times}+4) J_+\pm w)+1] [24 J_+ | J_{\times}|]^{-1} \\
    &\times \sqrt{\frac{\left(2 J_{\times}^2+1\right) w\pm \left(-4 J_{\times}^3 J_++6 J_{\times}^2+6 J_{\times} J_+-1\right)}{w}} \,.
\end{align}
Here, the plus sign applies to the right domain with $J_+ > 1 - \tfrac{1}{\sqrt 2}$, and the minus sign to the left domain with $J_+ < 1 - \tfrac{1}{\sqrt 2}$, and $w = \sqrt{4 J_{\times}^2 J_+^2 - 3J_{\times} J_+ + 1}$.

%%%%%%%%%%%%%%%%%%%%%%%%%%%%%%%%
\section{Discussion and Outlook}

We have unveiled a rich variety of magnetic textures that span the phase diagram of an extended classical Heisenberg model on the square-kagome lattice. Motivated by the possibility of stabilizing non-coplanar magnetic orders, we show that a minimal set of interactions needed to realize these involves introducing cross-plaquette interactions on top of the nearest-neighbor Heisenberg model, with either antiferromagnetic or ferromagnetic nearest-neighbor interactions, or a combination of both (ferromagnetic on bowtie bonds and antiferromagnetic on square bonds). A thorough classical Monte Carlo analysis reveals a plethora of non-coplanar states including a new type of cuboc order (dubbed cuboc3, Fig.~\ref{Fig:cuboc3_order}), as well as highly intricate incommensurate non-coplanar spirals. The underlying magnetic unit cells of these spin spiral states feature a highly complex structure and large sizes but, remarkably, we are able to provide a semi-analytical construction of these phases based on a symmetry-optimized parameterization. This renders possible obtaining explicit expressions (depending only on a small number of parameters) for their ground-state energy as a function of the coupling strengths, which in turn enables us to establish, with high precision, the phase boundaries between these complex spiral phases.  

Besides the ground state, we also study the thermo\-dynamics of non-coplanar states employing classical Monte Carlo simulations. By virtue of being chiral, non-coplanar states are expected to feature a symmetry breaking phase transition at $T\neq0$. In particular, for the cuboc order  we present the temperature evolution of the specific heat, the chiral order parameter, and its susceptibility, for different system sizes which manifestly exhibits signatures of a chiral phase transition. For the elementary model with only three symmetry inequivalent nearest-neighbor couplings, we show that, besides the isotropic point, within the entire region occupied by an extensively degenerate manifold of ground states one can always identify three distinct temperature regimes, namely, a high-temperature thermal paramagnet, an intermediate-temperature cooperative paramagnet, and a low-temperature coplanar regime selected via an order-by-disorder mechanism. We show that upon introduction of ferromagnetic cross-plaquette interactions of just one type, i.e., either $J_+$ or $J_\times$, the extensive degeneracy is reduced, but there still persists one zero mode per per triangle or per unit cell, respectively. This results in a $T\to0$ limiting value of specific heat which is less than one, however, the intermediate-temperature cooperative paramagnetic temperature regime disappears.

Since classical non-coplanar magnetic orders are characterized by a finite scalar spin chirality, it is plausible that, if quantum fluctuations are turned on and are successful in restoring spin rotational symmetry, e.g., in the extreme quantum limit of small spin $S=1/2$, the chiral symmetry breaking still persists and carries over in the resulting non-magnetic ground state~\cite{Bieri-2016,Hickey2016,Hickey2017}. Such quantum melting would give rise to novel chiral paramagnetic phases, characterized by a spontaneous breaking of time reversal and lattice symmetries, while preserving their product. One such example is the chiral quantum spin liquid, first elucidated by Kalmeyer and Laughlin~\cite{KL1987}, which hosts bulk semion excitations and a chiral gapless edge mode~\cite{Wen1990}. Furthermore, in our phase diagrams, the non-coplanar orders break lattice symmetries in such a way that simply restoring spin rotational symmetry would not fully restore all lattice symmetries \footnote{In other words, to use the language of \cite{Messio-2011}, the non-coplanar orders found here are not regular magnetic orders on the square-kagome lattice. A regular magnetic order would be a classical spin state which respects all lattice symmetries modulo global $O(3)$ spin transformations. Thus, restoring spin rotational symmetry would result in a fully symmetric state.}, potentially leading to descendent nematic chiral liquids for small values of spin~\cite{Lu-2016}. We thus provide a detailed symmetry analysis for the myriad of non-coplanar orders, paving the way for the systematic classification of descendant chiral spin liquid states. It should be noted here that, unlike the kagome lattice, the square-kagome has an {\it even} number of sites per unit cell. This leaves open the possibility of fully trivial paramagnetic phases appearing, as the Lieb-Schultz-Mattis-Hastings-Oshikawa theorem does not preclude a fully symmetric, yet topologically trivial, gapped phase~\cite{Lieb1961,Oshikawa2000,Hastings2004}. For the non-coplanar umbrella-like double cone states with $m\neq0$, interpolating between the ferromagnetic and octagonal orders, the proximity to ferromagnetism opens the exciting possibility of realizing in the corresponding quantum model the elusive spin nematic orders~\cite{Shannon-2006,Iqbal-2016,Jiang-2022}. The spin rotational symmetry breaking in these quadrupolar ordered states is described by a time-reversal invariant order parameter, given by a symmetric traceless rank-2 tensor~\cite{Andreev-1984} in contrast to a conventional dipolar order parameter which breaks time-reversal symmetry. The exploration of the aforementioned exotic phases in the corresponding quantum models employing state-of-the-art numerical quantum many-body approaches constitute an important direction of future research.
 
In the context of material realizations, it is worth noting that in Na$_6$Cu$_7$BiO$_4$(PO$_4$)$_4$[Cl,(OH)]$_3$~\cite{Liu-2022}, the $J_{\times}$ interactions are mediated by Cu-O-Na-O-Cu superexchange pathways, while the $J_{+}$ bonds pass through a chloride group but distances that prevent any Cu-Cl hybridization, and thus not triggering a superexchange. The presence of the nonmagnetic Na ions in the center of the octagons is likely to trigger a finite $J_{\times}$ interaction akin to the scenario realized in the kagome based materials kapellasite and haydeeite~\cite{Iqbal-2015}. One could think of possible chemical substitutions which are likely to enhance this interactions, e.g., by replacing Na with Cs which has a larger ionic radius. Another route towards strengthening the cross-plaquette couplings would involve preparing the corresponding sulfide version instead of oxides leading to Cu-S-Na-S-Cu type superexchange pathways, similar to the breathing chromium spinels~\cite{Ghosh-2019spinel}. In KCu$_6$AlBiO$_4$(SO$_4$)$_5$Cl~\cite{Fujihala2020}, it is the sulfate SO$_{4}^{2-}$ that occupies the octagon centers and one may consider the possibility of substituting it with a selenate group SeO$_{4}^{2-}$ to enhance both the cross-plaquette couplings. These would constitute interesting future explorations on the material synthesis front.

%%%%%%%%%%%%%%%%%%%%%%%%%%%%%%%%
\acknowledgments 
We thank Harald O. Jeschke for helpful discussions.
The Cologne group acknowledges partial funding from the DFG within Project-ID 277146847, SFB 1238 (projects C02, C03).
S.T. thanks the Center for Computational Quantum Physics at the Flatiron Institute, 
New York, for hospitality during the initial stages of this project.
M.G. thanks the Bonn-Cologne Graduate School of Physics and Astronomy (BCGS) for support.
The numerical simulations were performed on the JUWELS cluster at the Forschungszentrum Juelich and the CHEOPS cluster at RRZK Cologne.
 Y.\,I. acknowledges support from the Department of Science and Technology (DST), India through the MATRICS Grant No.~MTR/2019/001042, CEFIPRA Project No. 64T3-1, the ICTP through the Associates Programme and from the Simons Foundation through grant number 284558FY19. The research of Y.I. was supported, in part, by the National Science Foundation under Grant No.~NSF~PHY-1748958 during a visit to the Kavli Institute for Theoretical Physics (KITP), UC Santa Barbara, USA for participating in the program ``A Quantum Universe in a Crystal: Symmetry and Topology across the Correlation Spectrum'', IIT Madras through the Institute of Eminence (IoE) program for establishing QuCenDiEM (Project No. SB20210813PHMHRD002720), the International Centre for Theoretical Sciences (ICTS), Bengaluru, India during a visit for participating in the program ``Frustrated Metals and Insulators'' (Code: ICTS/frumi2022/9). The work of Y. I. and S. T. was performed, in part, at the Aspen Center for Physics, which is supported by National Science Foundation grant PHY-2210452. The participation of Y. I. at the Aspen Center for Physics was supported by the Simons Foundation. Y.~I.~acknowledges the use of the computing resources at HPCE, IIT Madras.

%%%%%%%%%%%%%%%%%%%%%%%%%%%%%%%%
\bibliography{shuriken}

%%%%%%%%%%%%%%%%%%%%%%%%%%%%%%%%
\newpage
\appendix

%%%%%%%%%%%%%%%%%%%%%%%%%%%%%%%%
\section{Monte Carlo simulations}
%%%%%%%%%%%%%%%%%%%%%%%%%%%%%%%%

Our Monte Carlo simulations are performed on finite lattices of $L \times L$ unit cells, i.e. $N = 6L^2$ spins, with periodic boundary conditions. Unless explicitly stated otherwise, we choose $L = 12 \ (N = 864)$. 

Local updates are performed with the Metropolis-Hastings algorithm. In order to resolve the thermal selection of ground states by subtle thermal order-by-disorder effects at very low temperatures, we employ a parallel tempering/replica exchange Monte Carlo scheme \cite{hukushima1996exchange, swendsen1986replica}, where 192 logarithmically spaced temperature points between $T_{\text{min}} = 10^{-4}$ and $T_{\text{max}} = 10$ are simulated simultaneously such that after every sweep (consisting of $N$ local update attempts), spin configurations of neighboring replicas are attempted to be exchanged with a certain probability. As a result, individual replicas make a random walk in temperature space and thus can escape local minima at low temperatures easily. We have carefully checked the thermalization of our parallel temperature scheme with feedback-optimized temperatures \cite{katzgraber2006feedback}.

For the thermodynamics of the $120^\circ$ phase (cf. Fig.~\ref{Fig:fss_120}) and the cuboc1 phase (cf. Fig.~\ref{Fig:fss_cuboc}), a conventional Monte Carlo scheme without parallel tempering and 192 linearly spaced temperature points between $T_{\text{min}} = 0.1$ and $T_{\text{max}} = 0.5$ is employed.

In all cases, measurements are performed over $5 \cdot 10^8$ sweeps after a thermalization period of $10^8$ sweeps.

%%%%%%%%%%%%%%%%%%%%%%%%%%%%%%%%
\section{Finite-size scaling}
%%%%%%%%%%%%%%%%%%%%%%%%%%%%%%%%

\begin{figure}[h!]
	\centering
	\includegraphics[width=1.0\linewidth]{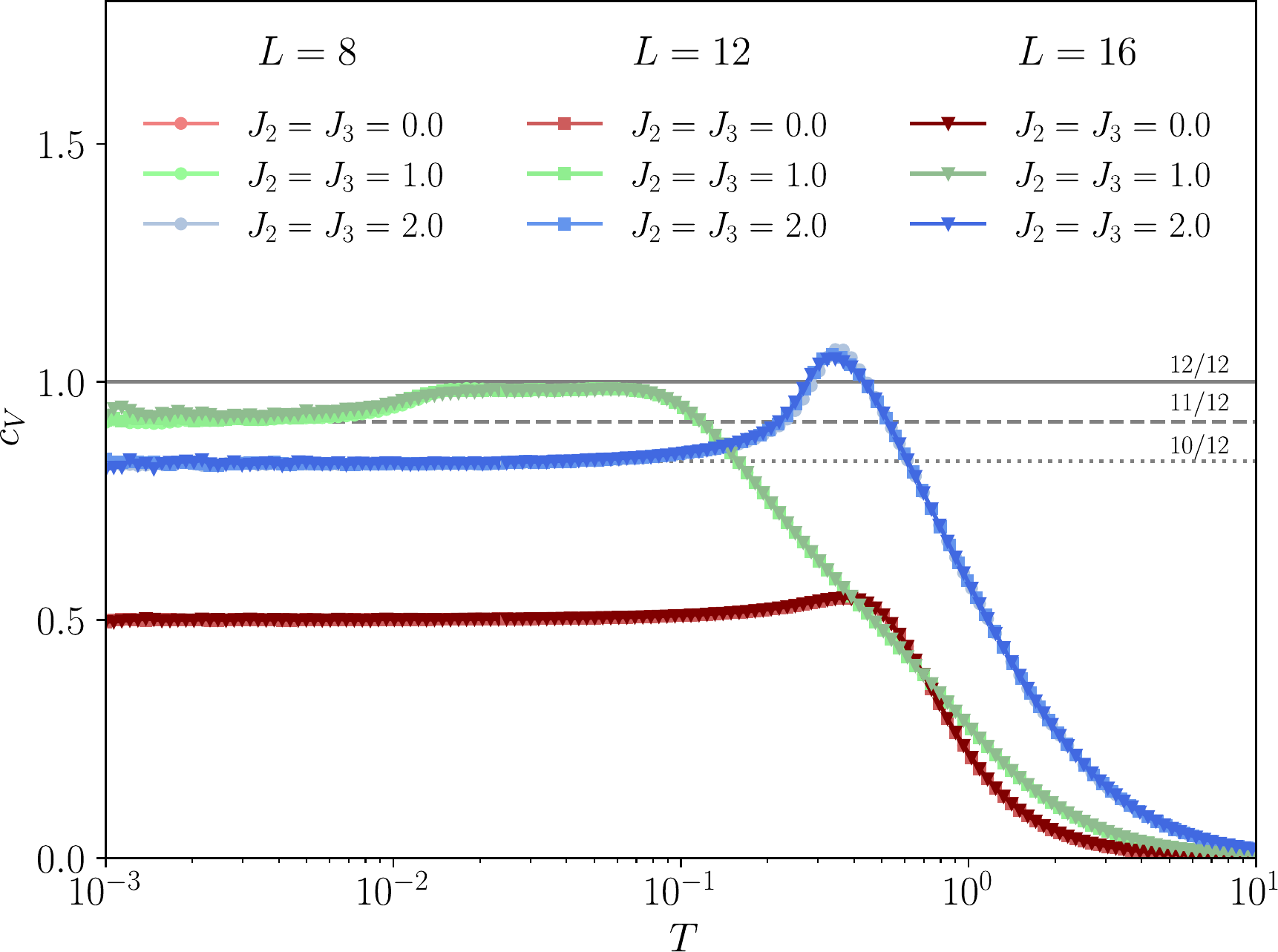}
	\caption{\textbf{Finite-size scaling of $c_V$ for selected parameters in the nearest-neighbor model.}
    Specific heat traces for $J_2 = J_3 = 0.0, 1.0, 2.0$ for different system sizes $L = 8, 12, 16$, showing no system size dependence.
	}
	\label{Fig:nn_fss}
\end{figure}

\begin{figure}[h!]
	\centering
	\includegraphics[width=1.0\linewidth]{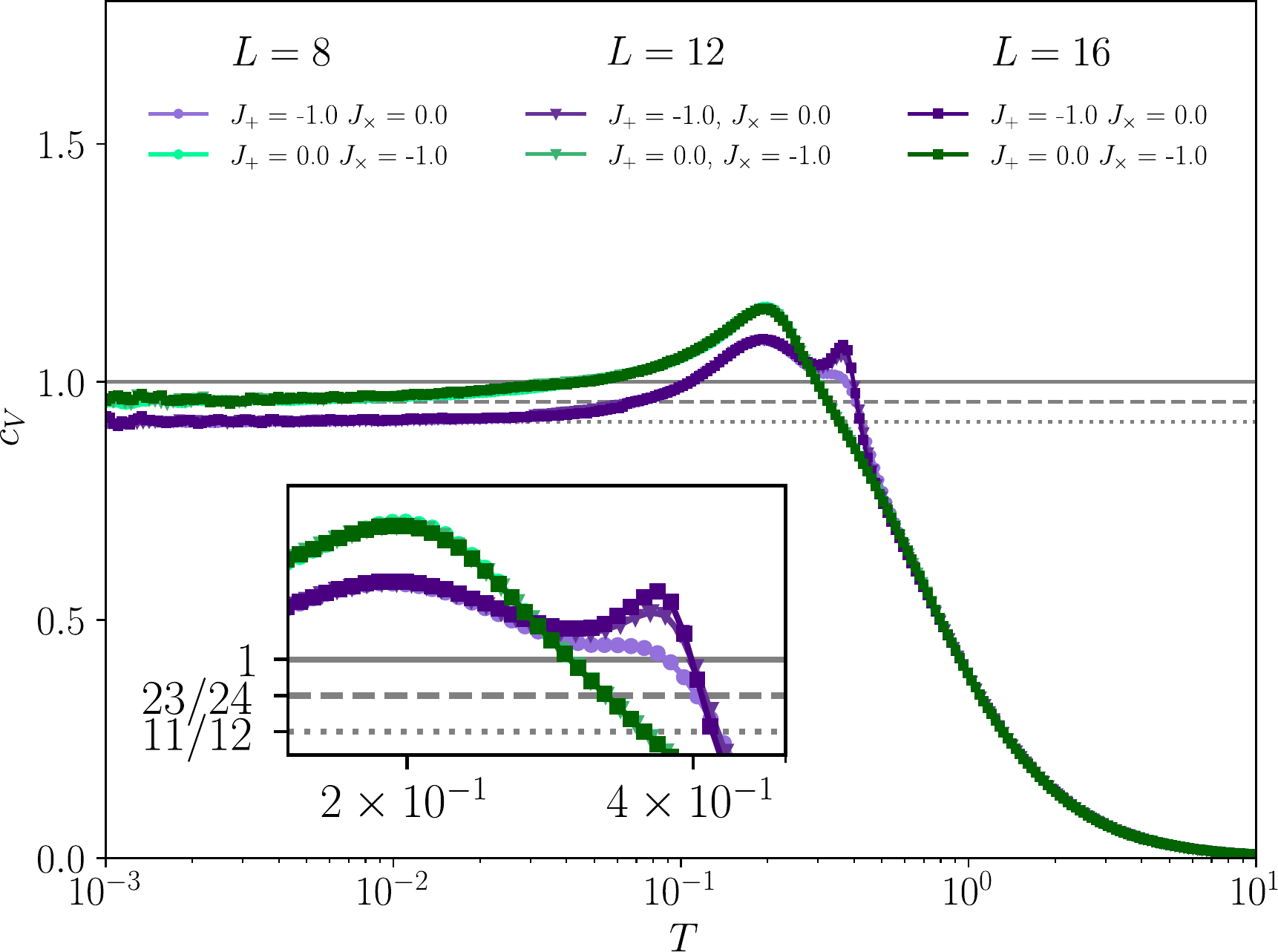}
	\caption{\textbf{Finite-size scaling of $c_V$ on the axes of the extended model.}
    Specific heat traces for $J_+ = -1.0, J_{\times} = 0.0$ and $J_+ = 0.0, J_{\times} = -1.0$ $(J_2 = J_3 = 1.0)$ for different system sizes $L = 8, 12, 16$.
	}
	\label{Fig:axes_fss}
\end{figure}

We present finite size behaviour for the data shown in Figs.~\ref{Fig:classical_phase_diagram}(b) and \ref{Fig:cv_on_axes} in the main text.
Fig.~\ref{Fig:nn_fss} shows data corresponding to the data shown in Fig.~\ref{Fig:classical_phase_diagram}(b), but for three system sizes, $L = 8, 12, 16$. For all three parameter sets $J_2 = J_3 = 0.0, 1.0, 2.0$, the curves fall onto each other for all three system sizes.
In Fig.~\ref{Fig:axes_fss}, the finite size behaviour corresponding to Fig.~\ref{Fig:cv_on_axes} is plotted. For $J_+ = 0.0, J_{\times} = -1.0$ $(J_2 = J_3 = 1.0)$, the curves, again, fall onto each other for all system sizes. In the case $J_+ = -1.0, J_{\times} = 0.0$, the feature at higher temperatures scales with increasing system size, while the feature at lower temperatures does not. As explained in Sec.~\ref{sec:physics_axes}, in the case of ferromagnetic $J_+$, the bow-tie spins align ferromagnetically, while the remaining spins are free to rotate as long as long as each triangle maintains a coplanar $120^\circ$ configuration. Thus, the feature scaling with the system size in this case can be assigned to the ferromagnetic ordering of the bow-tie spins. \\

%%%%%%%%%%%%%%%%%%%%%%%%%%%%%%%%
\clearpage
%%%%%%%%%%%%%%%%%%%%%%%%%%%%%%%%

\onecolumngrid
\begin{figure*}[ht!]
	\centering
	\includegraphics[width=1.0\linewidth]{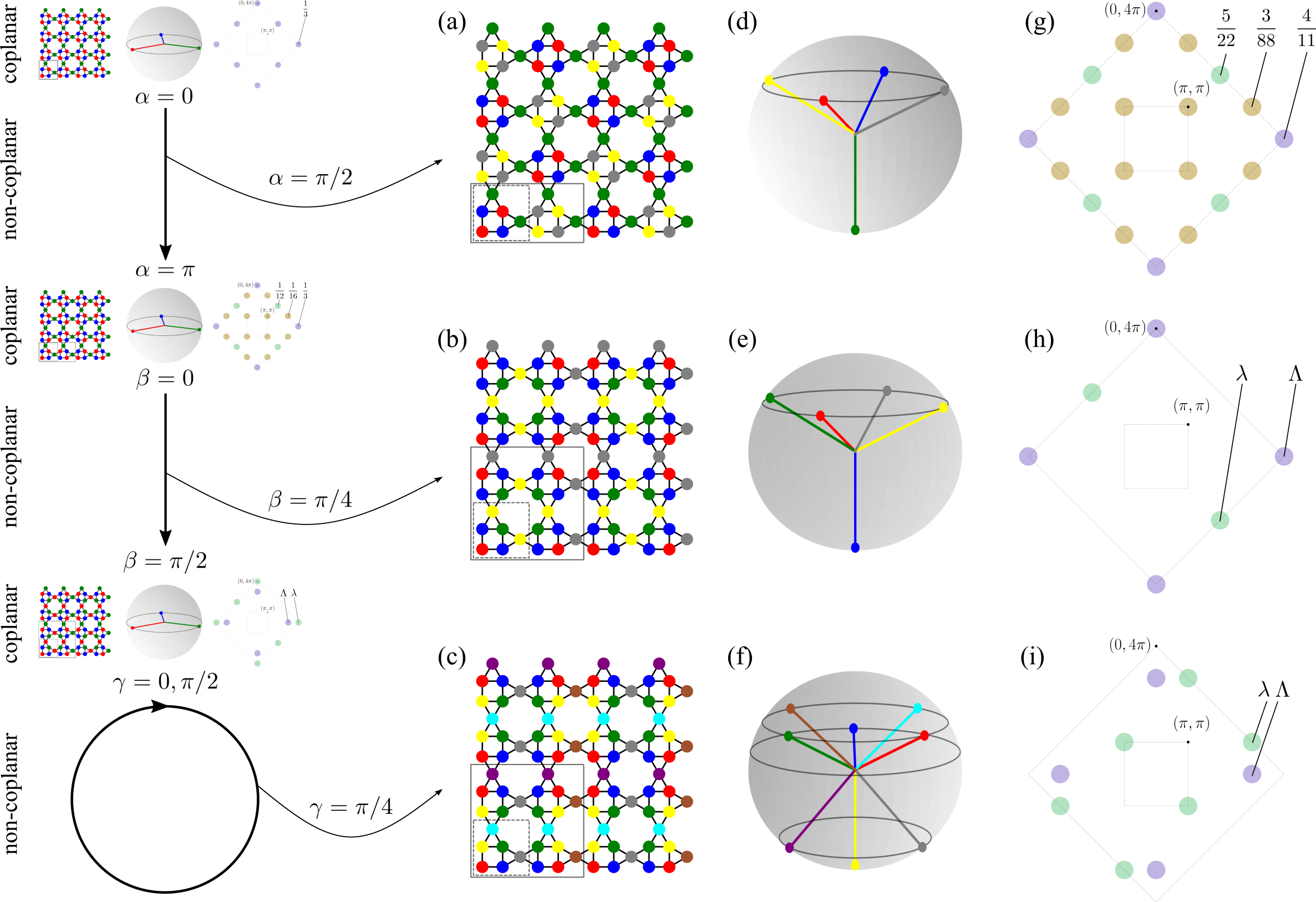}
	\caption{\textbf{Three families of non-coplanar orders arising from coplanar $120^\circ$ order.}
    In the $120^\circ$ phase, there are three one-parameter families of non-coplanar states, each described by a single angle $\alpha$, $\beta$, or $\gamma$, whose limiting cases yield the coplanar orders presented in Fig.~\ref{Fig:120_degree_order}, as sketched on the left. The first family, here depicted in (a)+(d)+(g) by its real space arrangement, a schematic common origin plot, and its structure factor, respectively, for $\alpha = \pi/2$, interpolates between the coplanar $\mathbf{q}=0$ state in Fig.~\ref{Fig:120_degree_order}(a)+(d)+(g) (limit $\alpha = 0$) and the coplanar state Fig.~\ref{Fig:120_degree_order}(b)+(e)+(h) (limit $\alpha = \pi$). The second family, here depicted in (b)+(e)+(h) by its real space arrangement, a schematic common origin plot, and its structure factor, respectively, for $\beta = \pi/4$, interpolates between the coplanar state in Fig.~\ref{Fig:120_degree_order}(b)+(e)+(h) (limit $\beta = 0$) and the coplanar state Fig.~\ref{Fig:120_degree_order}(c)+(f)+(i) (limit $\beta = \pi/2$). The ratio of the weight of the subdominant peaks $\lambda$ to the weight of the dominant peaks $\Lambda$ in the structure factor is $\lambda/\Lambda = 40 \%$.  The third family, here depicted in (c)+(f)+(i) by its real space arrangement, a schematic common origin plot, and its structure factor, respectively, for $\gamma = \pi/4$, has the coplanar state in Fig.~\ref{Fig:120_degree_order}(c)+(f)+(i) for both limits $\gamma = 0$ and $\gamma = \pi/2$ and non-coplanar states in between. The ratio of the weight of the subdominant peaks $\lambda$ to the weight of the dominant peaks $\Lambda$ in the structure factor is $\lambda/\Lambda \approx 39 \%$.
	}
	\label{Fig:120_noncoplanar}
\end{figure*}
\twocolumngrid

%%%%%%%%%%%%%%%%%%%%%%%%%%%%%%%%
\section{Coplanar and non-coplanar $120^\circ$ orders}
%%%%%%%%%%%%%%%%%%%%%%%%%%%%%%%%

The $120^\circ$ order found for ferromagnetic $J_+$ and $J_\times$ possesses a rather complex ground state manifold, with three distinct families of non-coplanar configurations connected via three coplanar configurations. Here, we give a more detailed description of the non-coplanar states and a brief discussion of the thermal order-by-disorder effects that select the coplanar configurations at finite temperatures.  

%%%%%%%%%%%%%%%%%%%%%%%%%%%%%%%%
\subsection*{Non-coplanar order}
\label{app:120_noncoplanar}
%%%%%%%%%%%%%%%%%%%%%%%%%%%%%%%%

The three coplanar ground states (a), (b) and (c) described in Fig.~\ref{Fig:120_degree_order} all satisfy the following rules.
\begin{itemize}
    \item  $120^\circ$   rule: any two spin vectors  of adjacent squagome sites form angles of $120^\circ$,
    \item  pairing rule: lattice sites with FM octagon-plaquette interactions (i.~e.~$J_+,J_\times <0$) are occupied by identical spin vectors.
\end{itemize}

Conversely, any spin configuration obeying these rules has an energy of $E_{120^\circ}=-1+\frac{1}{3}\left(J_++J_\times\right)$ and thus represents a ground state. However, in addition to the above three coplanar ground states, these rules are also satisfied by certain $1$-parameter families of, in general, non-coplanar spin configurations. These can be further characterized by the fact that the sequence of spin vectors in the squares of unit cells is either of the form $\left({\mathbf a}, {\mathbf b}, {\mathbf a}, {\mathbf b} \right)$ (1st family) or $\left({\mathbf a}, {\mathbf b}, {\mathbf a}, {\mathbf c}\right)$ (2nd family) or $\left({\mathbf a}, {\mathbf b}, {\mathbf c}, {\mathbf d}\right)$ (3rd family). It turns out that the 1st family connects the coplanar ground states (a) and (b) of Fig.~\ref{Fig:120_degree_order}, the 2nd one the coplanar ground states (b) and (c), and the third family loops (c) with itself,
see Figure  \ref{Fig:120_noncoplanar}. %Order by disorder calculations, however, show (cf. Fig.~\ref{Fig:obd_correction}) that coplanar states are selected over these non-coplanar states by a thermal order by disorder mechanism at small, but finite temperatures. 
Moreover, in the region of the deformed $120^\circ$ phase, there is another one-parameter family of non-coplanar ground states. For more details see the Mathematica files in the Supplement \cite{Supplement}.

%%%%%%%%%%%%%%%%%%%%%%%%%%%%%
%\section{Probabilities of different coplanar $120^\circ$ orders}
\subsection*{Order-by-disorder selection of coplanar order}
\label{app:120_histograms}
%%%%%%%%%%%%%%%%%%%%%%%%%%%%%%%%

\begin{figure}[h]
	\centering
	\includegraphics[width=1\columnwidth]{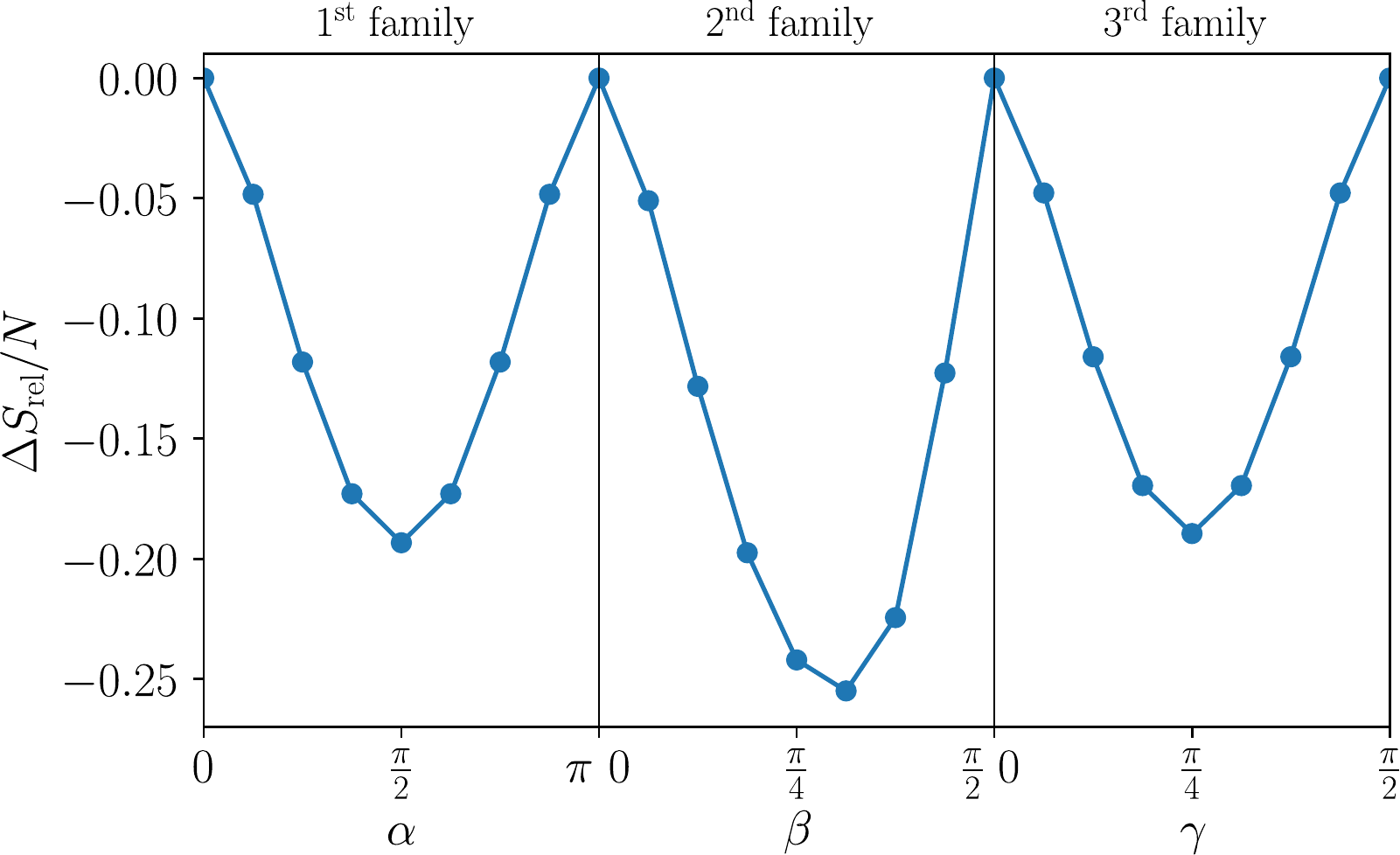}
	\caption{\textbf{Order by disorder correction.} The entropy correction $\Delta S_\text{rel}$ relative to the coplanar correction for the three one-parameters families of non-coplanar order in the $120^\circ$ phase as a function of the parameters $\alpha$, $\beta$, and $\gamma$ shows that the entropy is maximal for the coplanar $120^\circ$ orders $(\alpha = 0, \pi; \beta = 0, \pi/2; \gamma = 0, \pi/2)$.}
	\label{Fig:obd_correction}
\end{figure}

The energy of the three families of non-coplanar ground states, as well as the three coplanar ground states, are all identical. However, as these spin configurations are not related by a symmetry of the Hamiltonian, their free energy, the relevant quantity that needs to be minimized at finite temperature, will not be identical. The selection of a subset of ground states via such a thermal order-by-disorder mechanism is a well-studied phenomenon within the context of frustrated magnets. By expanding to quadratic order in classical fluctuations about a particular ordered ground state one can compute the lowest-order correction to the entropy. With the free energy $F=E-TS$, the states with the highest entropy will be selected at finite temperatures which, as shown in Fig.~\ref{Fig:obd_correction}, are precisely the three coplanar states. Interestingly, these all possess the exact same correction at this lowest quadratic order, meaning that the ultimate selection of the ground state must be via higher-order non-linear corrections. The calculation of such corrections are well beyond the scope of the current manuscript so we instead apply a numerical approach to tease out the preferred ground state.     

\begin{figure}[h]
	\centering
	\includegraphics[width=1\linewidth]{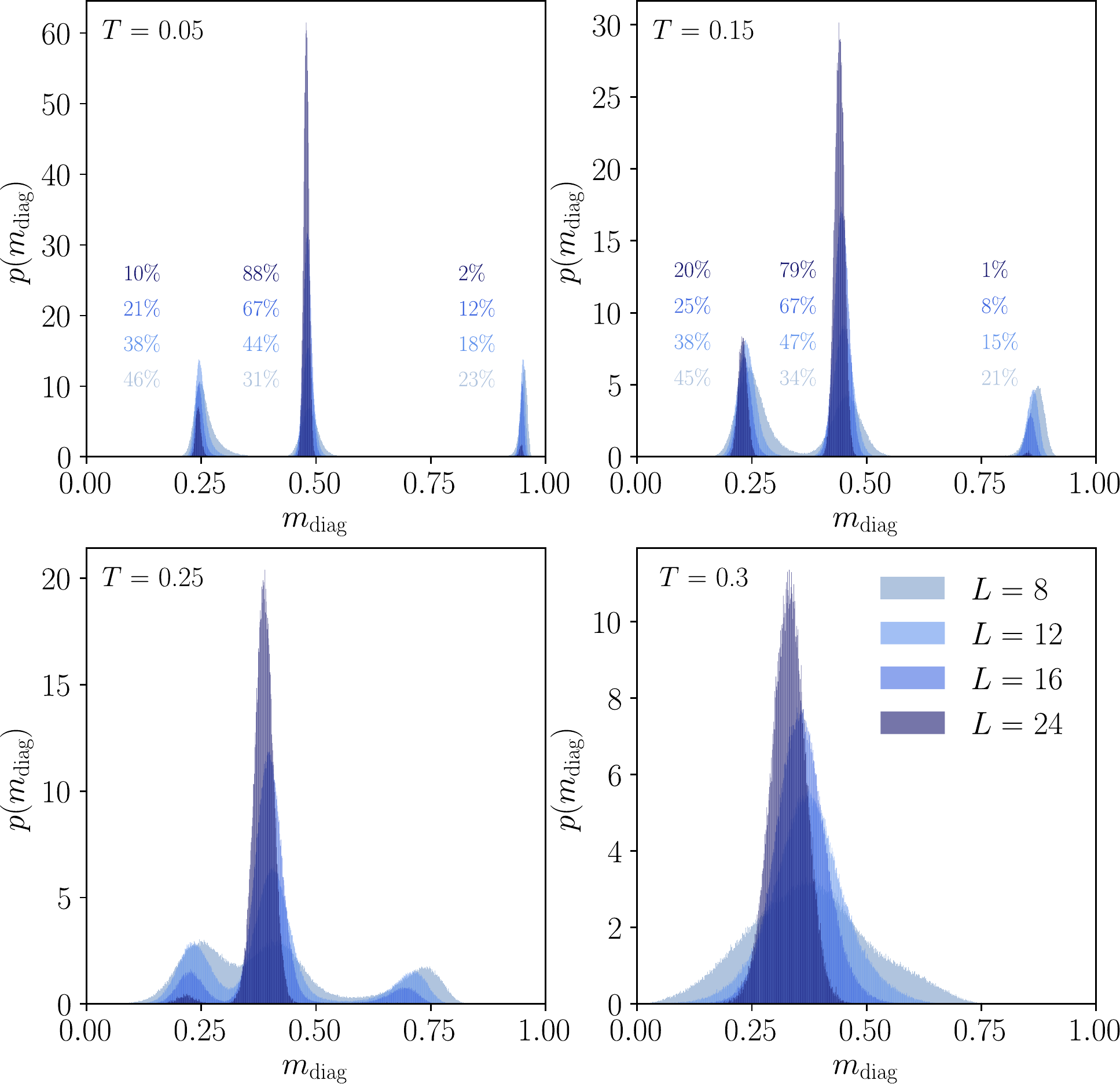}
	\caption{\textbf{Square diagonal magnetization distribution in the $120^\circ$ phase.} Histograms from Monte Carlo timeseries showing the equilibrium distribution of $m_{\text{diag}}$ in phase VI $(J_+ = J_{\times} = -1.0)$ for different system sizes at different temperatures. }
	\label{Fig:120_histograms}
\end{figure}

The three different coplanar $120^\circ$ orders,  shown in Fig.~\ref{Fig:120_degree_order}, each have a distinct value of $m_{\text{diag}}$. In Fig.~\ref{Fig:120_histograms}, we show histograms from Monte Carlo timeseries showing the equilibrium distribution of $m_{\text{diag}}$ within the $120^\circ$ phase $(J_+ = J_{\times} = -1.0)$ for different system sizes at different temperatures.  Below the phase transition at $T=0.27$ (cf. Fig.~\ref{Fig:fss_120}), the distribution of $m_{\text{diag}}$ splits up into three distinct peaks, corresponding to the three different coplanar $120^\circ$ orders. At the lowest temperature, $T=0.05$, and for $L=24$, the probability of $\mathbf{q}=0$ order (Fig.~\ref{Fig:120_degree_order}(a)) is $2\%$, whereas the orders Fig.~\ref{Fig:120_degree_order}(b) and (c) have probabilities of $88\%$ and $10\%$, respectively. We thus suspect that the order in (b) is the likely spin configuration favored by thermal fluctuations at finite temperatures, on the largest system sizes. Only for smaller system sizes, $\mathbf{q}=0$ order occurs with significant probability. Overall, the probabilities only slightly change between $T=0.05$ and $T=0.15$.

%%%%%%%%%%%%%%%%%%%%%%%%%%%%%
\clearpage
%%%%%%%%%%%%%%%%%%%%%%%%%%%%%%%%

\onecolumngrid
\begin{figure*}[ht!]
	\centering
	\includegraphics[width=1\linewidth]{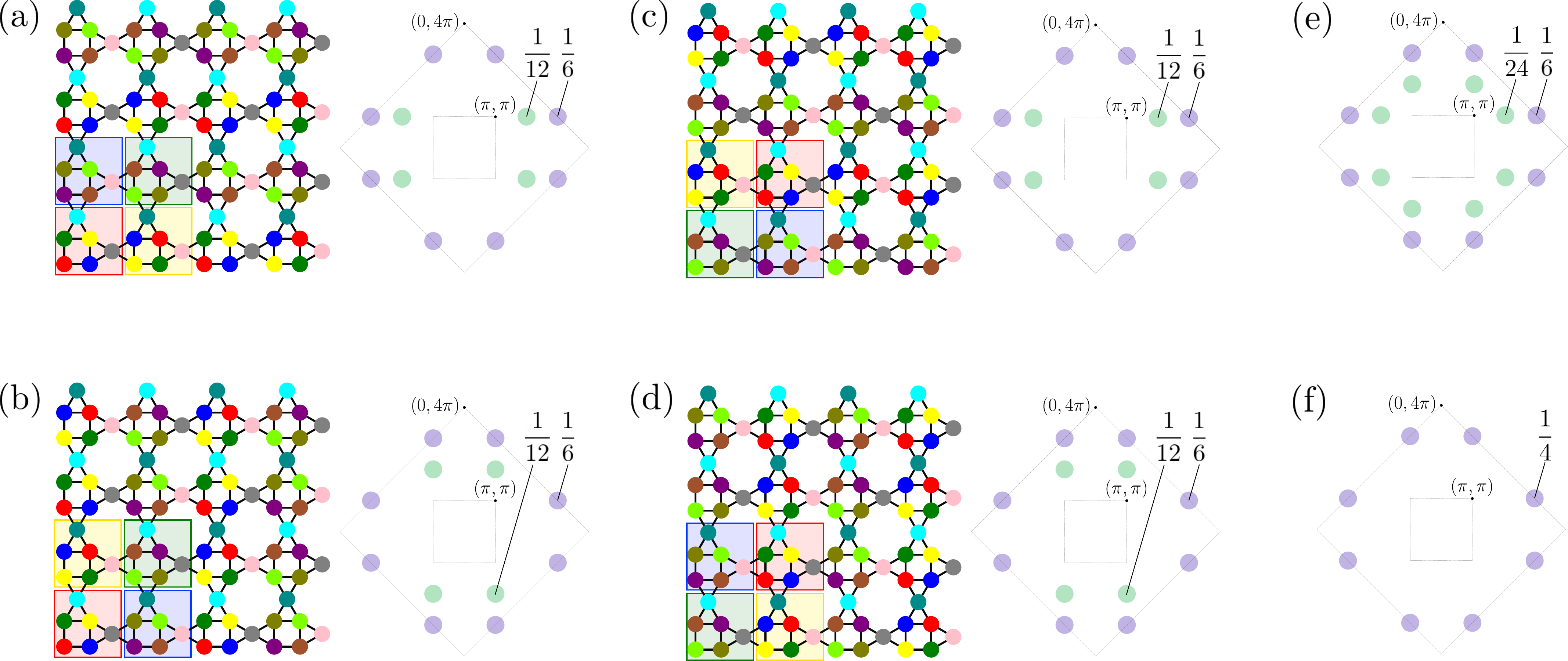}
	\caption{\textbf{Variants of cuboc1 order on the square-kagome lattice.} (a)-(d) show four different possible arrangements of cuboc1 order on the square-kagome lattice, each with its corresponding structure factor. The arrangement in (a) corresponds to the arrangement shown in Fig.~\ref{Fig:cuboc1_order} in the main text and has a clockwise placement of the building blocks highlighted in blue, green, yellow, and red, in the magnetic unit cell. The arrangement in (c) also has this clockwise placement, but in a different order, and has the same structure factor as (a). The other two arrangements, (b) and (d), on the other hand, have an anti-clockwise placement of blue, green, yellow, and red, resulting in a structure factor that is rotated by $\pi/2$ (two additional arrangements are possible in each case, but not explicitly shown). (e) Averaging the real space correlations of a clockwise and an anti-clockwise arrangement results in a structure factor with twice as much subdominant peaks. (f) The averaged real space correlations of all possible cuboc1 orders yields a structure factor with only main peaks on the edges of the extended Brillouin zone without further subdominant peaks.}
	\label{Fig:cuboc1_orders}
\end{figure*}
\twocolumngrid

%%%%%%%%%%%%%%%%%%%%%%%%%%%%%
\section{Different cuboc1 order realizations}
\label{sec:cuboc1_orders}
%%%%%%%%%%%%%%%%%%%%%%%%%%%%%%%%

There are several different possibilities to set up cuboc1 order on the square-kagome lattice. These can be best described by dividing the 24-site magnetic unit cell into four building blocks, each encompassing one geometric unit cell, with some specific fixed arrangement of spins -- indicated as blue, green, yellow, and red squares in Fig.~\ref{Fig:cuboc1_orders}, where four possible such arrangements are shown. These building blocks can be arranged clockwise or anti-clockwise, each in four different ways. The order parameter to distinguish between clockwise and anti-clockwise arrangements of blue-green-yellow-red is
\begin{equation}
    \mathcal{O} = \frac{1}{M} \sum_{\bowtie} \mathcal{O}_{ijklm}
    \label{eq:cuboc1_op}
\end{equation}
where
\begin{equation}
    \mathcal{O}_{ijklm} = \frac{1}{\sqrt{2}}\Vec{S}_i \cdot \left[ (\Vec{S}_j - \Vec{S}_k) \times (\Vec{S}_l - \Vec{S}_m)\right]
    \label{eq:cuboc1_op_susc}
\end{equation}
is summed over all skew bow-ties as shown in the inset of the lower panel of Fig.~\ref{Fig:fss_cuboc} and $M$ is the number of skew bow-ties. For cuboc1 order with a clockwise (anti-clockwise) arrangement of blue-green-yellow-red, $\mathcal{O}$ yields $+1$ $(-1)$. $\mathcal{O}$ and its absolute value, $|\mathcal{O}|$, from Monte Carlo simulations are shown in Fig.~\ref{Fig:fss_cuboc} as well as the associated susceptibility
\begin{equation}
    \chi_{\mathcal{O}} = \frac{M}{T}(\langle\mathcal{O}^2\rangle - \langle|\mathcal{O}|\rangle^2) \,.
\end{equation}

%%%%%%%%%%%%%%%%%%%%%%%%%%%%%
\section{Spiral phases for AFM interactions}
\label{sec:spiral}
%%%%%%%%%%%%%%%%%%%%%%%%%%%%%%%%

In this section we will briefly characterize the remaining spiral phases which are not described in the main text. In all cases, the procedure is based on the semi-analytical method described in Section \ref{sec:semi-analytical}. For all scenarios the nearest-neighbor couplings are set to $J_1 = J_2 = J_3 = 1.0$ and the numerical Monte Carlo ground states are taken at a temperature of $T = 10^{-4}$.  

\subsubsection{Spiral phase IV with {$D_{6}^{s,t}$  symmetry}}
\begin{figure}[t]
	\centering
	\includegraphics[width=1\columnwidth]{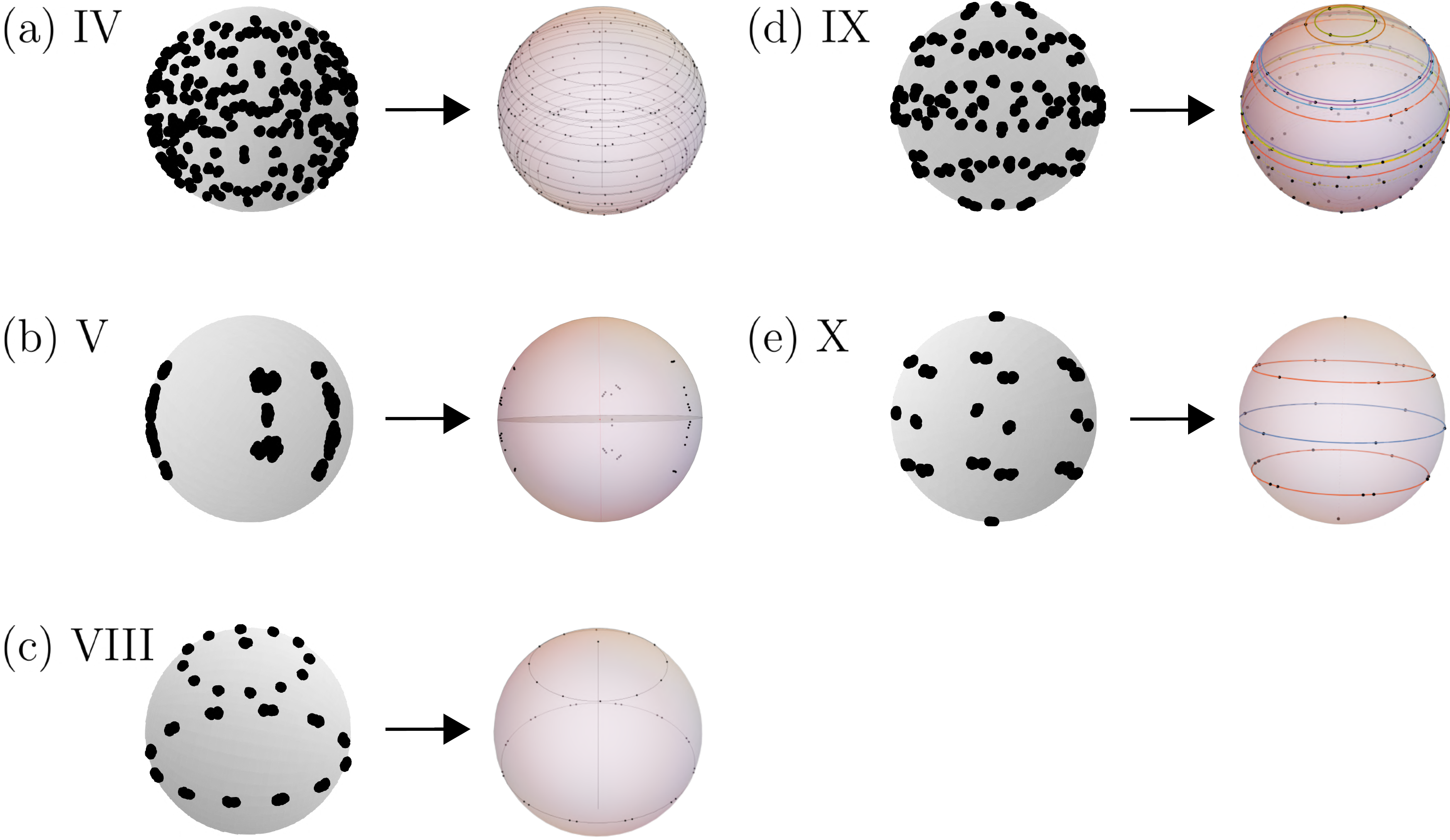}
	\caption{\textbf{Spin spiral phases.} Shown are common origin plots (left column) of the numerical Monte Carlo ground states of spiral phases IV (a), V (b), VIII (c), IX (d), and X (e), at $T = 10^{-4}$. Each common origin plot depicts the spin direction of all $N = 864$ spins on the unit sphere. The right column shows the result of reducing the number of spin directions by grouping spins that point to the same direction which eventually allows for the symmetry analysis and analytical description outlined in the text.}
	\label{Fig:spiral_phases}
\end{figure}

We consider the ground state for $J_+ = J_{\times} = -0.2$. On the left of Fig.~\ref{Fig:spiral_phases} (a), a common origin plot of the $N = 864 \ (L = 12)$ spin vectors of the numerical ground state (at $T = 10^{-4}$) is shown. After grouping those spins that point to the same direction, we are left with $M = 228$ unique spin directions which are shown on the right of Fig.~\ref{Fig:spiral_phases} (a). The symmetry group $G$ of this state is generated by two reflections and rotations of $\pi/3$ about the symmetry axis i.e.\ { $G = D_{6}^{s,t}$}, where the { superscripts imply additional mirror symmetries} on the $xy$-plane {and the $yz$-plane}, i.e.\ $s = \text{diag}(1,1,-1)$ and  $t = \text{diag}(-1,1,1)$. The energy per site can eventually be described as a function of 19 parameters and numerical minimization of this energy with $J_+ = -0.2$ then leads to $E_{\text{IV},\text{semi-analytical}} = -1.10525$ in accordance with the Monte Carlo result $E_{\text{IV},\text{MC}} = -1.1051(4)$
for the same parameters. 
For more details see the explicit description in the  Mathematica files of the Supplement \cite{Supplement}.
 
Specific heat traces for this spiral phase for different system sizes are shown in the top panel of Fig.~\ref{Fig:fss_cuboc_spirals}. 
However, due to the incommensurability of these spin spirals there are significant finite-size effects, 
making it hard to narrow down the location of the finite-temperature transition.

%%%%%%%%%%%%%%%%%%%%%%%%%%%%%
\subsubsection{Spiral phase V with {$\{\text{id},s\}$}  symmetry}
%%%%%%%%%%%%%%%%%%%%%%%%%%%%%%%%

For $J_+ = 0.2$ and $J_{\times} = -1.0$, there are $M = 47$ unique spin vectors left after grouping the $N = 864$ ground state spin vectors according to unique directions (cf. Fig.~\ref{Fig:spiral_phases} (b)). There is no exact rotation symmetry but a reflection symmetry with respect to the equatorial plane, thus the symmetry group of this phase is $\lbrace\text{id,s}\rbrace$. The energy per spin then is a function of 48 parameters and yields, after minimization, $E_{\text{V},\text{semi-analytical}} = -1.36719$ in accordance with the Monte Carlo result $E_{\text{V},\text{MC}} = -1.367(1)$ for the same parameters. 
For more details see the explicit description in the  Mathematica files of the Supplement \cite{Supplement}.

%%%%%%%%%%%%%%%%%%%%%%%%%%%%%%%%
\subsubsection{Spiral phase VIII with $D_{12}$ symmetry}
%%%%%%%%%%%%%%%%%%%%%%%%%%%%%%%%
The ground state for $J_+ = -1.0$ and $J_{\times} = 1.0$ (Fig.~\ref{Fig:spiral_phases} (c)) can be constructed depending on two polar angles $\alpha$ and $\beta$ and one azimuthal angle $\gamma$ as follows. Let $\Vec{S}_i \ (i = 1,\dots,6)$ be the six spin vectors of the primitive unit cell with $\Vec{S}_1$ and $\Vec{S}_2$ being the bow-tie spins. One can define 
\begin{align} 
\notag\Vec{S}_1  &= (0,0,1), \\
\notag \Vec{S}_2 &= (\sin \alpha, 0, \cos \alpha), \\
\notag \Vec{S}_3 &= (\sin \beta \cos \gamma, \sin \beta \sin \gamma, \cos \beta), \\
\notag \Vec{S}_4 &= (\sin \beta \cos \gamma, -\sin \beta \sin \gamma, \cos \beta), \\
\notag \Vec{S}_5 &= (-\sin \beta \cos \gamma, \sin \beta \sin \gamma, \cos \beta), \\
\Vec{S}_6 &= (-\sin \beta \cos \gamma, -\sin \beta \sin \gamma, \cos \beta).
\end{align}
With $\mu = (m,n,i)$ denoting the general index of a spin site on the lattice, any spin of the ground state can thus be written as $\Vec{S}_{\mu} = \sigma^n \rho^m \Vec{S}_i$  with
\begin{equation}
\sigma = \begin{pmatrix}
-1 & 0 & 0 \\
0 & -1 & 0 \\
0 & 0 & 1
\end{pmatrix}, \quad \rho = \begin{pmatrix}
\cos \tfrac{\pi}{6} & \sin \tfrac{\pi}{6} & 0 \\
-\sin \tfrac{\pi}{6} & \cos \tfrac{\pi}{6} & 0 \\
0 & 0 & 1
\end{pmatrix}. 
\end{equation}
The ground state energy per spin of this state is given as
\begin{align}
	\notag E_{\text{VIII}} = \frac{1}{12} &\bigg[2 J_+ (1 + \cos^2 \alpha) + 8 \cos \beta (1 +\cos \alpha)\\
    & \notag + 4(2+J_{\times})\cos^2 \beta + \sqrt{3}J_+ \sin^2 \alpha\\
    & \notag + 8 \cos \gamma \sin \alpha \sin \beta + 2\sqrt{3} J_{\times} \cos^2 \gamma \sin^2 \beta\\
    & \notag + 4 J_{\times} \cos \gamma \sin^2 \beta \sin \gamma - 8 \sin^2 \beta \sin^2 \gamma \\
    & - 2\sqrt{3} J_{\times} \sin^2 \beta \sin^2 \gamma\bigg] \,.
\end{align}

\begin{figure}[t]
	\centering
	\includegraphics[width=1\columnwidth]{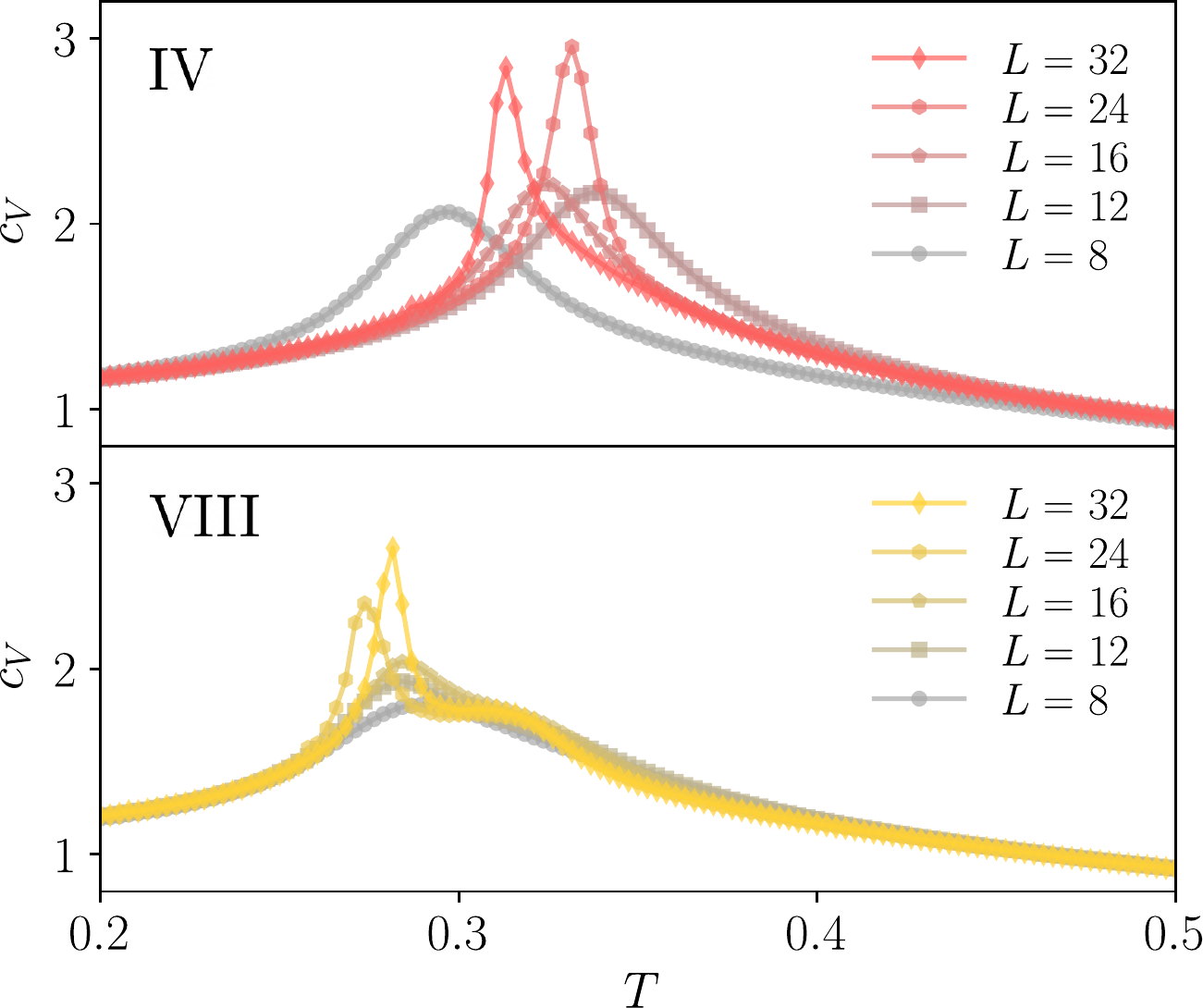}
	\caption{\textbf{Specific heat of spiral phases IV and VIII.} To exemplify the complicated behaviour of the specific heat of the spiral phases, we show $c_V$ traces for the spiral phase IV (top) and VIII (bottom). In both cases, the specific heat qualitatively changes drastically with the system size which indicates that the spiral ground state depends very sensitively on the system size.}
	\label{Fig:fss_cuboc_spirals}
\end{figure}

Specific heat traces for this spiral phase for different system sizes are shown in the bottom panel of Fig.~\ref{Fig:fss_cuboc_spirals}. 
However, due to the incommensurability of these spin spirals there are significant finite-size effects, 
making it hard to narrow down the location of the finite-temperature transition.

%%%%%%%%%%%%%%%%%%%%%%%%%%%%%%%%
\subsubsection{Spiral phase IX with $D_{3}^\sigma$ symmetry}
%%%%%%%%%%%%%%%%%%%%%%%%%%%%%%%%
For $J_+ = -0.2$ and $J_{\times} = 1.0$, grouping the $N = 864$ ground state vectors according to their direction leaves us with $M = 108$ unique spin vectors. These can be reduced further by identifying the symmetry group $G$ of this state, which is generated by a point reflection and rotations of $2\pi/3$ about the symmetry axis, i.e.\ $G = D_3^\sigma$, where the superscript $\sigma$ implies additional point reflection symmetry, i.e. $\sigma = \text{diag}(-1,-1,-1)$. The energy per spin is then given by a function of 26 variables and minimization yields $E_{\text{IX},\text{semi-analytical}} = -1.3272$ as compared to the Monte Carlo result $E_{\text{IX},\text{MC}} = -1.327(1)$ for the same parameters. 
For more details see the explicit description in the  Mathematica files of the Supplement \cite{Supplement}.

%%%%%%%%%%%%%%%%%%%%%%%%%%%%%%%%
\subsubsection{Spiral phase X with $D_{6}^s$ symmetry}
%%%%%%%%%%%%%%%%%%%%%%%%%%%%%%%%
For $J_+ = -0.05$ and $J_{\times} = 0.4$, there are $M = 32$ unique spin directions left after identifying the numerical ground state spin vectors that are close to each other. The symmetry group of this state is generated by a point reflection and by rotations of $\pi/3$ about the symmetry axis. The energy per site can be expressed as a function of three variables $\alpha_1, \alpha_2, z$ as
\begin{align}
	\notag E_{\text{X}} = \frac{1}{36} &\bigg[8 J_{\times} \left(z^2-1\right) \sin \left(\pi  \left(-\alpha_1+\alpha_2+\frac{1}{6}\right)\right) \\
 \notag &+2 J_{\times} \left(z^2-1\right) \cos \left(\pi  \left(\alpha_1-\alpha_2-\frac{5}{3}\right)\right) \\
 \notag &+2 J_{\times} \left(z^2-1\right) \cos \left(\pi\left(\alpha_1-\alpha_2+\frac{1}{3}\right)\right) \\
 \notag &-12 J_{\times} z^2-3 J_+-2 \sqrt{1-z^2}\cos \left(\pi  \left(\alpha_1+\frac{2}{3}\right)\right) \\
 \notag &+2 \sqrt{1-z^2} \cos \left(\pi\left(\alpha_1+\frac{5}{3}\right)\right) \\
 \notag &+2 \left(z^2-1\right) \cos \left(\frac{1}{3} \pi (3\alpha_1-3 \alpha_2-4)\right) \\
 \notag &+6 \left(z^2-1\right) \cos \left(\frac{1}{3} \pi  (3 \alpha_1-3
   \alpha_2+2)\right) \\
 \notag &+2 \left(z^2-1\right) \cos \left(\pi  \left(\alpha_1-\alpha_2-\frac{4}{3}\right)\right) \\
 \notag &+2 \left(z^2-1\right) \cos \left(\pi  \left(\alpha_1-\alpha_2+\frac{2}{3}\right)\right) \\
 \notag &+4 \sqrt{1-z^2} \left(\sqrt{3}
   \sin (\pi  \alpha_1)+\cos (\pi  \alpha_1)\right)\\
 &+12 \sqrt{1-z^2} \cos (\pi  \alpha_2) +12 z^2-24 z-12\bigg]. 
\end{align}
Minimization of this expression gives $E_{\text{X},\text{semi-analytical}} = -1.12813$ in agreement with the Monte Carlo result $E_{\text{X},\text{MC}} = -1.1280(3)$ for the same parameters. \\

%%%%%%%%%%%%%%%%%%%%%%%%%%%%%%%%%%%%%%%%%%%%%%%%%%%%%%%%%%%%%%%%%%
\section{Energy and magnetization cuts}
\label{sec:energy_cuts}
%%%%%%%%%%%%%%%%%%%%%%%%%%%%%%%%

We show Monte Carlo data of energy and magnetization for selected horizontal and vertical cuts through the phase diagrams of Figs.~\ref{Fig:phase_diagram_afm} (AFM nearest-neighbor interactions) and \ref{Fig:phase_diagram_fm} (FM nearest-neighbor interactions) in the main text. 

\subsection{AFM nearest-neighbor model}
\begin{figure}[t]
	\centering
	\includegraphics[width=1\columnwidth]{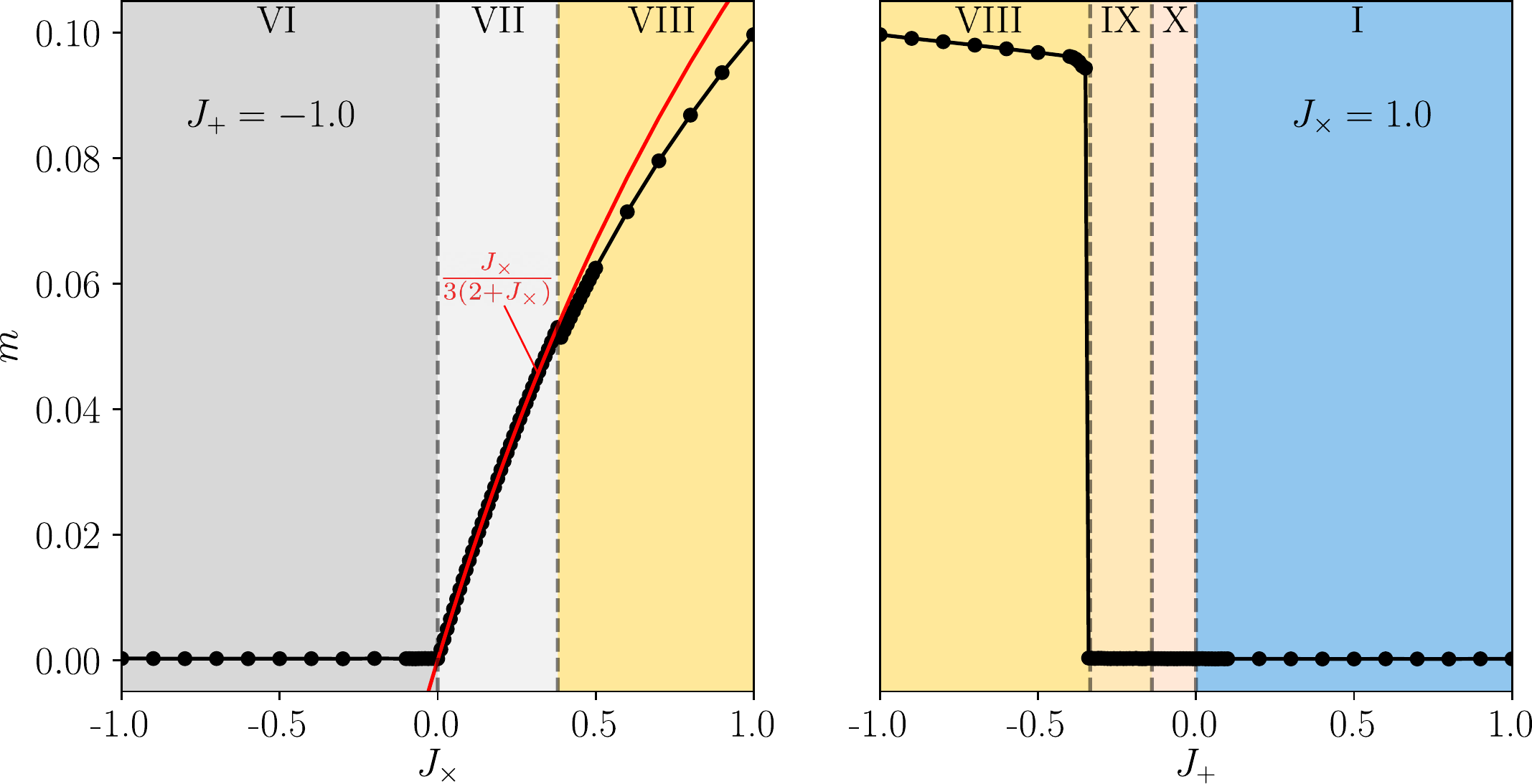}
	\caption{\textbf{Magnetization cuts for AFM nearest neighbor interactions.} 
	Magnetization cuts from Monte Carlo simulations for fixed $J_+ = -1.0$ (left) and $J_{\times} = 1.0$ (right) with the colorcode of the phases shown in Fig.~\ref{Fig:phase_diagram_afm} in the background. Only the deformed $120^\circ$ phase (light gray) and the spiral phase with $D_{12}$ symmetry (yellow) have non-zero magnetization. For the deformed $120^\circ$ phase the magnetization as a function of $J_{\times}$ can be expressed analytically as $m = J_{\times}/(3(2+J_{\times}))$ (red curve).}
	\label{Fig:mag_cuts_afm}
\end{figure}

Fig.~\ref{Fig:mag_cuts_afm} shows Monte Carlo magnetization cuts for fixed $J_+ = -1.0$, and $J_{\times} = 1.0$, respectively. The only two phases with non-zero magnetization are the $120^\circ$-d phase (VII) and the spiral phase VIII. In the deformed $120^\circ$ phase, the magnetization only depends on $J_{\times}$ and is given by $m_{120^\circ\text{-d}} = J_{\times}/(3(2+J_{\times}))$. 

\begin{figure}[t]
	\centering
	\includegraphics[width=1\columnwidth]{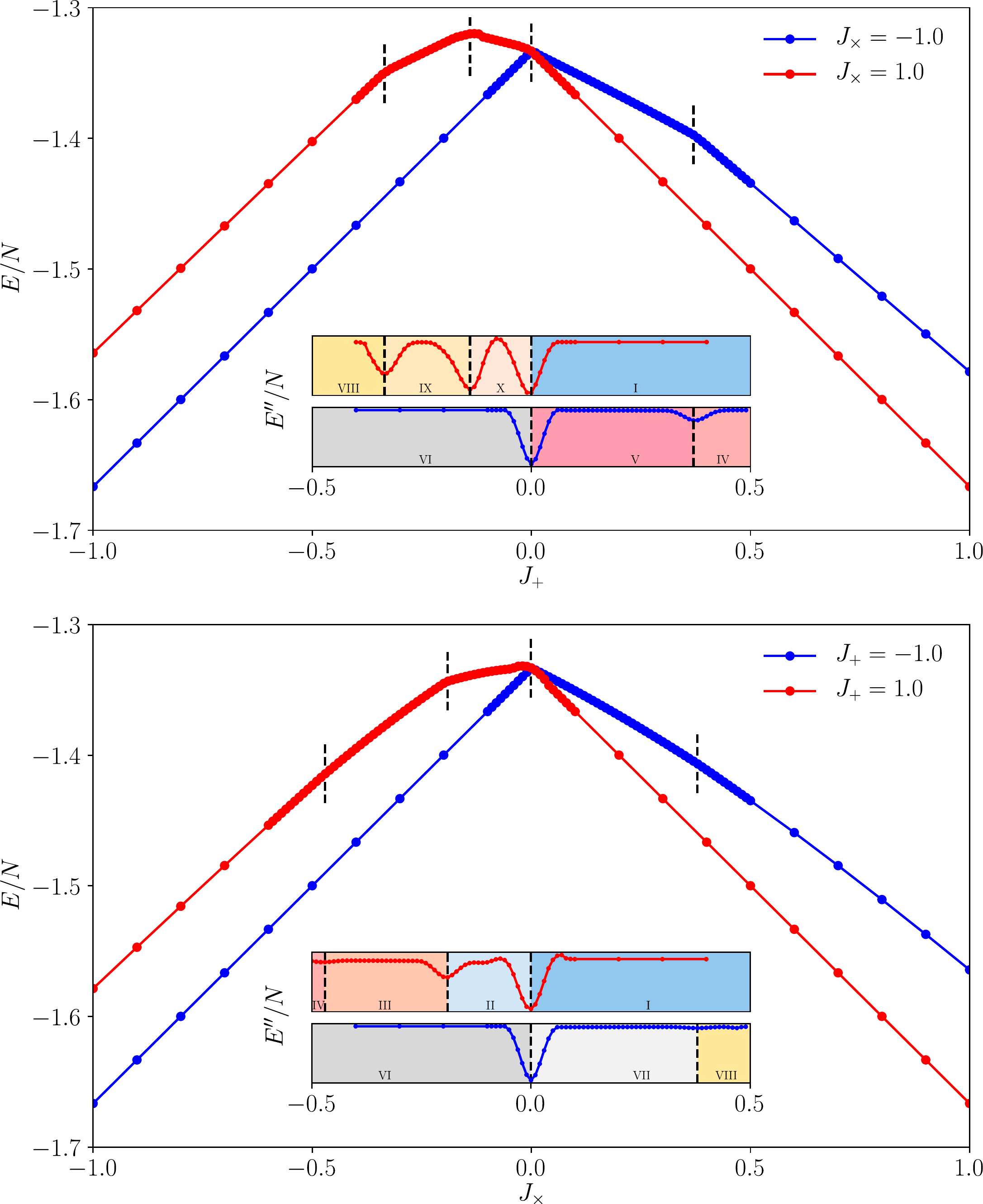}
	\caption{\textbf{Energy cuts for AFM nearest neighbor interactions.} Energy cuts from Monte Carlo simulations for fixed $J_{\times} = \pm 1.0$ (top) and $J_+ = \pm 1.0$ (bottom). The inset shows the second derivatives of the energy per spin with the phases shown in the phase diagram Fig.~\ref{Fig:phase_diagram_afm} in the background with the phase boundaries (dashed lines) obtained (semi-)analytically as described in the main text. The second derivatives of $E/N$ show sharp features exactly at all phase boundaries.}
	\label{Fig:energy_cuts_afm}
\end{figure}

In order to demonstrate the good agreement between the (semi-)analytically derived phase boundaries and Monte Carlo results, we show cuts of $E/N$ for fixed values of $J_+ = \pm 1.0$, and $J_{\times} = \pm 1.0$, respectively, from Monte Carlo simulations along with the corresponding second derivatives in Fig.~\ref{Fig:energy_cuts_afm}. The second derivatives of $E/N$ show sharp features exactly at all (semi-)analytically determined phase boundaries. 

%%%%%%%%%%%%%%%%%%%%%%%%%%%%%%%%
\subsection{FM nearest-neighbor model}
%%%%%%%%%%%%%%%%%%%%%%%%%%%%%%%%

\begin{figure}[t]
	\centering
	\includegraphics[width=1\columnwidth]{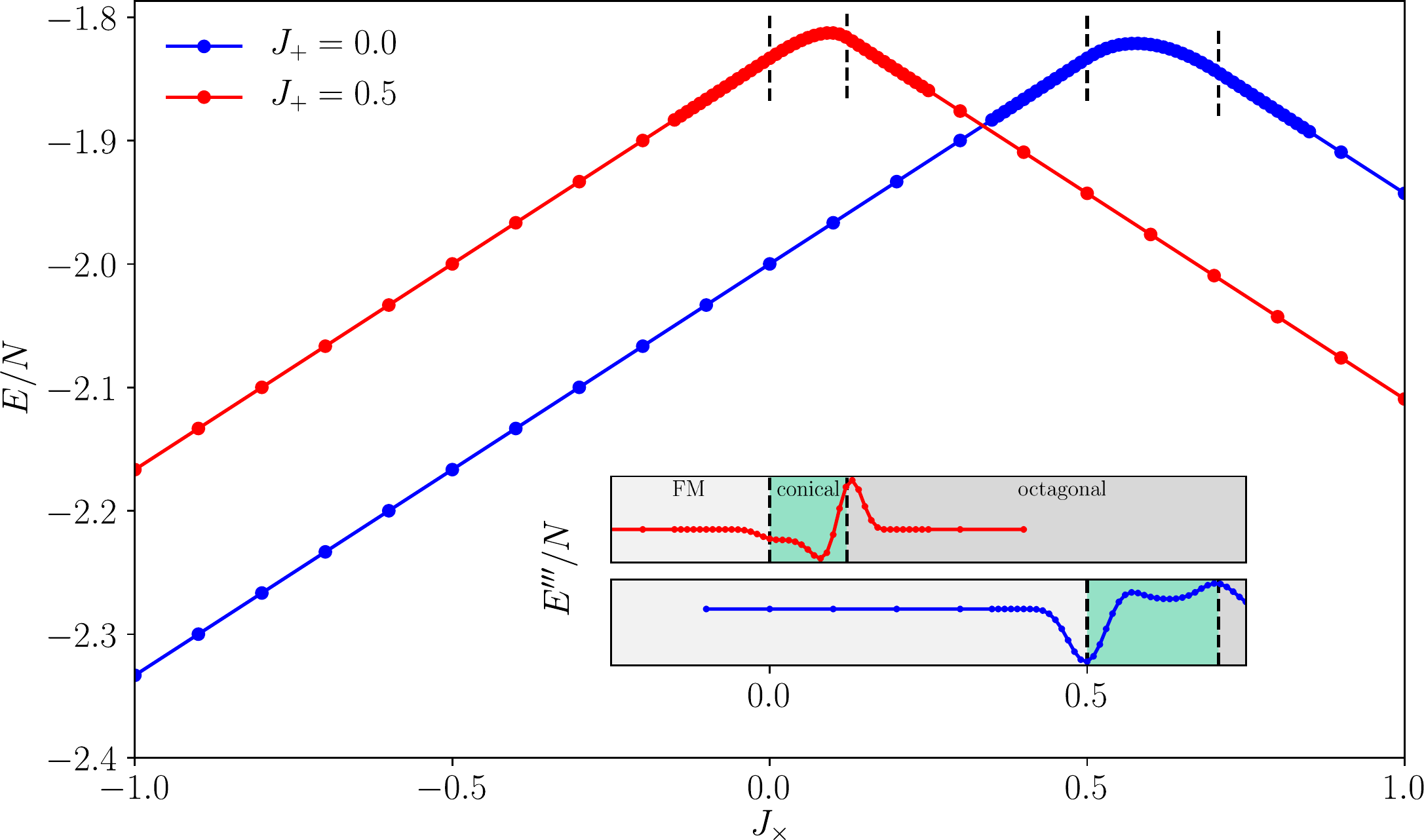}
	\caption{\textbf{Energy cuts for FM nearest neighbor interactions.} Energy cuts from Monte Carlo simulations for fixed $J_{+} = 0.0$ (blue) and $J_+ = 0.5$ (red). The inset shows the third derivatives of the energy per spin with the phases shown in Fig.~\ref{Fig:phase_diagram_fm} in the background with phase boundaries (dashed lines)  obtained analytically as described in the main text. The third derivatives of $E/N$ show sharp features exactly at all phase boundaries.}
	\label{Fig:energy_cuts_fm}
\end{figure}

Similarly good agreement between analytically determined phase boundaries and Monte Carlo results we find in the case of FM interactions. Fig.~\ref{Fig:energy_cuts_fm} shows cuts of $E/N$ for fixed values of $J_+ = 0.0$, and $J_+ = 0.5$, respectively, from Monte Carlo simulations along with the corresponding derivatives in Fig.~\ref{Fig:energy_cuts_afm}. Since the phase transition is one order higher than usual we consider the third derivatives of $E/N$. These show sharp features exactly at all (semi-)analytically determined phase boundaries.

\clearpage

\end{document}